\documentclass{report}
\usepackage{amsmath}  
\usepackage{float}
\usepackage{lscape}
\usepackage[toc,page]{appendix}
\usepackage{indentfirst}
\usepackage{blindtext} 
\usepackage{setspace} 
\usepackage{graphicx} 
\usepackage[margin=1in,letterpaper]{geometry} 
\usepackage{caption}
\usepackage[english]{babel}
\usepackage{fancyhdr}
\usepackage{cite} 
\usepackage[final]{hyperref} 
\usepackage{tensor}
\hypersetup{
	colorlinks=true,       
	linkcolor=blue,        
	citecolor=red,        
	filecolor=magenta,     
	urlcolor=blue
}

\begin{document}

\begin{figure}
\begin{center}
\includegraphics[width=0.9\textwidth]{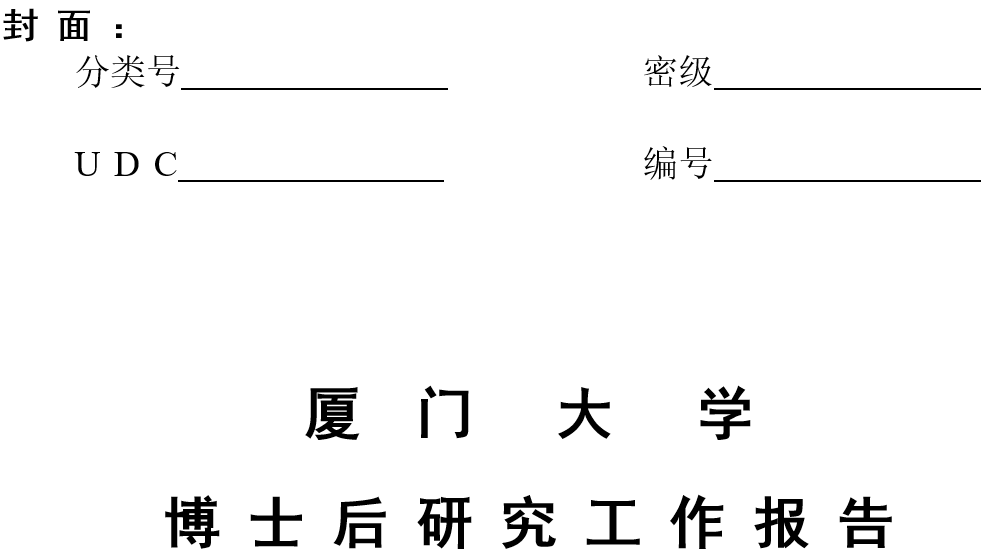}
\end{center}
\end{figure}
\vspace{0.5 cm}
\begin{center}
    \begin{LARGE}
\textbf{A model for the emission of SGRBs-GW from binary mergers} \\
\vspace{1cm}
By: Shad Ali (2021175031)\\ 
Supervised by: Prof. Tong Liu
\end{LARGE}
\end{center}

\vspace{0.5 cm}

\vspace{2.0cm}

\begin{figure}[H]
\begin{center}
\includegraphics[width=0.2\textwidth]{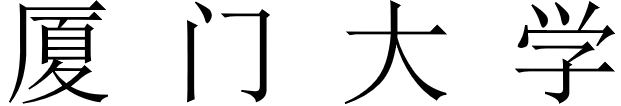}\\\today
\end{center}
\end{figure}

\newpage
\clearpage
\thispagestyle{empty}
\hfill
\newpage
 \begin{titlepage}
  \begin{large}
  \begin{center}
      \Large{A Model for the Emission of SGRB-GW from Binary Mergers}
  \end{center}
\vspace{2cm}

\begin{figure}[H]
\begin{center}
\includegraphics[width=0.8\textwidth]{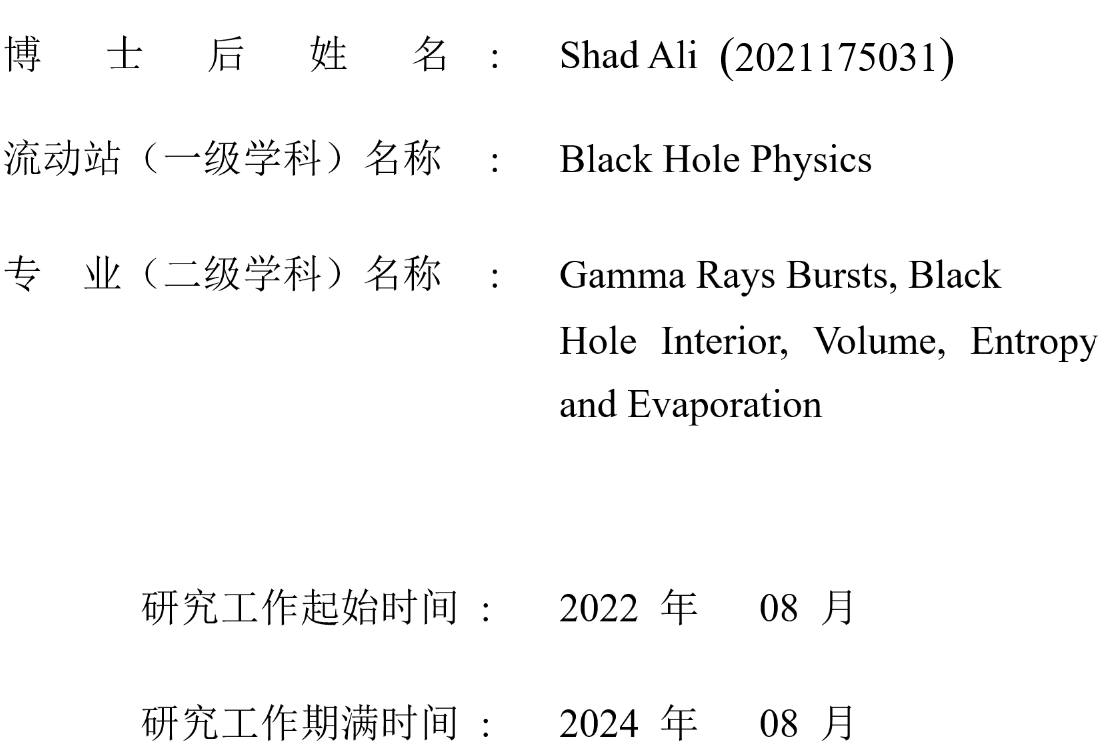}
\end{center}
\end{figure}
\vspace{6cm}
\begin{figure}[H]
\begin{center}
\includegraphics[width=0.2\textwidth]{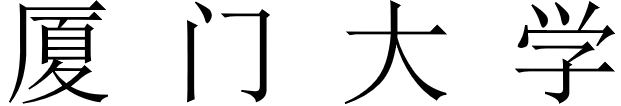}\\ \today
\end{center}
\end{figure}
\end{large}
\end{titlepage}
\newpage
\clearpage
\thispagestyle{empty}
\hfill
\newpage
\begin{figure}[H]
\begin{center}
\includegraphics[width=1.0\textwidth]{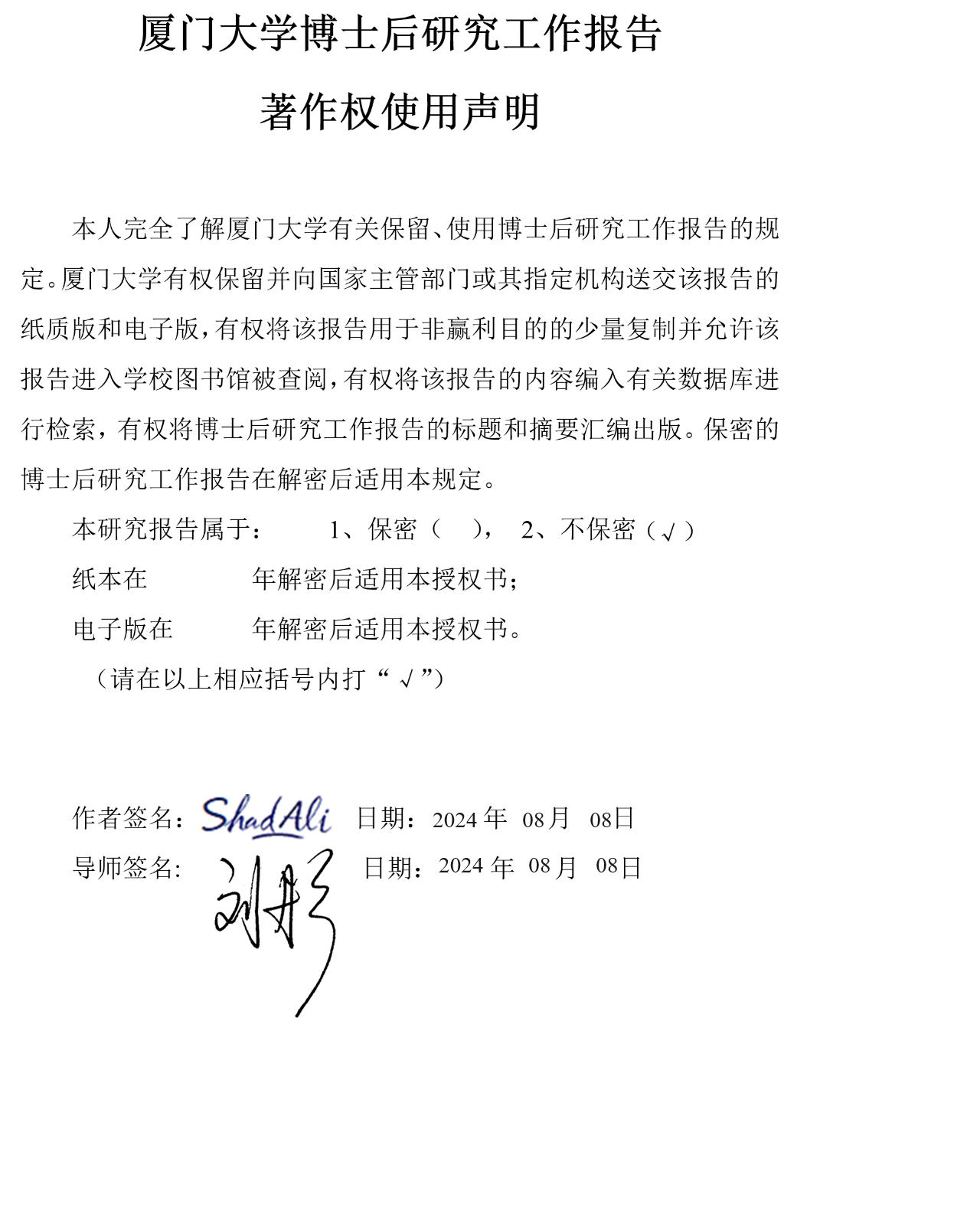}
\end{center}
\end{figure}
\newpage
\clearpage
\thispagestyle{empty}
\hfill

\newpage
\section*{Summary}
\addcontentsline{toc}{section}{Summary}
This report mainly consists of three parts divided into five chapters:

The first two chapters are related to work on Short GRB (SGRB) emissions from binary black hole mergers surrounded by a strong magnetic field. By introducing our model, we investigated the physics of the emission of  SGRBs from rotating and charged rotating BHs. A rapidly spinning, strongly magnetized neutron star (millisecond magnetar) is the primary source of strong magnetic fields ranging from $10^{13}~\rm to ~ 10^{16} G$. The decay of the magnetic field could power electromagnetic radiation, especially X-rays and gamma rays from NSs or NS-BH mergers as their primary sources. Considering the merger of compact bodies (NS-NS or NS-BH or BH-BH) to result in a rotating or charged rotating BH as a plausible source of SGRBs, we can obtain interesting results about their progenitor. 

Rotating BH with an accretion disk surrounded by a strong magnetic field establishes a global dipole moment to rotate the BH. For a particle in a circular orbit, any fluctuation from a circular path could lead to instability supported by a magnetic field. Such instabilities are mainly magneto-rotational (MRI) type instabilities. An MRI causes a random motion to the fluid displaced (faster toward the BH and slower outward) that generates fluctuations in both the magnetic field and the flow of the fluid. Due to the magnetic field, the outward-moving fluid is restricted by the magnetic field lines to ensure the rigid rotation of the fluid. As the rigid rotation increases, the field motion under the magnetic field assumes a cylindrical geometry along the z-axis. Due to these fluctuations and rigid rotations, each additional twist causes an increase in the cylindrical geometry. After many twists and restricted cross-sectional area, a narrow intense jet starts in the form of Poynting flux resembling the BZ mechanism. The nature of emitted flux strongly depends on the way that how the energy is generated in the disk. If the fluctuation is moderately large (long wavelength) then the outward force dominates and the MRI sets in. For the launch of a BZ jet, a poloidal magnetic field, and a fast-spinning BH are two key factors behind it. Accretion plays an important role in accumulating magnetic flux as well as spinning up the BH. 

In the case of charged rotating BH, Such events from a charged rotating BH with an accretion on to it could be more energetic and ultra-short-lived if the magnetic force dominates the accretion process because the attraction of ionized fluid with a strong magnetic field around the rotating BH further amplifies the acceleration of the charged particle via a gyromagnetic effect. Thus a stronger magnetic field and gravitational pull will provide an inward force to any fluid displaced in the radial direction and move it toward the axis of rotation with an increasing velocity. After many twists during rotation and the existence of restoring agents,  Such events could produce a narrow intense jet starts in the form of Poynting flux along the axis of rotation resembling the Blandford-Znajek (BZ) mechanism.

Using our model, we investigated both rotating and charged rotating BHs and obtained characteristic results (e.g., the remnant mass, magnetic field strength, luminosity, opening angle, viewing angle, and variation of viewing angle on the  SGRB luminosity detection) that have a nice coincidence with rare events having GW associated with EM counterparts. This study gives a new insight into events having strongly magnetized disks dominating the accretion process of energy extraction.

Chapter 3 is the basic study for understanding BH in GR and its interior. This introductory chapter is just added as the basis for the next chapter 4 where we reviewed the BH interior using the notion of  evaporation by entropy variation. In Chapter 4, we introduced the concept of black hole evaporation under Hawking radiation and found an interesting solution for the issue of information paradox. The basic idea of his work is based on a comparison between the interior quantum mode entropy of black hole, and Bekenstein Hawking entropy. As the interior quantum mode is directly related to the interior volume of the black hole so, determining the black hole's interior volume for existing quantum modes, one can investigate the quantum mode entropy.  Due to several problems, the interior volume of black has not been defined clearly. The main issues are the exchange of space-time coordinates and the existence of a black hole horizon. To define the interior volume one needs a space-like hyper-surface in the interior of BH. As the interior volume of a black hole is not the same as that of volume in flat space-time thus, a different technique is needed to investigate the BH interior volume. In 2015, Christodoulou and Rovelli defined the largest hyper-surface for the interior volume of BH bounding the maximal interior volume. After defining the largest space-like hypersurface, the interior volume of Schawarzschild BH is found to increase with Eddington time i.e., a time-dependent interior volume is determined. Similarly, this idea is also extended to other types of black holes.

A black hole of a given volume can have different observable internal configurations whose logarithm gives its entropy (a measure of the BH's initial state/information). So, using the volume relation, a time-dependent quantum mode entropy is also determined by Baocheng Zhang. This means that the proportionality relation of the black hole interior volume and quantum mode entropy with Eddington time is the main point to understanding the variation of information or statistical quantities in a BH. These investigations  exposed a great reality about the BH information and  evaporation that could be an interesting way to probe the problem of information paradox by comparing the interior and exterior entropy on BH.

This study of black hole interior volume and entropy gives a good concept for understanding the nature of black hole interior information and evaporation from its initial phases to and final. The evolution relation obtained from two types of entropy gives diverse understandings of the evaporation of BHs under Hawking radiation. Here the consideration of interior and exterior entropy will give the basis for probing the initial information of BH. The comprehensive approach advocated in this proposal will provide not only a new way of thinking about the information loss paradox but also guarantees to derive connections between BH formation and evaporation using different models by putting BHs in various dimensions. This research needs an extension for the astrophysical BHs by using modified gravity theories.

In chapter five, considering the phase transition of a d-dimensional $f(R)$ AdS black holes, we analyzed the effects of modified gravity factor $b$ associated with $f'(R)$ on the critical point parameters $(P_c, T_c, \nu_c, G_c)$, coexistence curves, difference in number densities of small and large black holes $(\frac{n_1-n_2}{n_c})$ and configuration entropy $S_{con}$. Following the analogy of black hole first order phase transition, we found our results in agreement with Reissner Nordstr$\ddot{o}$m and charge AdS black hole. Whereas the  black hole molecules crossing the critical point are found to have vanishing latent heat and become indistinguishable from outside observers. At the critical point, the pressure and temperature are independent of parameter $b$, whereas the free energy $G$ is dependent on it. A consistent result is also found for reduced parameters like specific volume, pressure, temperature, and free energy. Further analysis revealed that the parameter  $b$ doesn't have any effect on the coexistence curve, which shows that our results are consistent with that of Reissner Nordstr$\ddot{o}$m black hole and our trick is according to the fact that obeys the condition of Van der Waals phase transition. Finally, we also investigate the relation between the difference in number density and dimension $d$ of configuration entropy. Finally, using our results, the laws of BH Physics are also discussed.
\vspace{1cm}
\noindent\rule{16.5cm}{2.0pt}

\newpage
\clearpage
\thispagestyle{empty}
\hfill
\newpage
    
    


\newpage
\section*{Abstract}
\addcontentsline{toc}{section}{Abstract}

This report is divided into three main parts:

\begin{enumerate}
    \item The first two chapters discuss the emission of Short GRB (SGRB) from binary mergers surrounded by a strong magnetic field. By introducing our model, we investigated the physics of the emission of  SGRBs from rotating and charged rotating BHs. A rapidly spinning, strongly magnetized neutron star (millisecond magnetar) is the primary source of strong magnetic fields ranging from $10^{13}~\rm to ~ 10^{16} G$. The decay of the magnetic field could power electromagnetic radiation, especially X-rays and gamma rays from NSs or NS-BH mergers as their primary sources. Considering the merger of compact bodies (NS-NS or NS-BH or BH-BH), we can obtain interesting results.

    \item In the next two chapters, we reviewed the BH interiors to understand the nature of black hole interior information and evaporation from its initial to final phases via entropy variation. The evolution relation obtained from two types of entropy gives diverse understandings of the evaporation of BHs under Hawking radiation.

    \item The fifth Chapter is related to BH configuration (information) entropy and the thermodynamic phase transition of  $f(R)$ BH. Here, we consider a d$-$dimensional black hole (BH) in $f(R)$ gravity and analyze the effect of modified gravity on critical point parameters, the difference in number densities, and configuration entropy during the BH phase transition phenomenon. These results were also compared with charged AdS BH, the holographic dual of van der Waal's fluid, and hence the BH in modified gravity.
\end{enumerate}

\textbf{Keywords:} Black hole mergers, magnetic field, disks,  GW, GRB, thermodynamics, Hawking radiation, Interior volume, entropy,  evaporation, phase transition, and Co-existence curves.

\noindent\rule{16.5cm}{2.0pt}

\newpage
\clearpage
\thispagestyle{empty}
\hfill

\newpage
\section*{Publication List}
\addcontentsline{toc}{section}{Publication List}
\begin{enumerate}
    \item SGRBs Emission from the merger of binary black holes. \\\textbf{DOI:} \url{10.1016/j.jheap.2023.05.001}.   \\ (Journal of High Energy Astrophysics)
    \item Configuration entropy and thermodynamics phase transition of black hole in f(R) gravity.\\ \textbf{DOI: }\url{10.1016/j.hedp.2024.101105}\\  (Journal of High Energy Density Physics)
    \item The CR Volume for BH and the Corresponding Entropy Variation: A Review.\\ \textbf{DOI: }\url{10.1016/j.newar.2024.101709}.     \\(Journal of New Astronomy Reviews)
    \item Short-duration gamma-ray bursts from Kerr-Newman black hole mergers      (Under review in European Physics Journal C)
\end{enumerate} 

\noindent\rule{16.5cm}{2.0pt}

\newpage
\clearpage
\tableofcontents
\vspace{1.0cm}
\noindent\rule{16.5cm}{2.0pt}
\listoffigures
\vspace{1.0cm}
\noindent\rule{16.5cm}{2.0pt}

\newpage

\chapter{Short Gamma Ray Bursts from binary black holes merger}

In this chapter, we introduced a model related to astronomical events having the co-detection of GW associated with Short Gamma Ray Bursts (SGRBs). Study shows that the existence of magnetized accretion disks is responsible for creating the events of GWs associated with electromagnetic (EM) counterparts from binary BHs mergers. Our model leads from space-time instability to the emission of EM radiations as its counterparts. Starting from Maxwell's stress tensor, we found the condition for the growth of instability that is proportional to the angular velocity and is independent of the magnetic field strength. Considering the event GW150914 with the final product as a Kerr BHs in an equatorial plan, we discussed the motion of particles under an effective potential on circular orbits around it and determined its frequency, redshift factor, and epicyclic frequency. Next, considering the evolution of mass and angular momentum, we calculated the remnant mass, rotational energy, and the magnetic field strength acting along the axis of rotation. The accretion rate and its luminosity are determined by the restoring and shear forces of the magnetic field. The energy extraction efficiency of the flow is determined to be very low. Results show the presence of a weak transient caused by magneto-rotational instability with a strong poloidal magnetic field that causes turbulence in the accretion disk onto black holes.

\section{Introduction}

GRBs are the most powerful, rapid, and intense flashes of radiations (with energy in the range of Gamma rays) resulting from the merging and collapsing of compact astronomical objects were first discovered in 1960 by the Vela Satellite and published in 1973 \cite{Klebesadel:1973iq}. Their energy ranges from $~	100 keV - 1MeV$ with an average rate of one event per day over the whole cosmological distances in the sky \cite{Fishman:1995st}.  This prompt emission is followed by broadband afterglow emission with energies ranging from X-rays to radio bands \cite{vanParadijs:1997wr, Costa:1997obd, Frail:1997qf, Heng:2008nr}. The afterglow emission can be observed for several weeks to months \cite{Wijers:1997xu} for more detail, also see the Refs. \cite{Zhang:2018ond, Kumar:2014upa}. Based on their emission from their progenitors, GRBs are classified into two classes as; 

 Long Gamma-Ray Bursts (LGRBs) and their sub$-$class Ultra-LGRBs (ULGRBs) are claimed to be originated from the core collapse of massive stars and supernovae (SNe) that results in born a Black Hole (BH) or Neutron star (NS) in the center of parental body \cite{Woosley:1993wj, MacFadyen:1998vz, Woosley:2006fn, Kumar:2014upa, Perna:2018wft, Dagoneau:2020mbg} both of which needs different mechanism. In BH formation, the fallback of matter during the collapse process  triggers an instability in the BH hyper-accretion disk to power a relativistic jet from the envelope by neutrino-antineutrino annihilation mechanism to release the gravitational energy of BH. \cite{Beloborodov:2008nx, Hanami:1997yy, Popham:1998ab, Liu:2016olx,Janiuk:2010xs, MacFadyen:2004ab, Liu:2017rwh}. In the case of NS formation, the spin down of NS (with a period millisecond  rotation) with a strong magnetic field (millisecond magnetar) extracts the rotational energy by electromagnetic torque that can produce a long GRB, even super-luminous supernovae (SNe) with a duration greater than $2 sec$ and softer spectra \cite{Evans:2008wp, Duncan:1992hi}. Their variability (on a millisecond timescale for long duration amount of energy (about $ 10^{52} ergs$) supports the idea that compact objects of stellar masses power them. Whereas, Short Gamma Ray Bursts (SGRBs) originated from the merger of double NS or BH$-$NS \cite{Eichler:1989ve, Narayan:1992iy}. For more detail also see the latest review \cite{DAvanzo:2015kdp} and references therein. The  duration of these Bursts from the first BATSE catalog was found from $1-2 sec$. SGRBs also show their hardness in their spectral ratio \cite{Kouveliotou:1993yx}.
 
Mainly GRBs are emitted by four mechanisms that are either Neutrino Dominated Accretion Flow (NDAF) in case of high accretion rate at very high temperature \cite{Liu:2017kga, Wei:2022ygk}, Advection Dominated Accretion Flow (ADAF), in case of low accretion rates at low temperature\cite{Yuan:2014gma, Liu:2022cph, Narayan:2008bv}, Blandford Znajek (BZ) Mechanism \cite{Blandford:1977ds} to powers a jet due to Poynting flux from the central engine. This mechanism also determines the presence of a magnetic field around the BH as an essential factor in the whole process. Another mechanism is accretion from a Highly Magnetized disk by using the "Magnetic Tower mechanism" introduced by Lynden Bell \cite{Bell:1996ad} states that if the existence of magnetic pressure restores the medium displaced by turbulence from the accretion disk then the matter splays along the magnetic field lines to form a cylindrical geometry that further increases with the rotational field twists and finally after many twists, it creates a narrow sharp outgoing jet.

In literature, it is claimed that SGRBs originated from the coalescence of double NS or NS-BH. This scenario was first observed during the merging of Binary NS (BNS) event GW170817 \cite{LIGOScientific:2017vwq, Savchenko:2017ffs}. The co-detection of GW associated with a GRB provided new insight into the co-observation of gravitational and electromagnetic events in astrophysics and astronomy. By determining the merger rate, the co-detection rate of GW-SGRBs is determined. However, it is also claimed that in the presence of a strong magnetic field, a small fraction of SGRBs progenitor is the binary BHs mergers that were first observed during the event GW150914. It was stated that the orbital eccentricity has an excellent effect on the GW spectra and hence on the GW-SGRBs co-detection rate. The power spectra of SGRBs from compact binaries are greatly suppressed by low frequencies. Many other retractions constrain the GRBs detection \cite{Tan:2017hqa}. In the case of the high mass system, all three cases of double NSs, double BHs or NS-BH merger can be considered for SGRBs emission but their small opening angle and short time duration only gamma rays pointing to the earth are detected that is why a very small fraction of SGRBs associated to GWs could be observed. Besides improving detector sensitivity, it also needs an alternative mechanism to observe SGRBs' emission rate.

On September 14, 2015, the two detectors of the LIGO simultaneously observed a transient GW signal that swept upwards in the frequency range from 35 to 250 Hz with a peak gravitational-wave strain of {$ 1.0\times10^{-21}$}. This signal has matched the waveform predicted by general relativity for the inspiral and merging of BHs pair and the ringdown of the resulting single BH \cite{LIGOScientific:2016aoc}. The Fermi Gamma-ray Burst Monitor satellite observatory data shows that the event of GW150914 (from the pair of two coalescing BHs) is related to a gamma-ray burst that has localization consistent with the direction of event GW150914 \cite{Meegan:2009qu}. These radiations (above 50 keV) are claimed to originated from a weak transient source and lasted for about 1 sec \cite{Connaughton:2016umz, Janiuk:2016qpe}. This is the first event detection of some GRBs from a binary BH merger event. According to general relativity, the final stage of binary coalescence is Kerr BH. Einstein's equation also predicts a unique case of these quasi-circular inspirals as a function of the mass and spin of the two plunging BHs. Using the early-stage coalescence data, we introduced a model to verify the remnant mass, accretion rate, rotational energy, magnetic field, and luminosity during the emission process of electromagnetic radiation. A circumbinary disk is required for the interpretation of the EM counterparts from binary BH mergers. BH-BH binaries embedded in AGN disk or gas remaining from their progenitors are possible scenarios for the origin of such a circumbinary disk as discussed in Ref.  \cite{Khan:2018ejm}. Where the two merging BHs are considered non-rotating BHs with magnetic field $\sim 10^{12} G$. In comparison to the NDAF mechanism discussed in Ref. \cite{Janiuk:2016qpe, Janiuk:2017tbq},  the main point of our investigations is that we have considered this event to be caused by pair-instability due mass gap of the two hierarchical BHs and sketch the model given in the sub-section (\ref{plan}) below.

\subsection{Sketch of Plan}\label{plan}

A hyper-accretion disk around a stellar-mass BH is a plausible source of the central engine to jet the GRBs as claimed by many authors but the main question is how to set the turbulence inside the hyper-accretion disk that results in jet GRBs. In search of the fact, the magnetic-field-driven disturbance called Magneto-Rotational Instability (MRI) is claimed to actively operate in hyper-accretion disk and causes the angular momentum transport in it \cite{Velikhov:1959xx, Balbus:1991ay, Balbus:1998az}. Actually, the MRI produces random motion (in which the fluid moves faster toward the BH and slower outward) that generates fluctuations in both the magnetic field and the flow of the fluid. The magnetic field lines restrict the fluid displaced outward and ensure the rigid rotation of the fluid that increases along the field direction to assume a cylindrical geometry.  Due to these fluctuations and rigid rotations, each additional twist causes an increase in the cylindrical geometry. After many twists and restricted cross-sectional area, a narrow intense jet starts in the form of Poynting flux resembling the BZ mechanism. The nature of emitted flux strongly depends on the way that how the energy is generated in the disk. If the fluctuation is moderately large (long wavelength) then the outward force dominates and the MRI sets in. For the launch of a BZ jet, a strong poloidal magnetic field, and a fast-spinning BH are two key factors behind it. Accretion plays an important role in accumulating magnetic flux as well as spinning up the BH. As shown in Refs. \cite{Proga:2006ir, Liu:2012ek}, a magnetic barrier will build up when the radial magnetic force can counteract the gravitational force of the BH. These effects and the instabilities induced may be more effective in the growth of magnetic fields. The accretion disk (with a strong magnetic field) leads to further amplification of the magnetic field due to turbulent dynamo action inside the disk (up to the level that may be in excess of $10^{16}$ G) to start an intense BZ jet.

The turbulence driven by the magneto-rotational instability can be described by Maxwell's stress tensor that has diagonal and off-diagonal elements as
\begin{equation}\label{Maxeq}
T_{ij} = \epsilon_{o}\left( E_{i}E_{j} - \frac{{\delta_{ij}E}^{2}}{2} \right) + \frac{1}{\mu_{o}}\left( B_{i}B_{j} - \frac{{\delta_{ij}B}^{2}}{2} \right),
\end{equation}

with
\[\begin{matrix}
\delta_{\text{ij}} = & \begin{matrix}
\begin{matrix}
1 & i = j ,\\
\end{matrix} \\
\begin{matrix}
0 & i \neq j ,\\
\end{matrix} \\
\end{matrix} \\
\end{matrix}\]
So, for \(i = j\) Maxwell's stress tensor becomes
\begin{equation}\label{magpres1}
P_{mag}= \frac{1}{2\mu_{o}}B^{2},
\end{equation}

and for \(i\neq j\), it gives the shear stress
\begin{equation}\label{tenshear1}
T_{shear} = \frac{1}{\mu_{o}}B_{ij},
\end{equation}
it represents the deformation (shear) along off-diagonal elements. The stress \(P_{\text{mag}}\) always resists deformation and acts as returning/restoring force to the perturbed fluid elements. In a differential rotating system (like BHs), this tension force can lead the system to instability. The magnetic field is an effective source for this shear-type instability to generate turbulence in the accretion disk that results in an inflow or outflow of fluid \cite{Nattila:2021qag}. 

\begin{figure}
\begin{center}
\includegraphics[width=0.4\textwidth, height=1.6in]{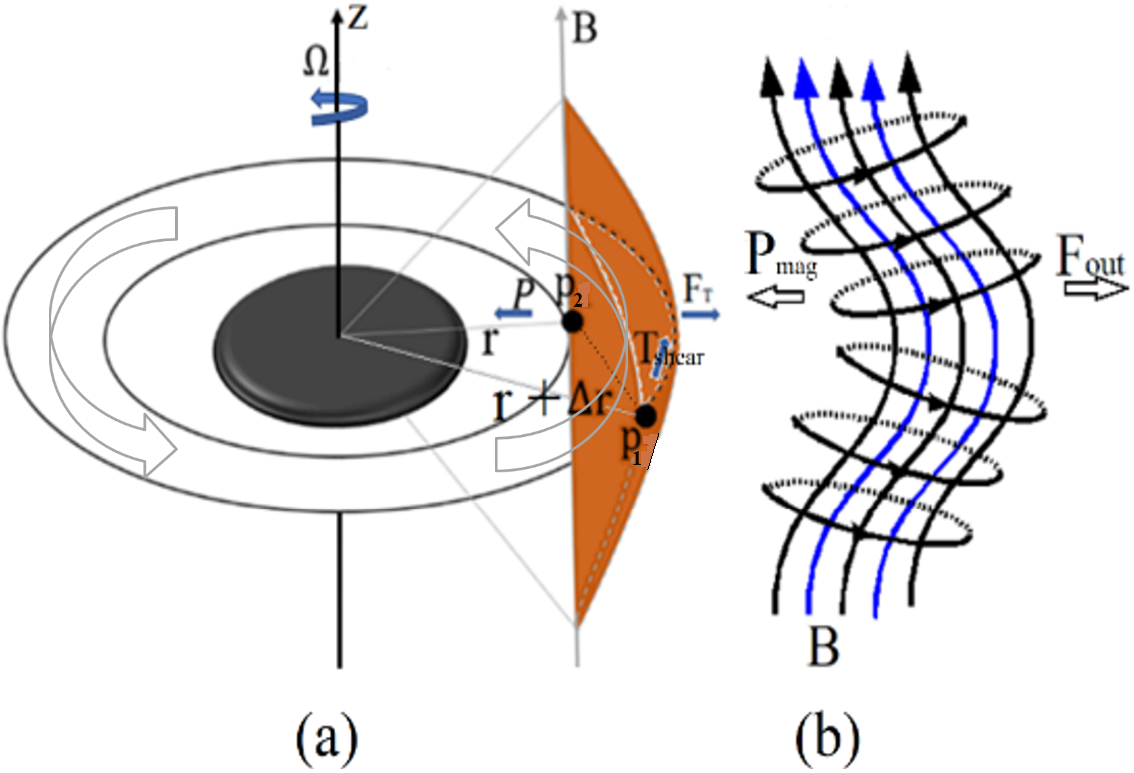}
\caption{\textbf{(a).} Sketch for turbulence due to a magnetic field. The fluid displaced from $p_1$ to $p_2$ due to instability feels a magnetic tension to restore the fluid while the shear magnetic field component spins up the BH. \textbf{(b).} The dynamics of the magnetic field under tension.}
\label{image-1.1}
\end{center}
\end{figure}

Let the total force \(F_{T}\) displaces the fluid to the outward direction in the accretion disk then the magnetic tension (\ref{magpres1}) acts in the opposite direction and is roughly written as 

\begin{equation}
    P_{mag}=\frac{\left(\frac{B^2}{4\pi \mu_o}\right)}{R},
\end{equation}

where $R$ is the radius of curvature of the magnetic field $B$. As a demonstration of the instability, we consider an axially symmetric steady rotating system with cylindrical geometry having vertical magnetic fields of magnitude $B$ as shown in Fig. (\ref{image-1.1}).

The global magnetic field in accretion disks is widely considered for angular momentum transport, jet formation, hole-to-disk interaction, and a case of study for understanding the instabilities that could originate the various kinds of active phenomena in the universe. Consider a rotating disk with conserved angular momentum proportional to \(r^{2}\Omega\), where \(r\) is the distance from the center of BH and \(\Omega\) is the angular velocity of a Keplerian-type disk. The efficiency of extracting energy (\(\varepsilon\)) from the rotating disk is equal to the ratio of the angular velocity of BH  to the angular velocity of the disk, i.e.,
\begin{equation}
\varepsilon=\frac{\Omega_f}{\Omega},
\end{equation}
here \(\Omega_f\) is the angular velocity of the field. This means the emission efficiency (\(\varepsilon\)) of the rotating disk depends on the inertia of the disk. For \(\Omega_f \simeq \frac{1}{2}\Omega\), the magnetic field is usually frozen in the disk so, it is obvious that \(\varepsilon\sim\frac{1}{2}\) is the maximum expected power out from a BH. The rest of the rotational energy will be used for the increase of entropy or the irreducible mass \cite{MacDonald1982}.

Consider Fig. (\ref{image-1.1}), let the particle be displaced in a radial direction from point \(p_{1}\) to \(p_{2}\) with a change in distance from \(r\) to \(r + \Delta r\) and the change in the angular momentum is \(\Omega^{'}\left( r + \Delta r \right) < \Omega\left( r \right)\) so, the particle will lag behind at \(p_{2}\). Following the conservation law, we can write \(\Omega > \Omega^{'}\) and hence the particle at the new position will rotate slowly. Similarly, for inward motion, the particle will speed up i.e., the particle has variable speeds while moving in or out due to instability.

Considering the astronomical event of GW150914 as a cause of the binary merger, we elaborate on some of its characteristics by using the data published in Refs. \cite{LIGOScientific:2016aoc, Connaughton:2016umz}. The structure of this chapter is such that in the next section (\ref{instability}), we explain the cause of the instability that led to the emission of the event GW150914. In section (\ref{potential}) the potential required for the body to power the jet of electromagnetic radiation and particle motion under this potential is discussed. The remnant mass and its energy and the associated magnetic field are discussed in section (\ref{Energy}) while the rate of accretion and luminosity are calculated in section (\ref{accretion}). Finally, the section (\ref{conclusion}) is dedicated to the conclusion and discussion of the work done.

\section{Demonstration for instability condition}\label{instability}

We discussed above that the particle of the fluid in an accretion disk at different radii are acted upon by different forces so, the particles of the fluid are moving at different speeds. We are trying to construct a condition for the instability that causes turbulence in the accretion disk. Let us consider the magnetic tension acting on the particle at \(p_{2}\) which can be written as
\begin{equation}\label{magpress2}
P_{mag} = \frac{B^2}{2\mu_o} =  \frac{B^2}{4\pi \mu_o R},
\end{equation}

It is a restoring magnetic force to the deformation caused by the outward force. \(R\) is the radius of magnetic field curvature and can be written as \(R = \frac{\lambda^{2}}{\Delta r}\), here \(\lambda\) is the wavelength of perturbation in the vertical direction and \(\sqrt{\frac{B^{2}}{4 \pi \mu_o}} \equiv c_{\text{Al}}\) is the speed of Alfve'n wave. So, the magnetic tension becomes

\begin{equation}\label{magpres3}
P_{mag} = \frac{B^{2}}{4\pi\rho}\frac{\Delta r}{\lambda^{2}} \approx \left( \frac{c_{Al}}{\lambda} \right)^{2}\ \Delta r,
\end{equation}
Now consider the displacement \(\Delta r\) at conserved angular velocity \(\Omega = \sqrt{\frac{\text{GM}}{r^{3}}}\) (for an unperturbed Keplerian disk) with $M$ as total mass, the outward centrifugal force increase rate for unit mass is \(\Delta F_{c}\sim\Omega^{2}\Delta r\) and due to this increased centrifugal force the gravitational force for unit mass will decrease by the amount
\begin{equation}\label{force}
\Delta F_{g} = F_{1} - F_{2}=\frac{\text{GM}}{r^{2}} - \frac{\text{GM}}{\left( r + \Delta r \right)^{2}},
\end{equation}
Here $F_1$ and $F_2$ are the forces at points $p_1$ and $_2$. Using binomial expansion for the second term on the right side of this equation, we get the form

\[\Delta F_{g} \approx \frac{2GM}{r^{3}}\Delta r \approx 2\Omega^{2}\Delta r,\]

So, the total inward force (\(F_{\text{T}}=\Delta F_c +\Delta F_g\)) acting on the perturbed particle is of order \(\sim 3\Omega^{2}\Delta r\). As we have to demonstrate the instability that causes the turbulence i.e., the outward force must be dominated during the whole process. So, the condition of instability can be written as

\begin{equation}
3\Omega^{2} > \left( \frac{c_{\text{Al}}}{\lambda} \right)^{2} \approx \lambda > \frac{c_{\text{Al}}}{\Omega},
\end{equation}

Now, the growth rate of the instability can be estimated by considering the time scale as
\begin{equation}
\tau = \frac{\pi \lambda}{2 c_{Al}} > \frac{1}{\Omega}\sim \frac{\pi}{2}\sqrt{\frac{r^{3}}{\text{GM}}},
\end{equation}
This shows that the growth rate depends on angular velocity and is independent of the magnetic field i.e., the greater the angular velocity, the smaller will be the growth time of the turbulence and vice versa. We can also say that the growth rate increases with angular momentum transport along the line of force that occurs with Alfve'n speed. Note that the instability sets in for perturbations with shorter wavelengths when the magnetic fields are weak.   Such a quantitative growth rate of the disk is also discussed in Ref. \cite{Balbus:1991ay} where the dispersion relation is shown to have the most rapid wave number in the thin disk where the growth rate is about \(75\% \Omega\).

\section{The effective potential of the particle around the black hole}\label{potential}

Generally, we know that the position of an event horizon around the BH is determined by its radius which can also be imagined as the breakup between the BH space-time and its surrounding space-time. Besides, a characteristic radius also existed near the BH ``called marginally stable circular orbit \({(r}_{\text{ms}})\)'' that determines the edge of the standard accretion disk and its efficiency of converting the accretion power to the radiation power denoted by \(\left( \eta \right)\). For getting these factors, one can start with the spacetime metric. Like in the case of rotating BHs, several authors have determined \(r_{\text{ms}}\) from Kerr BH geometry \cite{Wald:1984rga, Shapiro:1983du, Pugliese:2011xn}. The Kerr metric is rotating uncharged and axis-symmetric about the polar axis (independent of \(\phi\) ) and \(a\) is the angular momentum per unit mass measured in $cm$. For a Kerr BH, $\frac{J}{M}<1$ and \(a = 1 \Rightarrow J = M\) gives extreme Kerr metric. Setting \(\Delta = 0\), the horizon or largest root is
\begin{equation}\label{horizon}
r_{H} = \frac{r_g}{2}\left(1 + \sqrt{1 - a^{2}}\right),
\end{equation}

here $r_g=\frac{2GM}{c^2}$, An important feature of Kerr geometry is that one can separate variables by using the Hamiltonian Jacobi technique and construct geodesics for determining the equation of motion for a particle moving near the BH. Let us consider a test particle motion in an equatorial plane (a place where \(\theta = 90^{o}\) to the accretion disk). So, introducing the Lagrangian $(L)$ of Kerr geometry one can define the conserved energy and angular momentum $(p_t=\partial_{\overset{.}{t}}L=-E, p_{\phi}=\partial_{\overset{.}{\phi}} L=l)$ as a function of  \(\overset{.}{t}\) and \(\overset{.}{\phi}\) (here the dot $(.)$ the differentiation took over an affine parameter\(\lambda\)). Solving these two equations of conserved quantities for \(\overset{.}{t}\) and \(\overset{.}{\phi}\) and setting in normalization condition as \(g_{\alpha\beta}p^{\alpha}p^{\beta} = - m^{2} = \frac{L}{2}\), \(m = 1\), one gets the conservation law \cite{Aliev:2013jqz, Bardeen:1972fi}.

\begin{equation}
\frac{1}{2}{\dot{r}}^{2} + V_{\text{eff}} = 0,
\end{equation}
with
\begin{equation}
V_{eff} = - \frac{M}{r} - \frac{M\left( l - a E \right)^{2}}{r^{3}} + \frac{1}{2r^{2}}\left( l^{2} + \left( 1 - E^{2} \right)\left( {r^{2} + a}^{2} \right) \right),
\end{equation}
as the effective potential.  This equation shows the effective potential of particles in the radial direction on an equatorial plane and at some point on the radial path \(\dot{r}\) must vanish which will give the location of a stable circular orbit. At that point, the \(V_{\text{eff}}\) can be treated to give the conditions
\begin{equation}
V_{\text{eff}} = 0, \qquad \partial_{r}V_{\text{eff}} = 0,
\end{equation}
solving the two parts one can get the energy and angular momentum for the circular path as
\begin{equation}
E_{circ} = \frac{r^{2} - 2\ M\ r \pm a\ \sqrt{\text{Mr}}}{r\left( r^{2} - 3M\ r \pm 2a\ \sqrt{\text{Mr}} \right)^{\frac{1}{2}}},
\end{equation}
\begin{equation}
l_{circ} = \frac{\pm \sqrt{Mr}\left( r^{2} \mp 2a\ \sqrt{\text{Mr}}\ + a^{2} \right)}{r\left( r^2 - 3M r \pm 2a\ \sqrt{Mr} \right)^{\frac{1}{2}}},
\end{equation}

The vanishing denominator of this equation gives the radius of the particle's motion along a partially parabolic path satisfying the condition of \(v \ll c\) for the particle. For \(a = 0\) we get \(r = 3M\) which corresponds to a marginally stable circular photon orbit. This case was first discussed in \cite{Bardeen:1973gs}. in Ref. \cite{Bardeen:1972fi} and later by Novikov and Frolov in \cite{Novikov:1989}. The particle moving in the nearest unstable orbit of the BH will have a speed approximately equal to the speed of light. The radius of this circular orbit is

\begin{equation}
r_{particle} =2M[1+cos\{\frac{2}{3}cos^{-1}(\mp\frac{a}{M})\}],
\end{equation}

The maximum binding energies of these orbits around Kerr black compared to Schwarzschild BH are significantly higher i.e., \(\sim42\%>\sim 6\%\)  \cite{Kato:2008xy, Wald:1984rga}.

Consider a circular motion in an equatorial plane $(\theta=\frac{\pi}{2})$, where  an observer (in stationary far away space-time) is observing the angular velocity of a particle moving around the Kerr BH. By definition of angular velocity as 
\begin{equation}\label{angfre}
\Omega_{Kerr}=\frac{d\phi}{dt}=\frac{d\phi}{d\tau}\times\frac{d\tau}{dt}=\frac{U^{\phi}}{U^t},
\end{equation}
for such stationary space-time, the four-velocity of the system can be written as 
\begin{equation}
U^\mu=(U^t,0,0,U^\phi)=(U^t,0,0,U^t\Omega_{Kerr}),
\end{equation}

Using the normalization condition, we can write as 
\begin{equation}
g_{\mu\nu}U^\mu U^\nu=1=g_{tt} U^t U^t + 2g_{t\phi} U^tU^{\phi}+g_{\phi\phi} U^{\phi}U^{\phi},
\end{equation}

or
\begin{equation}\label{velo}
U^t=(g_{tt} +2 g_{t\phi} \Omega_{Kerr}+g_{\phi\phi} \Omega^2_{Kerr})^{-\frac{1}{2}},
\end{equation}

According to the Ref.  \cite{Carroll:2004st}, if the observer/particle is moving then the conventional Doppler effect will dominate the gravitational effects so, the condition of four velocity $U^\mu U^\nu=1$ will satisfy and for any observer, the frequency of photon moving along a null geodesic $x^{\mu}(\lambda)$ will be 
\begin{equation}
\omega_{obs}=U^\mu g_{\mu\nu} \frac{dx^\nu}{d\lambda}=U^t p_t+U^\phi p_\phi,
\end{equation}
From this, we can write as 
\begin{equation}
\omega_{obs}=U^t(-E+ \Omega_{Kerr} L),
\end{equation}
Using the Eq. (\ref{angfre}) and (\ref{velo}), we get

\begin{equation}
\omega_{obs}=\frac{-E+\Omega_{Kerr} L}{(g_{tt} +2 g_{t\phi} \Omega_{Kerr}+g_{\phi\phi} \Omega_{Kerr}^2)^{\frac{1}{2}}},
\end{equation}
For the case of an equatorial plane, the metric elements of Kerr BH become

$g_{tt}=1-\frac{2M}{R}, \qquad g_{\phi \phi}=-\left(\frac{a^2 r_g}{r}+a^2+r^2\right)$ \qquad  and \qquad  $\Omega_{Kerr}=\pm\frac{M^{\frac{1}{2}}}{r^{\frac{3}{2}}+a M^{\frac{1}{2}}}$

so, we get
\begin{equation}
\omega_{obs}=\left(\frac{r^2 (r-3 M)\pm2 a \sqrt{M} r^{3/2}}{\left(a \sqrt{M}\pm r^{3/2}\right)^2}\right)^\frac{1}{2},
\end{equation}
here the plus sign represents the notion in the direction of fluid motion and the negative sign shows the motion in the direction opposite to fluid motion (retrograde motion)

If the emitted frequency is $\omega_{emi}=1$, then one can define the Red-Shift factor $"g"$ associated with observed frequency as $g\propto\frac{1}{\omega_{obs}}$  as 
\begin{equation}
g=\frac{1}{z+1}=\frac{\left(a \sqrt{M}\pm r^{3/2}\right)}{\left(r^2 (r-3 M)\pm2 a \sqrt{M} r^{3/2}\right)^\frac{1}{2}},
\end{equation}
This equation is also discussed in Ref. \cite{Okazaki:1987st} to obtain the vertical and  epicyclic frequency for a perturbed disk in an equatorial plane as 

\begin{equation}
\begin{pmatrix}
         \partial^2 _t +\kappa^2 
     \end{pmatrix}
     \times
     \begin{pmatrix}
         v^r\\ 
        v^\phi 
     \end{pmatrix}
      =
     0, 
\end{equation}
where
\begin{equation}
\kappa^2=\frac{M(r^2-6Mr \pm8a(Mr)^{\frac{1}{2}}-3a^2)}{\left( \sqrt{M}\pm r^{3/2}\right)^2},
\end{equation}
For more detail see Okazaki et. al. \cite{Kato:2008xy, Okazaki:1987st}.

\section{Energy interpretation}\label{Energy}

\begin{figure}
\begin{center}
\includegraphics[width=0.65\textwidth, height=2.5in]{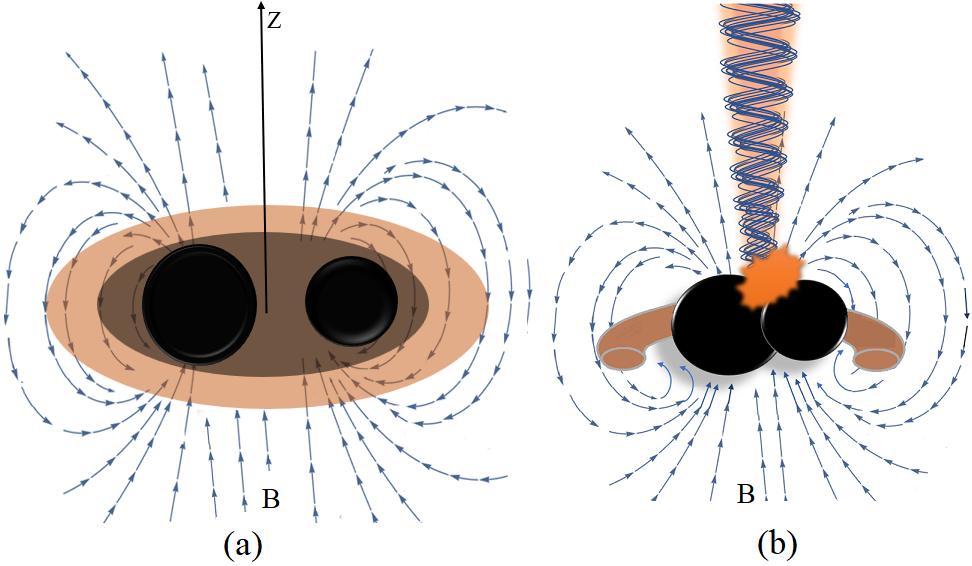}
\caption{\textbf{(a)}. Rotating disk around the plunging BHs with a strong magnetic field. \textbf{(b).} BH colliding and emission of the jet along the axis of rotation.}
\label{image-1.2}
\end{center}
\end{figure}

A Kerr BH can be generally described from its mass and angular momentum \cite{Blandford:1977ds, Wald:1984rga, Chandrasekhar:1985kt}. Let us consider the evolution of mass and angular momentum of a Kerr BH as 
\begin{equation}\label{M,J1}
\frac{dM c^2}{dt}=-P_L, \qquad \frac{dJ}{dt}=\tau,
\end{equation}
Here the quantity $\tau$ is the torque. In the hyper-accretion process, evolution should also have the contribution from mass and angular momentum i.e., $\dot{M} E_{ms}$ and $\dot{M} J_{ms}$, where $E_{ms}$ and $J_{ms}$ are specific energy and specific momentum of the material that is accreted at the innermost radius. Let a uniform disk of radius r rotate with angular velocity $\omega$ across which the current "I" flows down as shown in Fig. (\ref{image-1.2}). When a magnetic field $B$ is applied normally to the surface area of the disk, then the surface current will feel a force to rotate with torque as given by 
\begin{equation}
\Delta \tau =-r sin\theta \times I h\Delta B=-\frac{I}{2\pi}(A \Delta B),
\end{equation}
here $h=rd\theta$ is the thickness and $A=2\pi r^2$ is the surface area of the disk. So, the torque can be written as 
\begin{equation}
\Delta \tau=-\frac{I}{2\pi}\Delta \psi,
\end{equation}
$\psi$ is the upward magnetic flux along the pole of the disk. Using this torque, we can define the rotational power of the disk as 
\begin{equation}\label{MagP1}
\Delta P_{mag}=  -\omega _F \times \Delta \tau=\frac{\Delta \psi  I \omega _F}{2 \pi },
\end{equation}

If due to large inertia, the loading region (in which the magnetic field anchor) doesn't give rise to angular velocity then the power delivered by torque $(\Delta \tau=\frac{\Delta \psi  I}{2 \pi })$ and the angular momentum transportation will be the same as $\Delta P_{mag}$ i.e. we can write as
\begin{equation}
\Delta P_L=\frac{\Delta \psi  I \omega _F}{2 \pi },
\end{equation}

So, one can write the emitted power as 
\begin{equation}\label{emmitpow1}
\Delta P_{emitted}=\Delta P_{mag}=\Delta P_{L}=\frac{\Delta \psi  I \omega _F}{2 \pi },
\end{equation}

From this Eq. (\ref{emmitpow1}), One can derive the expression for the total BZ power by integrating both sides integrating as
\begin{equation}
    P_{mag}=- \frac{1}{2\pi}\int{\omega I d\psi},
\end{equation}
Where $\omega, \quad I, \quad$  $\psi$ are constants along the magnetic field surface and one can evaluate this equation on the horizon of the BH so, the above equation becomes
\textbf{\begin{equation}
    P_{mag}=- \frac{1}{2\pi}\int_h {\omega I d\psi},
\end{equation}}

Let the initial mass of a newborn Kerr BH is $M_o$ that goes to emits a portion of its energy as a result of turbulence, we can write the evolution process in terms of its mass and angular momentum from Eq. (\ref{M,J1}) using Eq. (\ref{MagP1}) as
\begin{equation}\label{Acc11}
\frac{dM}{dt}=\Omega _F\frac{ dJ}{dt},
\end{equation}
In the case of a Kerr BH the angular velocity of the field is defined as \cite{Misner:1973prb}
\begin{equation}
\Omega_F=\frac{d\phi}{dt}=\frac{a}{r^2 _+ +a^2}=\frac{J}{2M^3\left(1+\sqrt{1-\frac{J^2}{M^4}}\right)},
\end{equation}
Let $\left(1+\sqrt{1-\frac{J^2}{M^4}}\right)=x$, then we can write as 
\begin{equation}
J^2=M^4\left(2 x-x^2\right) \qquad \Rightarrow \qquad \frac{dJ^2}{dt}=2M^4(1-x) \frac{dx}{dt}+(2x-x^2)\frac{dM^4}{dt},
\end{equation}

So, from Eq. (\ref{Acc11}), we get
\begin{equation}\label{mas}
\frac{1}{M}dM=\frac{1-x}{2x^2}dx \qquad \Rightarrow \qquad M=M_o \sqrt{\frac{x_o}{x}} ,
\end{equation}
after the integration, we used the series expansion of the exponential term, to get Eq. (\ref{mass}). Let us consider the first observed event of $GW150914$ recorded aLIGO that is claimed to be a Kerr BH with total initial mass $\sim 65 M_\odot$ and final spin $0.67^{+0.05} _{-0.07}$ \cite{LIGOScientific:2016vbw}. A BH  retarded by the emission process will leave the final remnant (irreducible) mass as 
\begin{equation}\label{Nmass}
M=0.954 M_o=62.01M_\odot,
\end{equation} 
here the initial spin just after merging of the component BHs is approximately  $0.736$. It is greater than the final spin. It means that at the instant of BHs collision, the energy release due to angular momentum transport lowers the spin of the remnant mass. The subsequent spin down of the BH leads it to a stable point that must be comparable to the spin of a particle at ISCO. This spin-down of BH also represents a significant amount of energy emission as rotational energy. The energy extracted from BH is $M_o-M=3M_{\odot}$ can be treated as energy loss by binary to power the jet as discussed in \cite{LIGOScientific:2016vbw}. It means that about $ 3M_\odot$ of initial mass has been taken out to power the jet from a moderately rotating BH. According to Ref. \cite{Wald:1984rga}, this emitted energy is $4.616\%$ mass-energy of the binary merger used to power the Poynting flux along the direction of\textbf{ }the\textbf{ }twisted magnetic field \cite{Blandford:1977ds}, while the remaining energy will be associated with the irreducible mass or to increase the entropy of BH. The timescale for the GRB emission can be defined as the ratio of BH power to the power output \cite{Lee:1999se}. i.e.,
\begin{equation}
    t=\frac{black \quad hole \quad power}{power \quad output \quad from \quad the \quad surface}=\frac{M c^2}{B^2 R^2 c},
\end{equation}
using the value of $R=\frac{2GM}{c^2}$ and $M=0.954M_o$, we get the magnetic field as  

\begin{equation}\label{magfield}
B=0.512\left(\frac{ c^5}{M_o t G^2}\right)^{\frac{1}{2}}\sim 3.4\times 10^{15}G,
\end{equation}
here $t= 1-sec$ is used \cite{Connaughton:2016umz, Janiuk:2016qpe, Janiuk:2017tbq}. It is a strong poloidal magnetic field to jet the GRB from these binary merging BHs via the BZ mechanism.  

\section{Accretion Rate and Luminosity\label{accretion}}

To discuss the accretion power for the emission of GW150914 Ref. \cite{LIGOScientific:2016aoc}, we consider the property $M_1 > M_2$ and the mass ratio $q=\frac{M_2}{M_1}<1$ which led us to consider a spin of constant magnitude. Let us investigate the accretion rate of a steady-state accretion disk with surface density $\sum$ and width as $\Delta r$. So, the conserved angular momentum can be expressed as \cite{Blandford:1977ds}

\begin{equation}\label{Acc2}
\dot{M} \partial_r({r^2 \Omega_{disk}})\Delta r =\Delta T _{ann}+\partial_r T _{shear} \Delta r,
\end{equation}
here $T_{ann}$ and  $T_{shear}$ are the components of the torque exerted by the annular ring and shear force as discussed in section (\ref{plan}) and $\Omega_{disk}$ is the angular velocity of the disk. To solve this equation, we can evaluate the accretion rate in the event $GW150914$. As the torque exerted by the annular ring is given by 

\begin{equation}\label{Ann}
\Delta T_{ann} = r(2\pi r J_r B_z)\Delta r= r^2 B_\phi B_z\Delta r,
\end{equation} 
 where the magnetic field density $J_r$ is given by 
 
 \begin{equation}
 J_r = -\frac{B_{\phi }}{2 \pi },
 \end{equation}
  Now, let the torque exerted by the shear force be as \cite{Pringle1981}
 
 \begin{equation}
T _{shear}=2\pi r v \sum (r\partial_r \Omega_{disk}) r,
 \end{equation}
 
 with $r\partial_r \Omega_{disk}=\frac{B_r B_\phi}{4\pi v \rho}$ so, we get 
 
  \begin{equation}
T _{shear}=2\pi r^2 \frac{\sum}{\rho} \frac{B_r B_\phi}{4\pi } =  r^2 \frac{c}{\Omega}B_r B_{\phi } ,
 \end{equation}

 here in the first step, we used $H=\frac{\sum}{2\rho}=\frac{c_s}{\Omega_{disk}}$.  As the accretion rate is determined by the magnetic braking along the z-axis so the accretion rate along the z-axis is much larger than that along the radial direction so let us consider the transformation $B_r=\frac{7r\Omega B_z}{3c}$ as $B=B\hat{z}$ and integrating, we get
 \begin{equation}\label{shear1}
 \partial_r T _{shear}=  7 r^2 B_z B_{\phi },
 \end{equation}
 Using Eq.(\ref{Ann}) and (\ref{shear1} in  Eq. (\ref{Acc2}), we get

\begin{equation}
\dot{M} \partial_r({r^2 \Omega_{disk}}) = 8 r^2 B_z B_{\phi },
\end{equation}
As $B_\phi =2r \Omega_{disk} \frac{B_z}{c}$ so, we get

\begin{equation}
\dot{M} \partial_r({r^2 \Omega_{disk}}) = \frac{16 r^3 B_z^2 \Omega_{disk}  }{c},
\end{equation}
 separating variables and integrating we get
 \begin{equation}\label{BZacc}
     \dot{M}=\frac{4 r^2 B_z^2}{c},
 \end{equation}
 As from Maxwell's equation one can get the magnetic field  proportional to $\frac{1}{r}$ (i.e. Ampere's law) so, the accretion rate is independent of $r$. Considering $r\sim 1.396\times10^5 cm$, we get the accretion rate as a function of horizon radius $r$ and the poloidal magnetic field $B$ 
 
 \begin{equation}\label{Nacc}
 \dot{M}=3\times 10^{-2} \left(\frac{B_z}{10^{15}G} \right )^2 M_{\odot} erg.sec^{-1}>>\dot{M}_{Edd},
 \end{equation}
That is much greater than the Eddington accretion rate $\dot{M}_{Edd}\sim 10^{-16}\left(\frac{M_{BH}}{M_{\odot}}\right)^2 M_{\odot} sec^{-1}$. As the poloidal magnetic field is of the order $\sim 10^{15}$ that suggests that the jet is powered by either the BZ mechanism or magnetized NDAF. The NDAF is discussed by \cite{Janiuk:2016qpe} for the slowly rotating BH but the existence of a strong magnetic field $(\sim 10^{15})$ supports the domination of the BZ mechanism. As the final BH is moderately rotating that will increase by forming a rotating core where the subsequent accretion onto the BH would further increase the spin. The existence of a strong poloidal magnetic field along the axis of rotation acts as a restoring agent and will trap the spin energy but due to the rigid rotations, it increases along the cylindrical geometry to power a Poynting flux and reassembles the BZ accretion flow.  This result from Eq. (\ref{BZacc} ) shows a higher accretion rate as compared to that obtained in Ref. \cite{Lee:1999se}. The main difference is due to the magnetic field and horizon size that gives consistent results in our simulations.

The luminosity found in this mechanism is 
\begin{equation}
L_{BZ}=f(a_*)c r^2 \frac{B_z}{8\pi}=1.8 \times 10^{49} erg.sec^{-1},
\end{equation}
we used $a=0.67$ as the BH final spin \cite{LIGOScientific:2016aoc}, to calculate the spin parameter \small {$f(a_*)=1-\sqrt{\frac{1}{2}(1-\sqrt{1-a^2}))}=0.067$) and use this value with other conventional units to get luminosity\textbf{.} This luminosity is much greater than the Eddington luminosity $\left(L_{Edd}\equiv1.25\times 10^{46} \left(\frac{M}{10^8 M_{\odot}}\right) erg. sec^{-1}\right)$ for the binary merging BHs Ref. \cite{Kato:2008xy}. This result shows the existence of a compact object in the center of the GRB's source as its progenitor. Using the NDAF mechanism, such a luminosity is also investigated in \cite{Janiuk:2016qpe}, where the binary BHs are considered with moderate spin parameters.  From such a moderately rotating BH under the effect of the strong magnetic field, one could get the maximum energy extraction efficiency of the flow as $\eta_{max} =\frac{L_{BZ}}{\dot{M}c^2}\sim 0.067 \%$. 

In comparison to this work, Janiuk et.al. \cite{Janiuk:2016qpe, Janiuk:2017tbq} proposed a model of a collapsing massive star to form a BH merging with another BH in a close binary that led to the event of GW150914. The gamma-ray burst was considered to be powered by weak neutrino flux produced in the star remnant’s matter with moderate spin rather than by the BZ mechanism. The GW is considered to originate from the merger collapsed core and the companion BH. For the spin 0.8 of each component BH, the angles between the spins and normal to orbital plan are $45^o$ and $180^o$ while, for the $90^o$ angle between the binary, the spin magnitudes are considered to be smaller than 0.7 and 0.8. At the maximum values of 0.2 and 0.3, the spins are assumed to be aligned with orbital angular momentum. The GRB fluence rate (amount of energy received over the time duration of the GRB) in the range of 1keV to 10 MeV  is $2.8\times10^{-7} erg.cm^2$. The implied source luminosity was found to be $1.8\times10^{49} erg.sec^{-1}$ with remnant mass $62M_{\odot}$, final spin $0.7$, and disk mass $15M_{\odot}$. In this model, the power and luminosity due to the BZ process are completely neglected due to ignoring the BH magnetization and lower spin. It was stated that the moderate spin affects the topology of the magnetic field which remains confined in the plasma torus and hence no magnetically driven wind resulting in low luminosity. In these simulations, $t=GM/c^3$  is used with the scale of $1000Min=0.304sec$. This time duration in contrast to LIGO observed time off course shows the event of long GRBs as compared to $t=1sec$ of aLIGO data. that is also claimed by them in their manuscript. The same result of luminosity is also obtained in Ref. \cite{Connaughton:2016umz}.

\section{Conclusions and Discussion\label{conclusion}}

Nowadays, a new discussion is in progress about the emission of SGRBs associated with the GW whose first event was simultaneously observed on $14^{th}$ Sep. $2015$ by two aLIGO detectors. The transient gravitational wave signals with a frequency $35-250$Hz were found. The detected wave signal was claimed to originate from the merger of two BHs with masses  $36^{+5} _{-4}M_{\odot}$ and  $29^{+4} _{-4}M_{\odot}$ whereas the final remnant mass is $62^{+4}_{-4}M_{\odot}$  with $3.0^{+0.5}_{-0.5}M_{\odot}$ is radiated as gravitational as well as electromagnetic energy \cite{LIGOScientific:2016aoc, Janiuk:2016qpe}. This electromagnetic signal is also observed by the Fermi Gamma-ray Burst Monitor satellite observatory to originate from a weak transient above $50 keV$, $0.4 sec$ after the GW and event lasting for $\sim 1$-sec long \cite{Connaughton:2016umz}. So, this event was the first confirmation of the co-detection of gravitational waves and GRB emission from the binary BHs as observed by two aLIGO detectors and the Fermi Gamma-ray Burst Monitor satellite observatory.  Later several observations were made to analyze the characteristics and physical properties via different models.

We introduce a model to sketch the GRB emission during the event of binary BHs mergers. The outline of this model is based on Magneto-rotational instability that causes turbulence in the accretion disk onto the final BH. As a demonstration of the model, we consider the first observed event of GW150914 and obtained results consistent with previous research.\textbf{ }According to general relativity, the two merging BHs will result in a Kerr BH as their final product so, we consider a Kerr BH of total mass $65M_{\odot}$ that is formed during the merging of binary $36^{+5} _{-4}M_{\odot}$ and  $29^{+4} _{-4}M_{\odot}$ BHs having an accretion disk on to it  where the MRI causes random motion of fluid to produce fluctuations in both the magnetic field and in the flow of the fluid. The fluid displaced outward is restricted by the magnetic pressure $\frac{B^2}{2\mu_o}$ and ensures its rigid rotation but if the outward shear force dominates due to large fluctuations then the MRI sets in that further amplifies the magnetic field due to turbulent dynamo action inside the accretion disk. After getting a large amount of matter inside the disk, it becomes gravitationally unstable and an intense mass accretion onto the central BH will occur through the gravitational torque that causes the prompt emission of GRBs and other EM radiations or GW. 

To demonstrate this model, starting from Maxwell's tensor, we introduced the magnetic pressure (restoring force to the fluid displaced) and shear stress to the accretion disk. Using these stresses, we found the condition for a growth rate of instability as a function of the angular velocity of the disk and independent of the magnetic field strength. The greater the angular velocity greater will be the growth rate of instability. As from the general relativity, the remnant of binary BHs is a Kerr BH so, we generally discussed the motion of particles in an equatorial plan around a Kerr BH in a circular orbit. Using the condition for a circular orbit, we discussed the motion of the test particle on geodesics in effective potential and calculated the energy $E$ and angular momentum $L$ as constants of motion.  For the same situation, we also investigated the angular frequency and epicyclic frequency, and  redshift factor observed by an observer at distance $r$. Actually, the investigation of these quantities is the limitations for the exact circular motion of a particle beyond which it will face instability that is the main purpose of this study to locate the GRB event.

Next, we consider the evolution of mass and angular momentum to investigate the remnant mass of BH as given in Eq. (\ref{mass}). Using the numerical values of the component BHs along with conventional units, our investigation gives the remnant mass to be $62.01M_\odot$ with $\sim 3M_{\odot}$ is used to power the jet from the binary merger that is exactly the same as observed in \cite{LIGOScientific:2016aoc, Connaughton:2016umz}. Using this remnant mass the magnetic field is found to be $~3.4\times 10^{15}G$ which is of course enough to power up the GRBs from binary BHs with a maximum accretion rate of $3\times 10^{-2}\left(\frac{B_z}{10^{15}G}\right)^2 M_{\odot} sec^{-1}>>\dot{M}_{Edd}$.  This accretion rate is much greater than that predicted by Lee et.al in Ref. \cite{Lee:1999se}. The main difference is due to the magnetic field $(\sim10^{15} G)$ and horizon size $(\sim 1.396\times10^5 cm)$ that gives consistent results in our simulations while they used magnetic field and horizon radius of order $\sim10^{14} G$ and $\sim 10^6$ cm  respectively. The associated BZ luminosity is found as $1.8 \times 10^{49} erg.sec^{-1}$. Finally, using the calculated accretion rate and luminosity, we determined the accretion efficiency to be very low $(\sim 0.067\%)$.  Compared to other models, this model uses the actual data as determined in Refs. \cite{LIGOScientific:2016aoc, Connaughton:2016umz} to give the most consistent and valid results for the proposed event.

SGRBs are often taught to be originated from the binaries of NS (or one member of the binary must be NS). This observation of GW150914 reverted the notion of NS existence in merging binaries and proved that SGRBs could be originated from the final product of binary BHs having a magnetized accretion disk to power up the Poynting flux along the axis of rotation. For such magnetically dominated outflow, the leading model is the BZ mechanism that needs a large-scale magnetic field connecting the BHs. As we have evaluated the magnetic field and the accretion rate for the final product as Kerr BH so, the spinning property of BH is a plus point to increase the rotation and produce jet along the cylindrical geometry and paginate GRB via BZ mechanism. 

\vspace{1.0cm}
\noindent\rule{16.5cm}{2.0pt}\\

\newpage
\chapter{Short-duration gamma-ray bursts from Kerr-Newman black hole mergers}

Black hole (BH) mergers are natural sources of gravitational waves (GWs) and are possibly associated with electromagnetic events. Such events from a charged rotating BH with an accretion on to it could be more energetic and ultra-short-lived if the magnetic force dominates the accretion process because the attraction of ionized fluid with a strong magnetic field around the rotating BH further amplifies the acceleration of the charged particle via a gyromagnetic effect. Thus a stronger magnetic field and gravitational pull will provide an inward force to any fluid displaced in the radial direction and move it toward the axis of rotation with an increasing velocity. After many twists during rotation and the existence of restoring agents,  Such events could produce a narrow intense jet starts in the form of Poynting flux along the axis of rotation resembling the Blandford-Znajek (BZ) mechanism. We investigated a charged rotating BH and obtained characteristic results (e.g., the remnant mass, magnetic field strength, luminosity, opening angle, viewing angle, and variation of viewing angle on the  SGRB luminosity detection) that have a nice coincidence with rare events having GW associated with EM counterparts. This study gives a new insight into events having strongly magnetized disks dominating the accretion process of energy extraction.

\section{Introduction}\label{sec:intro}

Gamma-ray bursts (GRBs) are the most luminous and short-lived explosions e.g., \cite{Klebesadel:1973iq, Fishman:1995st, Piran:2004ba, Meszaros:2006rc, Zhang:2018ond}. This prompt emission is followed by a long-lived afterglow emission with energies ranging from X-rays to radio bands e.g., \cite{Costa:1997obd, Frail:1997qf}, vanParadijs1997, Wijers:1997xu, Heng:2008nr, Kumar:2014upa. They can be classified as long-duration GRBs (LGRBs), $T_{90}>2 ~\rm s$; e.g., \cite{Kouveliotou1993, Woosley1993, MacFadyen:1998vz, Fruchter:2006py, Woosley:2006fn, Perna:2018wft, Dagoneau:2020mbg} and short-duration GRBs (SGRBs), $T_{90}<2 ~\rm s;$ e.g. \cite{Eichler:1989ve, Narayan:1992iy, DAvanzo:2015kdp}. LGRBs are claimed to be emitted by the collapses to form a neutron star (NS) or a black hole (BH). The collapsar model is widely accepted as the standard model to describe the LGRB and ultra-LGRB emission e.g., \cite{Hanami:1997yy, Popham:1998ab, Beloborodov:2008nx, Evans2008, Janiuk2010, Liu2017, Song:2018vup}, whereas, SGRBs are expected to originate from NS-NS or BH$-$NS mergers e.g., \cite{Belczynski:2006br, Lee:2007js, Liu2012, Liu:2015zta, Fong:2013iia, Giacomazzo:2012zt, Ruiz:2016rai, Savchenko2017}.

The double BHs mergers were not expected to power GRBs until a plausible SGRB associated with the binary BH (BBH) merger event GW150914 was first predicted by the Fermi-GBM \cite{Connaughton2016} as a weak transient above $50 ~\rm keV$ with a delay of $0.4 \rm s$ after the gravitational wave (GW) event with false alarm of $0.0022$. This weak transient from the BBH merger lasted for about $1 \rm s$. In the source frame, the masses of the component BHs were observed as $ m_1=36^{+5}_{-4}~M_{\odot}$ and $ m_2=29^{+4}_{-4}~M_{\odot}$. After the merger, a remnant mass equal to $ M=62^{+4}_{-4}~M_{\odot}$ was observed whereas the mass $3.0^{+0.5}_{-0.5}~ M_{\odot}$ is radiated as GWs associated with an unexpected electromagnetic event. The final BH spin parameter inferred from general relativity (GR) was $ a=0.67^{+0.05}_{-0.07}$. The source of this event is constrained to an annulus section of $610 ~ \rm {deg^2}$, primarily in the southern hemisphere at the luminosity distance $410^{+160}_{-180} \rm Mpc$  corresponding to red-shift $z=0.09^{+0.03}_{-0.04}$  e.g., \cite{LIGOScientific:2016aoc, LIGOScientific:2016vlm}.

After predicting the event GW150914 as the progenitor of GW-GRB co-emission, many researchers put forward different models to explain the physics of this event e.g., see \cite{Zhang:2016rli, Loeb:2016fzn, Perna:2016jqh, Liu2016, Woosley:2016nnw, Janiuk2017, Ali:2023zva}. A hyperaccretion disc around a stellar-mass BH is considered one of the plausible central engines of GRBs e.g., for a review see \cite{Liu2017}. The jets launched by the hyperaccretion disc are possibly powered by the neutrino annihilation process \cite{Popham:1998ab, Narayan2001, Kohri:2005tq, Chen:2006rra, Kawanaka:2007sb, Liu:2014xya, Liu:2012qca} or BZ mechanism \cite{Blandford1977}. It is worth noting that the neutrino annihilation process is effectively only for the central BH mass less than about $50~M_\odot$ \\cite{Liu2021}. Janiuk et al. \cite{Janiuk2017} studied the hyperaccretion disc. It is proposed that a collapsing massive star (making a BH) merges with another BH in a close binary leading to a weak transient due to the accretion onto the final BH with a moderate spin. The main misunderstandings in Janiuk's work are due to the time assumption, mass, and magnetic field of BBHs. First, Janiuk relayed the time assumption of $t=2000\rm M(\approx0.6 ~\rm s)\neq 1 ~\rm s$. Second, their simulated data represents a mass of $62M_{\odot}$  rather than a total mass of $65M_{\odot}$. Third, the power and luminosity available through the Blandford–Znajek process at the horizon and in polar regions are considered negligible due to low spin and magnetization. Ignoring the magnetic field strength,  they found the emitted power $\dot{E}=1.1\times 10^{50} ~\rm erg ~s^{-1}$  for $a=0.8$ and $3.7\times 10^{51} ~\rm erg ~s^{-1}$  for $a=0.9$. The observed luminosity is much smaller as compared to these luminosities \cite{Connaughton2016}. If the central BH rotates with a poloidal magnetic field threading to its horizon may drive a powerful relativistic jet by a process resembling the BZ mechanism \cite{Liu2016, Zhang:2016rli}.
Similarly, Khan \cite{Khan:2018e} also performed magnetohydrodynamic simulations using an accretion disc onto the non-spinning BH binaries and investigated several properties of the event GW150914. We examined the event GW150914 by considering a BBH merger creating a KNBH as their final product with a magnetized disc emitting the rotational energy via narrow Poynting flux. We investigated their characteristic properties that could help the reader understand the physics of a jet possibly emitting from the merging binaries with a strong magnetized disk.

The structure of this chapter is such that in section \ref{Model}, we present our model with a sketch of jet emission from a KNBH with a strong magnetic field around it that could be presented by Maxwell's EM equation. Such magnetic fields around a spinning BH could be the main source of GW associated with short-lived EM counterparts. In section \ref{sec: Ins}, we discuss the basic condition to set an instability in surrounding matter that led to turbulence. In section \ref{sec: Energy}, we discussed the role of the gyromagnetic effect, and energy extraction from the final BH. This is the main part of adopting our model and the basic condition of instability in a suitable relation for the energy interpretation from the final BH. In section \ref{sec: accretion}, we considered the first observed event GW150914 in our model and obtained the luminosity associated with the emission of energy, the opening angle of the jet, and the effect of the viewing angle on the luminosity are discussed. Finally, some remarks and conclusions derived from this work are summarized in section \ref{sec: conclusion}.

\section{Model}\label{Model}
Angular momentum and charge of BH retained by the magnetosphere  are quantities creating a strong magnetic field around a spinning BH. The merger of such BHs could be naturally the progenitor of a new spinning BH with a stronger magnetic field that could extract the rotational energy from BH as an intense burst in a short time scale rather than by the accretion flow. \cite{Proga2006} and \cite{Liu2012} have shown that a magnetic barrier will build up while the radial magnetic force counteracts the gravitational pull of the BH. Due to this reason, the rotational energy emission from the final BHs will have less chance to break the magnetic field along the radial direction. This energy of BH will act as a source of instabilities that tends to propagate along the axis of rotation due to a strong magnetic barrier and turbulent dynamo action while the gravitational pull along the radial direction. It could also provide further amplification of magnetic field $(\rm up~to \sim 10^{16} ~\rm G)$ to excite electromagnetic waves in the form of a BZ jet e.g., \cite{Blandford1977,deSouza:2009ne}. The fact is that a charged spinning BH with a strong magnetic field and its dynamics could play an important role in accumulating magnetic flux e.g., \cite{Balbus:1991ay}. 

If at least one of these merging BBHs has some charge that is possibly retained by the magnetosphere of the BH, then it will strengthen the magnetic field around by induction as can be seen from the Maxwell equation
\begin{equation}
\nabla \times\textit{\textbf{E}} =-\partial_t\textit{\textbf{B}},
\end{equation}
where ${\textbf{E}}$ is the electric field intensity of the BH charge and ${\textbf{B}}$ is the magnetic induction intensity. Using Stokes theorem, we can write this as
\begin{equation}\label{E->B}
 \oint_\Gamma \textbf{E} ds =- \partial_t (Magnetic \quad flux \quad through \quad S),
\end{equation}
where $\Gamma$ is the closed curve bounding the surface $S$. The integral on the left is the electromotive force (EMF) and on the right is the rate of change of the magnetic flux, while the negative sign $(-)$ shows the direction of EMF and magnetic flux concerning each other. 

Among several mechanisms, our model used Magneto-rotational instability (MRI), which relates significantly to turbulence in accretion disks. A hyper-accretion disk around a stellar-mass BH is a plausible source of the central engine to jet the GRBs as claimed by many authors but the main question is how to set the turbulence inside the hyper-accretion disk that results in jet GRBs. In search of the fact, the magnetic-field-driven disturbance called Magneto-Rotational Instability (MRI) is claimed to operate in hyper-accretion disks actively. It causes the angular momentum transport  \cite{Balbus:1991ay, Velikhov:1959xx, Balbus:1998az}. The MRI produces random motion (where the fluid moves faster toward the BH and slower outward) that generates fluctuations in both the magnetic field and the flow of the fluids. In such cases, the magnetic field lines restrict the outward displaced fluid and ensure the rigid rotation of the fluid that increases with time along the field direction to assume a cylindrical geometry. Due to these fluctuations and rigid rotations, each additional twist causes an increase in the cylindrical geometry. After many twists and restricted cross-sectional area, a narrow intense jet starts like a Poynting flux resembling the BZ mechanism.


By using Maxwell's stress tensor, the turbulence driven by the MRI can be described as \cite{Ali:2023zva}
\begin{equation}\label{Maxeqn}
 T_{ij} = \epsilon_{o}\left( E_{i}E_{j} - \frac{{\delta_{ij}E}^{2}}{2} \right) + \frac{1}{\mu_{o}}\left( B_{i}B_{j} - \frac{{\delta_{ij}B}^{2}}{2} \right).
\end{equation}
A BH can be completely described by its mass $ M$, angular momentum $ J$, and charge $ q$. The first two quantities have been measured for stellar and supermassive BHs with various observations, but the charge is often neglected and implicitly set identically to zero. However, both classical and relativistic processes can lead to a small non-zero charge of black holes \cite{Zajacek:2019kla}. The charge of BH could be associated with the surrounding material making a disk. Due to the rotation property of the remnant BH, it is greatly expected that the charge contribution could be added to the magnetic fields so, to get a clearer insight into the turbulence, we decompose Maxwell's stress tensor into diagonal and off-diagonal components as
\begin{equation}
 \delta_{ij} =
\begin{cases}
1,& \text{if }i=j\\
0,& \text{if }i\neq j.
\end{cases}
\end{equation}
So, for \(i = j\), Maxwell's stress tensor becomes
\begin{equation}\label{magpres0}
P_{mag}= \frac{1}{2\mu_{o}}B^{2},
\end{equation}
where \( P_{\text{mag}}\) is the restoring force on the perturbed fluid elements. For \( i\neq j\), it gives the shear stress
\begin{equation}\label{tenshear}
 T_{shear} = \frac{1}{\mu_{o}}B_{ij}.
\end{equation}

In a differential rotating system, this tension force can lead the system to instability. The magnetic field is an effective source for this shear-type instability leading to turbulence in the accretion disc resulting in an inflow or outflow of fluid \cite{Nattila:2021qag}.

\begin{figure}
\begin{center}
\includegraphics[width=0.4\textwidth, height=2.5in]{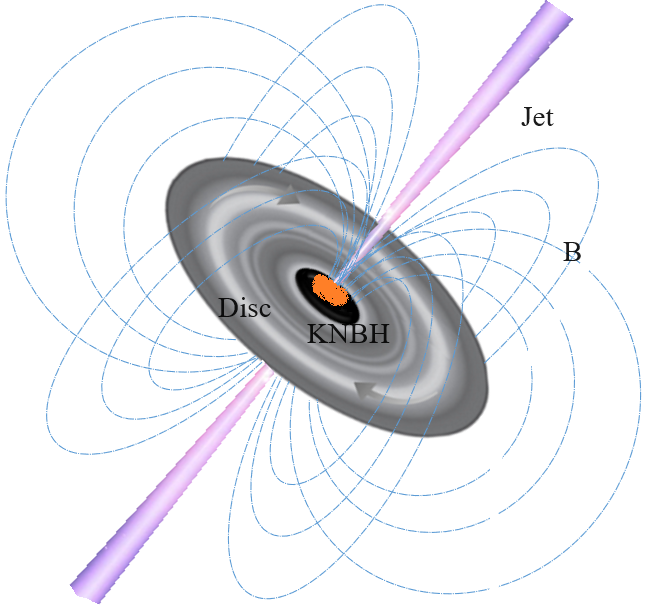}
\caption{A sketch of KNBH with magnetized accretion disc emitting a narrow GRB by BZ jets.}
\label{image-2.2}
\end{center}
\end{figure}

\begin{figure}
\begin{center}
\includegraphics[width=0.3\textwidth, height=1.5in]{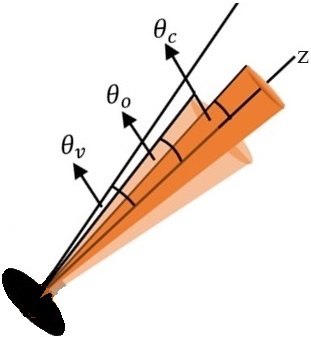}
\caption{A close look at the jet along the axis of rotation from KNBH with magnetized accretion disc. The core angle $\theta_c$, opening angle $\theta_o$, and angle $\theta_v$ (the angle that the outgoing photon makes with the normal to the jet surface.) are shown from the axis of rotation \cite{Perna2019, Salafia2015, Hayes2020}}
\label{image-2.3}
\end{center}
\end{figure}

\section{Basics condition for instability}\label{sec: Ins}
Before demonstrating the remnant mass and the emitted energy from the merger system, we need some boundary conditions for instability caused by the merging process to overcome the system's stability. Various forces act upon the particles of the fluid in an accretion disc at different radii, so the fluid particles are moving at variable speeds \cite{Ali:2023zva}. As a result, the restoring force $ P_{\rm mag}$ along the gravitational forces on the fluid provides the potential required in a circular motion of the disc with an angular velocity $\Omega$ (or a calm Keplerian disc) and the shear force is responsible for seeding the instability at the instant of merging. This means that if a particle of the fluid is displaced from its position, then it will be acted upon by three forces i.e., gravitational and restoring magnetic tension toward the original position of the particle and the shear force along the diagonal but overall motion of the particle will be retained along the circular path. As long as the mass accretion increases, the instability increases, but due to rotational motion and magnetic restoring force, the displaced fluid tends toward the axis of rotation. Thus the crucial role of the magnetic field in shaping instability and fluctuation needs to be considered for understanding the basic process. Let the restoring magnetic tension acting on a particle is
\begin{equation}\label{magpres2}
 P_{\rm mag} = \frac{B^2}{4\pi \rho R}\approx \left( \frac{c_{\rm Al}}{\lambda} \right)^{2}\ \Delta r,
\end{equation}
where $ R= \lambda^{2}/\Delta r$ is the radius of magnetic field curvature with \(\lambda\) as the perturbation wavelength in the vertical direction and $ c_{\rm Al} \equiv \sqrt{B^{2}/4 \pi \rho}$ as the speed of Alfv\'{e}n wave. As the Alfv\'{e}n propagates in the direction of the magnetic field. So, the motion of the ions/particle and the perturbation of the magnetic field are transverse to the direction of propagation \\cite{McIntosh2011}. This means that if the propagation of the magnetic field is in the radial direction the emitted energy (flux) will be perpendicular to it. From Eq. (\ref{magpres2}), the displacement \(\Delta r\) at conserved angular velocity $\Omega = \sqrt{ GM/r^{3}}$, the growth rate of centrifugal force for a unit mass is \(\Delta F_{c}\sim\Omega^{2}\Delta r\). It will result in the decrease of the gravitational force on unit mass by the amount
\begin{equation}\label{grav_force}
 \Delta F_{g} \approx \frac{2GM}{r^{3}}\Delta r \approx 2\Omega^{2}\Delta r.
\end{equation}

Thus the total inward force acting on a single perturbed particle is \(\sim 3\Omega^{2}\Delta r\). For an instability to cause turbulence, the outward force must be dominated over the magnetic tension i.e., we can use the inequality between total force and Eq. (\ref{magpres2}) to represent the domination of magnetic tension,
\begin{equation}
3\Omega^{2} > \left( \frac{c_{\text{Al}}}{\lambda} \right)^{2} \approx \lambda > \frac{c_{\text{Al}}}{\Omega},
\end{equation}

Here, one can use angular velocity and Alfv\'{e}n waves speed values to get a typical wavelength value that must satisfy the instability condition. In the case of rotating BH the angular velocity is defined as 
\begin{equation}
    \Omega=\sqrt{\frac{G M}{r^3}}\left(1+\frac{a_{\ast}}{(8 \hat{r}^3)^{\frac{1}{2}} }\right)^{-1}
\end{equation}
here, 
$$a_{\ast}=\frac{c^2 a}{G M}=\frac{c J}{G M^2}, \qquad \rm \qquad a=\frac{J}{M c}$$
and $\hat{r}=\frac{r}{r_g}$. So, we can write as
\begin{equation}
    \Omega=2 \sqrt{\frac{G M}{r^3}}\left(4+\frac{\sqrt{2} c J}{G M^2}\right)^{-1}
\end{equation}
The condition of instability becomes
\begin{equation}
    \lambda>\frac{ \left(4+\frac{\sqrt{2} c J}{G M^2}\right)c_{Al}}{2 \sqrt{\frac{G M}{r^3}}}
\end{equation}
Let us consider the event $GW150914$ with $r=10^6 \rm cm$ and other typical values, we can estimate the instability condition as
\begin{equation}
    \lambda=(1.1\times 10^{-4}  \rm sec) c_{Al}
\end{equation}
here the $c_{Al}$ depends on the mass density and the magnetic field strength. Near the magnetosphere, the Alfv\'{e}n speed could rise to $10^8 \rm cm/s$ with a wavelength up to $10^4 \rm cm$. Here, it is important to take  $v_A<<~c$ because at $v_v\approx c$  the Alfv\'{e}n wave becomes an ordinary electromagnetic wave.  The growth rate of instability from the time scale is
\begin{equation}
 \tau = \frac{\pi \lambda}{2 c_{Al}} =\frac{\pi}{2\Omega}\approx 1.7 \times 10^{-4} \rm sec > \frac{1}{\Omega}.
\end{equation}
This equation shows that the growth rate of the instability depends on angular velocity and is independent of the magnetic field, i.e., the greater the angular velocity, the smaller the growth time of the turbulence, and vice versa. The magnetic field is an agent of seeding instability in the accreting fluid of the disc. At peak angular velocity, the instability grows  in a very short time scale. Note that the instability sets in for perturbations with shorter wavelengths when the magnetic fields are weak.

\section{Gyromagnetic effect and energy interpretation}\label{sec: Energy}

The Schwarzschild and Reissner-Nordstr\"{o}m radii of a charged BH are
\begin{equation}
 r_{s}=\frac{2 G M}{c^2}~ {\rm and}~ r_q=\frac{\sqrt{G} q}{c^2},
\end{equation}
where $M$ is the mass of BH. Comparing the above two equations, we can obtain the upper limit of the charge $q_c$ as
\begin{equation}
q_c=2 \sqrt{G} M=(6.1\times 10^{31}\rm {e.s.u})\left(\frac{M}{10 M_{\odot}}\right)\sim 2 \times 10^{22}\rm C ,
\end{equation}
Such a limited quantity of BH's charge would significantly modify the magnetic field geometry to set the electric field that also contributes to the magnetic field via the gyromagnetic effect as given in Eq. (\ref{E->B}). Due to BH charge contribution to the magnetic field during the gyromagnetic effect, we can follow the track of evaluating two quantities for describing a charged spinning BH as
\begin{equation}\label{M,J}
\frac{dM c^2}{dt}=-P_L~ \qquad{and}\qquad ~ \frac{dJ}{dt}=\tau,
\end{equation}
where $P_L$ is the power due to angular momentum transportation, $\tau$ is the torque exerted by the magnetic field normal to the disc's surface, and the negative sign $(-)$ shows the power loss. In the hyperaccretion process, evolution should have the contribution from mass and angular momentum i.e., $\dot{M} E_{\rm ms}$ and $\dot{M} J_{\rm ms}$, where $\dot{M}$ is the accretion rate and $E_{\rm ms}$ and $J_{\rm ms}$ are the specific energy and angular momentum of the material accreted at the innermost radius, respectively. For KNBH, \cite{Zhang:2016rli} investigated the magnetic dipole moment that drives Poynting flux during the merging process and claimed that the dissipation of Poynting flux at large radii would power SGRB if the BH has a charge of $10^{-5}-10^{-4}$, and if the BH charge is $\sim 10^{-9}-10^{-8}$ that will drive an FRB \cite{Liu2016}. Let the disc of radius $r$ rotate with an angular velocity $\omega_{F}$, the change in torque and the rotational power of the disc is
\begin{equation}\label{MagP}
\Delta \tau=-\frac{I}{2\pi}\Delta \psi\quad \Rightarrow\quad \Delta P_{\rm mag}= -\omega _F \times \Delta \tau=\frac{\Delta \psi I \omega _F}{2 \pi },
\end{equation}
here $I$ is the current due to the flow of charge and $\Delta \psi$ is the change in magnetic flux due to induction. If the loading region is unable to give rise to angular velocity due to large mass and inertia, then the power delivered by torque and the power of angular momentum transportation $\Delta P_L$ will be the same as $\Delta P_{\rm em}$ from the disc, i.e.,
\begin{equation}\label{emmitpow}
\Delta P_{ em}=\Delta P_{\rm mag}=\Delta P_{\rm L}=\frac{\Delta \psi I \omega _F}{2 \pi}.
\end{equation}
By integrating, Eq. (\ref{emmitpow}) could give the total BZ power as
\begin{equation}\label{T.Pow}
P_{\rm mag}=- \frac{1}{2\pi}\int{\omega_F I d\psi}\quad\Rightarrow \quad P_{\rm mag}=- \frac{1}{2\pi}\int_h {\omega_F I d\psi}.
\end{equation}
The subscript $h$ denotes the horizon. As $\omega$, $I$, and $\psi$ are constants along the magnetic field surface, it is easy to evaluate Eq. (\ref{T.Pow}) on the horizon of the BH.

Consider the predicted merging event GW150914 resulting in a KNBH emitting some of its energy due to turbulence supported by the magnetic field of ionized fluid. So, combining the evolution relations of mass and angular momentum from Eq. (\ref{M,J}) using Eq. (\ref{MagP}), we get
\begin{equation}\label{Acc1}
 \frac{dM}{dt}=\Omega _F\frac{ dJ}{dt},
\end{equation}
here, the angular velocity of the field in Kerr Newman geometry can be written as \cite{Misner1973}
$$\Omega_F=\frac{d\phi}{dt}=\frac{a}{r^2 _+ +a^2}$$
\begin{equation}\label{ang_vel}
=\frac{J}{2 M^3 \left(1-\frac{q^2}{2 M^2}+\sqrt{1-\frac{J^2}{M^4}-\frac{q^2}{M^2}}\right)},
\end{equation}
where $r_+=M+\sqrt{M^2-a^2-q^2}$ is the KNBH horizon radius. KNBH has a non-zero angular momentum (spin) that imparts rotational energy to the BH, which can be tapped into. The magnetic field of such BH amplifies with the differential rotation. The charge-to-mass ratio $({q}/{M})$ in Eq. (\ref{ang_vel}) is the gyromagnetic ratio of KNBH denoted by $\gamma_{\rm KNBH}$ (the ratio of the magnetic moment to the angular momentum) of the Dirac electron \footnote{A standard electron that can be described by the Dirac theory also known as Larmor radiation that plays an important role in the emission of GRBs from a KNBH.}. Specifically, it is related to the radiation emitted by charged particles moving in the strong magnetic field around the BH.

For a KNBH, the magnetic dipole moment is $\mu=qa$, and angular momentum is $J=Ma$ so, $\gamma_{\rm KNBH}={q}/{M}>\gamma_{\rm classical}$. The underlying fact of the gyromagnetic effects is that the nuclear and the electronic spins, as well as their orbital angular momenta, generate a magnetic moment parallel to the angular momentum of constant magnitude that can be transformed into a rotating coordinate system by Newton's second law of motion or Schrodinger's equation gives Larmor's theorem \footnote{The effect of a uniform magnetic field (produces periodic effects on a system of spins or particles) can be transformed away by going to a rotating coordinate system, provided the angular momentum of the system in the field direction is a constant of motion \cite{Heims:1962sjg}.}. 

\begin{figure}
\begin{center}
\includegraphics[width=0.6\textwidth, height=2.4in]{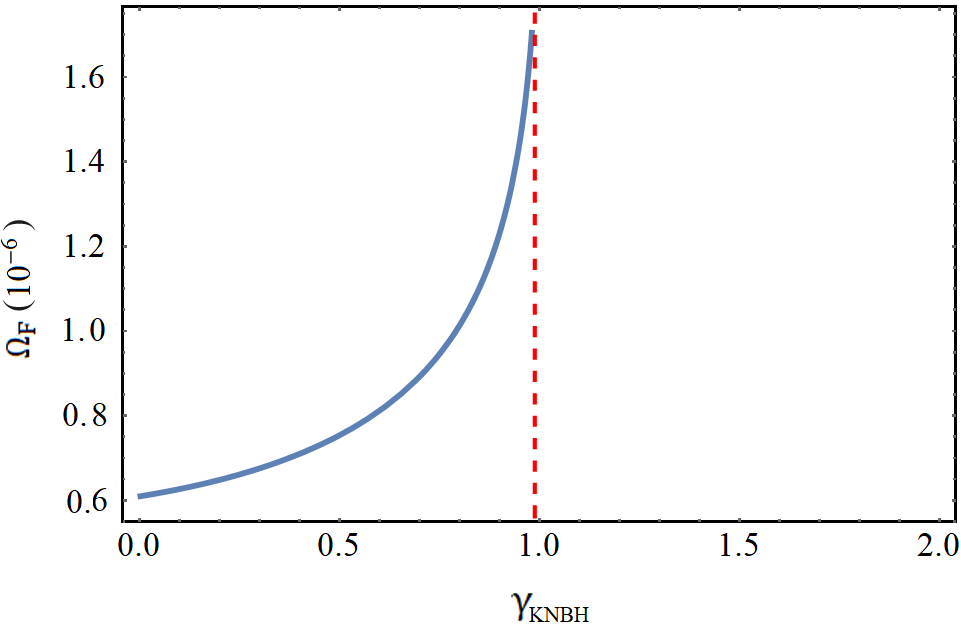}
\caption{The plot of $\gamma_{\rm KNBH}$ vs. $\Omega_{F}$ for a charged rotating BH with $M=65 {M}_{\odot}$ and $a=0.67$.}
\label{image-2.4}
\end{center}
\end{figure}

In the context of magnetic fields, the gyromagnetic ratio determines the strength of the magnetic field produced by a spinning charged particle hence, it supports our statement that the BH charge will enhance the magnetic field of BH during its rotation. While solving Eqs. (\ref{Acc1}) and (\ref{ang_vel}), one can't eliminate $M$ and $J$ to get a suitable solution. Hence, considering a fixed contribution of gyromagnetic ratio in Eq. (\ref{ang_vel}) and neglecting the complex factors due to the smaller value of $\gamma_{\rm KNBH}$ for the evaluation of KNBH can lead us to evaluate remnant mass by using the evaluation relation of $M$ and $J$. It is also a fact that the gyromagnetic ratio of KNBH and Dirac electron is always $<2$ e.g., \cite{Qiu:2021qrt}. The relation of gyromagnetic effect vs. angular frequency $(\Omega_{F})$ is also drawn in Fig. (\ref{image-2.4}). The plot shows an increase in the value of $\gamma_{\rm KNBH}$ with a small increase of the $\Omega_{F}$ value, but as the $\gamma_{\rm KNBH}$ crosses a certain limit, the $\Omega_{F}$ value increases without any significant increase in $\gamma_{\rm KNBH}$ value. It shows a maximum effect of the gyromagnetic ratio on the $\Omega_{F}$ and the magnetic field strength. The gyromagnetic ratio approaches its maximum value i.e., $\sim 1$, which means the charge becomes neutral in a very short interval \cite{Liu2016}.

\begin{equation}
J^2=M^4\left(2 x-x^2\right)+\gamma,
\end{equation}
and by differentiating, we get
\begin{equation}
2J\frac{dJ}{dt}=2M^4(1-x) \frac{dx}{dt}+(2x-x^2)M^3\frac{dM}{dt},
\end{equation}
hence, Eq. (\ref{Acc1}) gives
\begin{equation}\label{mass}
\frac{dM}{dt}= \frac{J}{2 x M^3}\frac{ dJ}{dt}\quad \Rightarrow \quad M \approx M_o \sqrt{\frac{x_o}{x}},
\end{equation}
here after the integration, we used the series expansion of the exponential term, to obtain Eq. (\ref{mass}). This relation could give the remnant mass merging binary. Let us consider the first predicted event GW150914 to be a KNBH surrounded by a magnetized accretion disk. During the merging process, such a BH is retarded due to the energy extraction by a weak transient and leaves the remnant (irreducible) mass of
\begin{equation}\label{Nmass1}
M=0.954 ~M_o=62.02~M_\odot,
\end{equation}
here the initial spin exactly after merging BHs (approximately 0.736) is greater than the final spin $(0.67)$. The difference may be due to the energy loss and the gyromagnetic effect during the merging process i.e. at the instant of BHs collision, the energy loss due to angular momentum transport lowers the spin of the remnant mass but the rotating ionized fluid supports the magnetic field via induction to maintain the stable spin. The subsequent electromagnetic radiation emission leads to spin-down of the BH to a stable spin that must be comparable to the particle's spin at ISCO. The amount of rotational energy extracted during this process is $M_o-M=3~M_{\odot}$ can be treated as energy loss by BBH during joint emission of GW- SGRBs. It means that $\sim 3~M_\odot$ from the initial mass has been extracted to power the jet of $4.616\%$ mass-energy from a slowly rotating BH \cite{Wald:1984rga}. This amount of energy is used to power the Poynting flux along the direction of the twisted magnetic field e.g., \cite{Blandford1977, Lee:1999se}. The remaining energy will be associated with the irreducible mass to increase the entropy of BH. The timescale for the GRB emission can be defined as the ratio of BH power to the power output. i.e., \cite{Lee:1999se, LIGOScientific:2016aoc}

\section{Jet Luminosity and its variation}\label{sec: accretion}

Bing Zhang \cite{Zhang:2016rli} stated that if at least one BH in binary merger carries a certain amount of charge then the inspiral generates a current loop circuit inducing a dipole moment to power a Poynting flux. The orbital decay rate due to the GW emission is investigated as
\begin{equation}
\frac{da}{dt}=-\frac{2c}{5}\hat{a}^{-3},
\end{equation}
where $\hat{a}$ is the distance normalized to $2 r_{s}$ and $c$ is the speed of light. The rapid evaluation of orbital separation (before merging) is considered to lead to a change in magnetic flux  powering the Poynting flux. Next, by using the magnetic dipole moment, a relation between wind luminosity and GW luminosity is obtained as
\begin{equation}
L_{\rm w} \approx 0.4 L_{GW}\hat{q}^2 \hat{a}^{-10},
\end{equation}
where $L_{\rm GW}\approx 3.6\times10^{56} \hat{a}^{-5}~\rm erg ~s^{-1}$ and ${\hat{q}}$ is the dimensionless parameter $\hat{q}\ll 1$. As for a charged rotating BH $\hat{a}<1$, the luminosity value gets $<< L_{\rm GW}$ incompatible with the observed value \cite{LIGOScientific:2016vlm}. At large radii, the dissipation energy of Poynting flux in the accretion flow would power SGRB if the BH has a charge of order $\hat{q}\sim 10^{-5} - 10^{-4}$. The main contradiction is that the electrodynamics description of the magnetosphere is left incomplete and, for simplification, the two BHs were supposed to have equal masses circulating in exactly circular inspiral. Furthermore, the magnetic field is taken as equivalent to that of a magnetar. Whereas in the case of the GW150914 event, the mass ratio of the BH is $\sim 0.80$.
Similarly, \cite{Liu2016} proposed a model showing that FRBs originate from the collapse magnetosphere of KNBH where the closed orbits of charged particles in the magnetosphere are unstable. The main focus of this model was the magnetosphere instability of a KNBH and its possible consequences that result in FRBs and their potential afterglow emission. In the present work, we focus on the predicted merger event GW150914, where the charge (in the form of a gyromagnetic effect) has a crucial role in amplifying the magnetic field. Using mass values, we calculated the magnetic field strength and fixed our model to investigate the accretion rate and luminosity for this event as
\begin{equation}
L _{\rm BZ}=f(a_*)\frac{B^2 r_s^2 c}{8\pi}\approx 9.2 \times 10^{50} ~{\rm erg ~s^{-1}},
\end{equation}
here, $a=0.67$ is the BH's final spin the spin parameter $f(a_*)=1-\sqrt{\frac{1}{2} \left(1+\sqrt{1-a^2}\right)}=0.067$, and, $ r_{\rm s} \sim 10^6 \rm cm$ is used with other conventional units to obtain this luminosity value. We see that the obtained luminosity is greater than the observed luminosity. This may be due to several reasons.  One can impose the beaming condition as \cite{Frail:2001qp}
\begin{equation}
L_{ob}=f_b L_{BZ}=1.8 \times 10^{49} {\rm erg ~s^{-1}},
\end{equation}
here, $f_b=1-cos\theta_o=0.0196$ with jet opening angle $\theta_o=11.36^o$, which is consistent with the range of detectable GRBs opening angle. Compared to NS mergers, most BBH mergers may have no observable EM counterparts due to either there being no detected EM counterparts or having no transient at all. For several undetected EM counterparts, there may be several reasons like, some event may damp/ suppress the EM counterpart before its emission ( e.g., the absence of tidally disrupted material that is the most direct source of accretion to power a jet, availability of insufficient energy from the engine, absence of accretion disc, BH may be chargeless, feeding of materials from its accretion by BH) while some factors that make it undetected (including GW150914) even if it is emitted too (e.g., jet opening angle, orbital inclination (i.e., if the jet is not in the line of sight and if the jet is in the line of sight then the viewing angle of the observer has a great effect on the jet luminosity and hence its detection), or weakness of GBM system and its large localization uncertainty). In the case of GRBs from GW150914, the jet might not be emitted along the line of sight \cite{Janiuk2017a, Morsony:2016upv}. Moreover, for any hypothetical jet from BBH mergers, there is no ejecta to interact with (especially when there is a strong magnetic field); hence, the probability of observing EM radiation along the off-axis direction is much lower \cite{Perna2019, Morsony:2016upv}. The gamma rays will likely be beamed during the GW-GRBs emission, while the GW emission is close to an isotropic propagation in all directions. Thus, it is difficult to rule out any particular model for a single event with both GW-EM emissions. Even if a model suggests a large flux, it is possible that the jet was beamed away from Earth, thus undetectable for GBM \cite{Li:2016iw}. This could be the case for many undetected events. Besides these, the existence of other transients and objects like asteroids during the event has also shown their effect on the EM event detection \cite{Kim:2021hhl}. 
From the relativity, the observed luminosity at a given frequency is related to emitted luminosity as \cite{Padovani:1999mw}
\begin{equation}
L_{\text{ob}}=4 \pi  \delta^4 L_{em}
\end{equation}
Here, the Doppler boosting factor (relativistic Doppler factor) is $\delta(\Gamma, \theta) =[{\Gamma (1-\beta  cos\theta_v)}]^{-1}$ \cite{Granot:2002za, LIGOScientific:2017zic}.  As compared to stationary sources, the Doppler factor modifies several measurable quantities for the observation. Using the values of $L_{em}=9.2\times 10^{50} \rm erg ~ s^{-1}$, $\beta=\frac{v}{c}=0.53$ \cite{LIGOScientific:2016aoc}, and $\Gamma$ range from $3.974-10.69$, we manipulate the relation between the observed luminosity and viewing angle as shown in Fig. (\ref{image-2.5}). 

The luminosity vs. viewing angle curve of Fig. (\ref{image-2.5})  shows the apparent dynamics of the jet in the event GW150914. The intrinsic structure of the jet is the distribution of the angular energy within the polar angle. In contrast, the apparent structure of the jet is the observer's frame dependent and hence can be discussed by the Lorentz transformation. Using relativistic beaming one could explain the difference between these two distributions by using the Lorentz factor and the viewing angle $\theta_v$ \cite{Salafia2015}. In the above Fig. (\ref{image-2.5}), $\theta _c=0.055~\rm rad\sim 3.15^o$ (point A) is the core angle for $\Gamma = 3.974$, $\beta \sim 0.53$ with the corresponding luminosity $L=9.2 \times 10^{50} \rm erg~s^{-1}$. From point A to point B, the curve shows the typically detectable luminosity with a viewing angle of $1.23 ~ \rm rad \sim 70.5^o$ to the jet direction, $\Gamma = 6.93$, $\beta \sim 0.53$ with the corresponding luminosity $L=1.017 \times 10^{50} \rm erg~s^{-1}$. If the jet is emitted along the axis of rotation then at point C, we get the luminosity of $1.8\times 10^{49}\rm erg~s^{-1}$ with a viewing angle of $0.42\rm ~ rad \approx 24^o$ with $\Gamma = 10.69$. As $\Gamma >1$, the observer receives emission from the whole emitting region (head of the jet) even if the $\theta_v$ is outside the confines of the intrinsic jet \cite{Hayes2020}. For the luminosity variation, the range of the Lorentz factor is  $3.974- 10.69$, where the maximum value gives the core luminosity at a small viewing angle and the minimum value represents the observed luminosity of the jet at a luminosity distance of $410~\rm MPc$. here we can also see that $\theta _v\propto \frac{1}{\Gamma}$. Since the event GRB 170817A detection, the threshold distance is about $80$ MPc with a threshold signal-to-noise ratio  $\leq 8$. So, it seems that if the GW150914 event may have occurred but due to the lack of the GBM system the detection may not be observed \cite{Li:2016iw}. At a smaller angle, the luminosity is at its peak equivalent to core luminosity but as the $\theta_v$ value increases the luminosity decreases (represents a Gaussian curve). This shows that even if this event is detected, the observed luminosity $1.8\times 10^{49} \rm erg~s^{-1}$ is the minimum luminosity that could be detected from a source at a distance of $410~ \rm MPc$. It is also claimed that the cosmological effects are small at such distances. Therefore, considering negligible cosmological redshift and apparent spectrum one can greatly understand the possible restrictions for many of the BBH events with no detectable EM counterparts.

\begin{figure}
\begin{center}
\includegraphics[width=0.6\textwidth, height=2.2in]{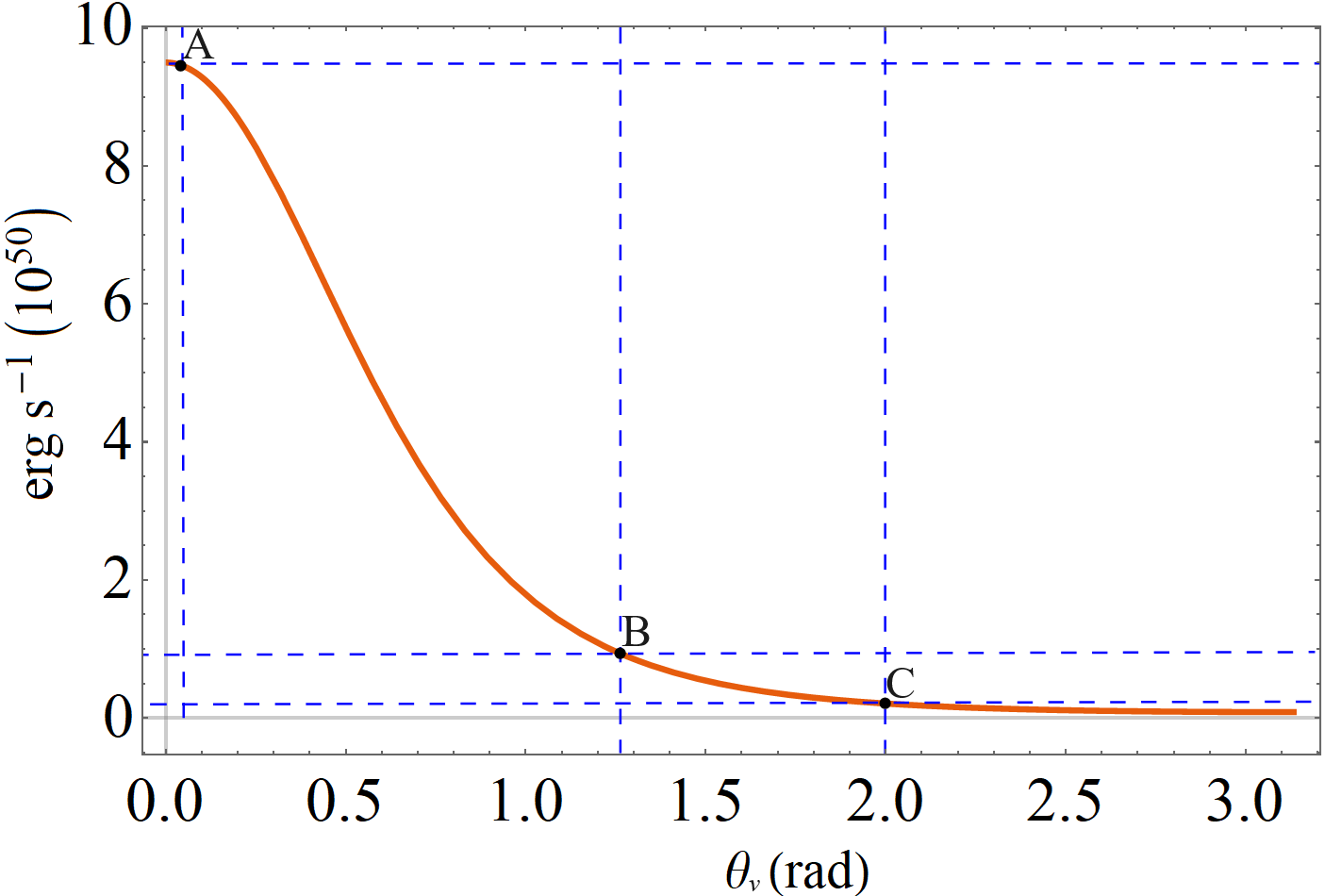}
\caption{The plot of observed luminosity vs. viewing angle of the GRB jet from the event GW150914. Points A, B, and C show the luminosity at different viewing angles and the jet Lorentz factor.}
\label{image-2.5}
\end{center}
\end{figure}

\section{Remarks and Discussion}\label{sec: conclusion}

SGRBs are generally associated with the mergers of NS-NS or NS-BH binaries whereas, the BBH mergers were not expected to power GRBs until the possible weak transient from the event GW150914 (the first predicted weak transient might be associated with GWs from the BBH merger) \cite{Connaughton2016}. If this event has happened from BBHs, then it will be a revolutionary trend in BH and GW physics. In this chapter, we presented a possible scenario for the emission of co-emission of GW associated with the EM counterpart. This work could be a new insight into the new events that could not be explained by existing models.

We examine GW150914 as the BBH merger event where before merging the mutual attracting of a binary caused the surrounding matter to get inspiral around by making a hyperaccretion disc around the final BH. The matter gets ionized due to fraction and finally magnetized with rotation. Such magnetic fields are greatly affected by BH charge via the gyromagnetic effect. As a result, a strongly magnetized disc is generated. In such discs, the MRI-type instability causes turbulence with the magnetic fields serving as an agent to extract rotational energy. The existence of charge and its contribution to the magnetic flux is just the same as the total EMF in the system equal to the total change in magnetic flux through the disk's surface. From such a BH, the emission of energy along the radial direction has less chance due to the magnetic field barrier so, exactly after merging the extractible energy will be pushed toward the axis of rotation due to centripetal force and acting as a source of instability in the final BH. Next, further amplification of the magnetic field due to turbulent dynamo action$(\sim 10^{16} ~\rm G)$ excites the charged particle in the form of electromagnetic waves as a BZ jet e.g., \cite{Blandford1977,deSouza:2009ne}. The fact is that a charged spinning BH with a strong magnetic field and its dynamics could play an important role in accumulating magnetic flux e.g.,\cite{Balbus:1991ay}. So, the angular momentum and charge are two main factors for creating a strong magnetic field around spinning BHs that could extract rotational energy in the form of an intense ultra-short-lived burst. 

Considering the instability of displacing the fluid, we obtained the magnetic pressure and shear force from the decomposition of Maxwell's tensor. These components clearly express the dynamics of the fluid in the accretion disc. Using these forces with gravitational force, we obtained the basic condition for MRI-type instability where the growth rate depends on angular velocity and is independent of the magnetic field strength. This means the magnetic field is an agent of seeding instability in the accreting fluid. Once the limitations to the stability are overcome, the instability sets in resulting in the emission of rotational energy. These limitations are just to overcome the potential and conserved quantities of motion for every single particle to maintain the stable circular motion of the fluid. 

If the mass $M$, angular momentum $J$, and charge $q$ of a BH are satisfied with the basic condition of instability then one can demonstrate the energy and other quantities related to the extraction process of GRBs. The BH charge is crucial during the evaluation process in amplifying the magnetic field via the gyromagnetic effect. An analytic result from the ${\Omega_F-\gamma_{\rm KNBH}}$ relation is shown in Fig. (\ref{image-2.4}). This figure shows an increase in the $\gamma_{\rm KNBH}$ value with a small increase in the $\Omega_{F}$ value but, as the $\gamma_{\rm KNBH}$ crosses a certain limit, the $\Omega_{F}$ value increases without any significant increase in $\gamma_{\rm KNBH}$ value that signifies the maximum effect of the gyromagnetic effect on $\Omega_{F}$ and the magnetic field strength. The maximum value of the gyromagnetic ratio is $\approx 1$, which means the charge becomes neutral as $\Omega_{F}$ approaches its peak value (that shows a very short interval of time). By considering the charge effect other than the gyromagnetic effect, one can't get the exact solution for Eq. (\ref{Acc1}) due to a problem in the elimination of mass $M$ and angular momentum $J$ of a KNBH. So, considering a fixed contribution of the gyromagnetic effect in the evaluation of KNBH could lead us to calculate the remnant mass from the assessment of $M$ and $J$.

Considering the event GW150914 as the merging of BBHs that possibly emits a portion of its rotational energy in the form of GW-EM energy. We evaluate the mass and angular momentum of the final BH with a gyromagnetic effect and investigate remnant mass equal to $ M=62^{+4}_{-4}~M_{\odot}$ where the initial spin exactly after merging is found to be $0.736$ and the final spin at the time of predicted GW-EM radiation emission was $0.67$. Initially, the BBHs spun fast but at the instant of BBH's collision, the energy loss due to angular momentum transport lowers the spin of the remnant mass. Still, the rotating ionized fluid supports the magnetic field via induction to maintain the spin of the irreducible remnant mass. The energy extraction leads to a spin-down of the BH that is comparable to the particle's spin at ISCO. The amount of rotational energy extracted during this process is predicted to be $\sim 3 ~M_{\odot}$ in the form of GWs associated with a weak transient. We investigated the magnetic field of order $\sim 10^{15}~\rm G$ to create turbulence. This is enough magnetic field to power the Poynting flux along the axis of rotation. In all this situation, we can't neglect the effect of BH's charge with the existence of a strong magnetic field that is a key factor in collimating the jet and extracting the rotational energy from BH. The maximum luminosity emitted during this process is found $9.2\times 10^{50} ~{\rm erg ~s^{-1}}$ greater than the observed luminosity $1.8 \times 10^{49} {\rm erg~s^{-1}}$ shows that if the jet is emitted along the axis of rotation but not along the line of sight. For this purpose, we analyzed the luminosity variation with the viewing angle as shown in Fig. (\ref{image-2.5}) and other effective factors. The luminosity of the emitted jet decreases with the increase of viewing angle and Lorentz factor. The viewing angle for the observed luminosity is found to be $~24^o$.  Examining the geometry and using the definition of viewing angle \\cite{Chen2019}, one can predict the inclination angle of $166^o$. Most of the BBH mergers have no EM counterparts due to several restrictions but for a jet along the axis of rotation with a strong magnetic field, it must have no ejecta with which the emitted jet could interact so, there is a high probability of  detection if the emitted jet is along the line of sight and as the jet goes away from the line of sight, its luminosity decreases and hence probability of its detection. 

\vspace{1.0cm}
\noindent\rule{16.5cm}{2.0pt}

\newpage
\chapter{An Introduction to Black Hole Physics}

The most striking feature of space-time is a black hole and the most interesting object in the study of gravitational physics. A black hole is considered the final stage of a star's life, after losing its energy. Classically, a black hole is a compact object with strong gravitational attraction even though light also can't come out from it, i.e. a place of no return for anything in it, due to its strong gravitational field. They are characterized by their boundary called event horizon. The existence of any object entering the BH will vanish at a point called the black hole singularity. The observer outside the horizon can not see this point due to the existence of the event horizon. This is also a fact behind the incomplete study of the black hole interior or a secret to our physical world. Due to insufficient information about the interior of a black hole, a singularity may be considered the cutoff in the geometrical structure of space and time. Inside a black hole, the space-time is coupled to infinity. It is thought that singularity is a 1-dimensional point occurring at the center of the black hole, for detail see the schematic fig. ({\ref{image-3.1}}) shown below.

\begin{figure}
\begin{center}
\includegraphics[width=0.5\textwidth]{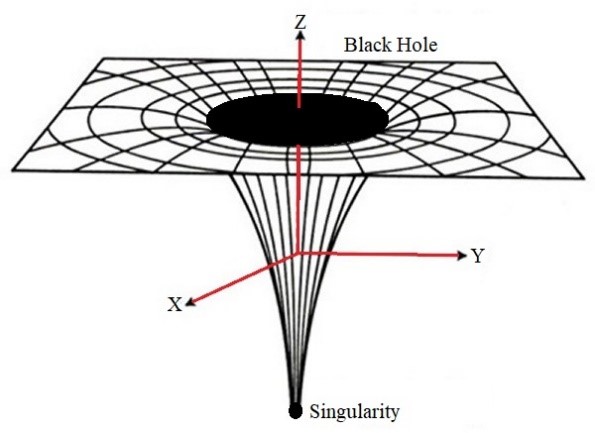}
\caption{Black hole singularity position}
\label{image-3.1}
\end{center}
\end{figure}

The theory of general relativity not only allows us to understand the point of singularity, but it is also unavoidable in our actual life  \cite{ruffini1971}. To completely understand the singularity, one needs to understand the space and time in its accrual meanings. We will discuss some literature related to black holes to understand the basic concept related to this study.

\section{Theory of relativity and black hole}

The special theory of relativity is proposed by Einstein in 1905 \cite{Weber, Einstein, John, rindler1991}. The bases of this theory are based on the Galilean principle of relativity, which states that the fundamental laws of physics are the same for all reference frames moving with uniform motion. The characteristic feature of the special theory of relativity is that it takes into account the speed of light $c$, the structure of space-time, and the equivalence of acceleration and gravity. This theory is mainly summarized in two postulates as below.
\begin{itemize}
\item The laws of physics are the same in all inertial frames reference (frames moving with uniform velocity). This postulate is also called the principle of relativity, and
\item The speed of light $c$ free space is a universal constant.
\end{itemize}

The consequences of this theory include dilation of time, length contraction, the equivalence of mass energy, bending of light, and the prediction of black holes, wormholes the birth of the universe as a result of the Big Bang. This theory explains the concept of formulation for physical laws without gravity, which is also a failure of this theory because one can't consider zero gravity at any orientation or can't consider the relative motion of two bodies in different frames of reference. It also deals with the motion of bodies under the uniform speed relative to each other. The major problem for this theory is how to explain the bodies in two different frames of reference. This means that the notion of including only an inertial frame is not sufficient for completing this theory. For example, the twin paradox. This paradox was experimentally proved in 1971.

After realizing the cases of failure in the theory of special relativity, Einstein in 1915 urged himself to change this theory in the General theory of relativity (simply GR), so that the above problems have solutions in presenting the theory of general relativity. The GR is the generalization for the theory of special relativity. It gives a unified description of gravity, space-time particularly curvature space-time, which is directly related to the energy-momentum of the body having the radiation and mass. It is experimentally verified theory by several tests \cite{Wald:1984rga}. This theory is proved experimentally in many ways e.g. perihelion procession of Mercury's orbit,  the deflection of light rays by the Sun,  the gravitational red shift of light, the equivalence principle, the detection of gravitational waves, and frame-dragging phenomena. The first three tests of these are based on the solution of the Einstein field equation for a spherically symmetric field. In the theory of general relativity, the mass is an important factor for the determination of gravitational effects. The relation is specified by a system of partial differential equations called Einstein Field Equation as given below,
\begin{equation}\label{fieldeq}
G_{\mu \nu }=R_{\mu \nu }-\frac{R g_{\mu \nu }}{2}=\frac{(8 \pi  G) T_{\mu \nu }}{c^2}
\end{equation}

Where $(\mu,\nu)=1,2,3,4$,.  $G_{\mu\nu}, ~R_{\mu\nu}$,  $g_{\mu\nu}$, $G$,  and  $T_{\mu\nu}$ are the Einstein tensor, Ricci tensor, metric tensor, Newton gravitational constant, and the energy-momentum tensor respectively. This equation tells us about the relation of curvature tensor and energy-momentum tensor, that is how the curvature tensor reacts in the presence of energy and momentum \cite{Wald:1984rga, Weber, Carroll:2004st}.  This equation is a counter piece of this theory of general relativity, General relativity theory gives a complete structure and properties of space-time and matter. Solving the Einstein equation completely is not possible, but it can be simplified in different specific forms. The vacuum solution can be obtained by using $R=0$. Consider a $4d$ Riemann space-time with Minkowski type metric signature $(-,+,+,+)$ see \cite{Weber, Carroll:2004st, Misner:1974qy}. Thus the metric takes the form
\begin{equation}\label{genmetric}
\text{ds}^2=\text{dx}^{\mu } \text{dx}^{\nu } g_{\mu \nu }
\end{equation}
using this relation, the Christoffel symbol can be calculated as 
\begin{equation}\label{chrstoffel}
\Gamma ^{\mu }_{\alpha \beta }\text{:=}\frac{1}{2} g^{\mu \sigma } \left(g_{\beta \sigma ,\alpha }-g_{\alpha \beta ,\sigma }+g_{\sigma \alpha ,\beta }\right)
\end{equation}
So, it is easy to define the geodesics of the Riemannian space 
\begin{equation}\label{geoeq}
\frac{\text{Dx}^2}{\text{D$\tau $}^2}\text{:=}\frac{d^2x^{\alpha }}{d\tau ^2}+\Gamma ^{\alpha }{}_{\mu \nu } dx^{\mu } dx^{\nu }=0
\end{equation}
By partial differentiation, the Riemannian tensor can be defined as
\begin{equation}\label{remtensor}
R^{\mu }_{\nu \alpha \beta }\text{:=}2\left(\partial _{[\alpha }\Gamma ^{\mu }{}_{|c}+\Gamma ^{\mu }{}_{\gamma [\alpha }\Gamma ^{\gamma }{}_{|\nu |\beta ]}\right)
\end{equation}
 for detail see also \cite{Wald:1984rga, Carroll:2004st, Misner:1974qy, Stephani:2004ud}. The indices can be summarized as below:\\
Symmetric indices are written in parenthesis 
$$(\alpha \beta) \text{:=}\frac{\alpha \beta +\beta \alpha }{2}$$
Anti-symmetric indices are written in brackets,
$$[\alpha \beta ]\text{:=}\frac{\alpha \beta -\beta \alpha }{2}$$
The generalization of indices is written as,
$$(\alpha \beta \gamma) \text{:=}\frac{1}{3} (\alpha \beta \gamma +\beta \gamma \alpha +\gamma \beta \alpha )$$
The indices written in the vertical strokes $|\alpha\beta|$ are not included in the anti-symmetric process. If the curvature is anti-symmetric, then it will have two pairs of commuting indices. The anti-symmetric terms will vanish, i.e.
$$R_{(\alpha \beta)  \mu \nu}=R_{(\mu \nu)  \alpha \beta }=0 \qquad R_{[\alpha \beta \mu \nu ]}=0 \qquad R_{\mu \nu \alpha \beta }=0$$
Let us take an example to explain the symmetric and anti-symmetric indices. Consider the indices to be $a,b,c$, etc. For two symmetric pairs, we have $ab=ba$, whereas anti-symmetric pairs are $ab=-ba$. Note that here it is important to write that algebraic symmetries must be valid $R_{ab}=R_{ba}$. For a $6\times 6$ matrix, there are over all $20$ independent components. Some main features of field equation are as follows
\begin{itemize}
\item Like all field equations, this equation is also a partial differential equation.
\item This equation is a tensor equation, where the coordinate system is independent from the law of nature.
\item This equation satisfies the poison equation of Newtonian gravitational theory.
\item The mass-density analog in this equation is the energy-momentum tensor. If one needs to find the simplest solution for the above field equation, then it can be concluded that $R=-8 \pi  G T_{\mu\nu}$. For the flat space-time $T_{\mu \nu}\approx0$, then one gets:
\end{itemize}
\begin{equation}\label{ricciten}
R_{\mu \nu }=0
\end{equation}
Which is also called the vacuum solution for the Einstein field equation. To find the complete solution of Einstein's field equation is very complicated, so it will be easy to find the solutions in free space-time. The gravitational fields which are most important in our daily life, are produced by the sun, earth, and moon. These gravitational fields are produced due to slowly rotating nearly spherical mass distribution, so these fields are spherically symmetric. Spherically symmetric fields are simple to understand. So, firstly, we will discuss these fields. In the following section, we will discuss the vacuum solution (also called the Schwarzschild solution) of the Einstein field equation and its effects on space-time.

\section{Schwarzschild (Static) Black Hole Solution}

The metric (which is a tensor equation of second order under the coordinate transformation) is a field describing the nature of space-time and gravity. The simplest form of which was first computed by Karl Schwarzschild from Einstein's field equation for a $4d$ space-time. Schwarzschild solution is also called the vacuum solution and it describes a spherically symmetric static black hole.  The line element of the Schwarzschild space-time in spherical coordinates (${{r,t,\theta,\phi}}$) is given by \cite{Weber, Wald:1984rga, Carroll:2004st, Stephani:2004ud, Gary}.
\begin{equation}\label{schmetric}
ds^2=-\left(1-\frac{2 M}{r}\right) dt^2+\left(1-\frac{2 M}{r}\right)^{-1} dr^2+r^2 \left(d\theta ^2+\theta  \sin ^2 d\phi ^2\right)
\end{equation}

Where $M$ is the mass of a Schwarzschild black hole with a horizon at $r_s=2 M$. Schwarzschild black hole is characterized by its mass $M$. For simplicity, we consider $c=\hbar=G=1$. This line element shows that the singular points are at $r=2M$ and $r=0$ i.e. the origin of flat space in polar coordinates. At these points, the coefficients become infinite. All rays of light travel parallel to each other, such rays make the event horizon (a null surface). All the light cones in the region of the event horizon of the black hole are bent towards the center $r=0$, instead of future timelike infinity. A future-directed light signal ends up at singularity before reaching the future null infinity. This means that at any space-like slice with a black hole, there is always exists a future-directed non-space-like curve, which vanishes at singularity before reaching future infinity. As a result, they remain in the interior of a black hole. Thus, the black hole bifurcation (combining two) is possible to form a single one in late time, but the reverse is not possible. As for a single black hole do not bifurcate reversibly so, this property of black hole leads to the theorem called area theorem of the event horizon. According to this the area never decreases.
\begin{equation}\label{arealaw}
\delta A_{BH}>0
\end{equation}

In Schwarzschild's solution, we will urge to know actually what happens at $r=2M$. Here one can't ignore the effect for $r=2M$. In general relativity, it is important to make a difference between the curvature singularity (divergence of space-time) and coordinate singularity (which is a place where the curvature is the same but the metric component makes a bad choice). If we compute the square of the Riemann tensor, then from metric Eq. (\ref{schmetric}), we get
\begin{equation}\label{remtensor1}
R_{\mu \nu \rho \sigma } R^{\mu \nu \rho \sigma }=\frac{48 G^2 M^2}{r^6}
\end{equation}

From this, we see that for $r=0$ is not only a coordinate singularity but also behaves as a curvature singularity, whereas $r=2M$ is just a coordinate singularity, not a curvature singularity. A curvature singularity results in the breakdown in space-time. Such a type of singularity can't be removed. In contrast, the coordinate singularity is not so important, it can be removed by changing the coordinate by transformation so that it will give a finite value at that point \cite{Heinicke:2015iva}.

\subsection{Tortoise Coordinates}

The Schwarzschild metric describes the motion of particles outside the horizon  $(r>r_s)$ i.e. in the exterior of a black hole but can not describe for $(r<r_s)$. It would be meaningful if one could find the coordinates, which are not singular at the horizon but they can also be extended to the interior of the horizon. This is convenient by using the ingoing or outgoing radial null geodesics. A good choice for this is the use of tortoise coordinates. Let the outgoing and ingoing geodesics be defined for Eq. (\ref{schmetric}) as $ds^2=0$, and $\theta=\phi=comstant$ so that, the above metric yields,
\begin{equation}\label{dszero}
\left(1-\frac{r_s}{r}\right)\left(\frac{dt}{d\lambda }\right)^2-\left(1-\frac{r_s}{r}\right)^{-1}\left(\frac{dr}{d\lambda }\right)^2=0
\end{equation}

From this equation, one can analyze the slope of the light cone in the direction of increasing or decreasing $r$. When $r\rightarrow r_s$ then the angle of the light cone decreases and as the geodesic diverges from $r_s$  the angle of the light cone increases as shown in Fig. (\ref{image-3.20}) above.
\begin{figure}
\begin{center}
\includegraphics[width=0.5\textwidth]{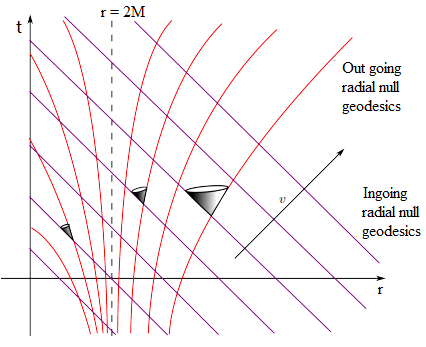}
\caption{In Schwarzschild coordinates the light cone appears to close up as $r\rightarrow 2M$ and for light cone opens when $r > 2M$}
\label{image-3.20}
\end{center}
\end{figure}
Consider the Schwarzschild matrix given in Eq. (\ref{schmetric}), and define the transformation coordinates as

$$\nu\rightarrow t+r^* \qquad \mu\rightarrow t-r^*$$
\begin{equation}\label{metintor1}
ds^2=\left(1-\frac{r_s}{r}\right)\left(dt^2-d{r^*}^2\right)-r^2 d\Omega
\end{equation}
Where the tortoise coordinate is
\begin{equation}\label{r*}
r^*=r+2M ln|\frac{r}{2M}-1|\Rightarrow \frac{dr^*}{dr}=\left(1-\frac{2M}{r}\right)^{-1}
\end{equation}
In these coordinates for $r^*\rightarrow\infty$, when $r=r_s$. Let us consider a set of photons each of them is assigned to a number $\nu$. This number remains constant during the motion of the photon. We choose the number as a new coordinate. Then for radial ingoing and outgoing photons, we can write as
$$\nu= t+r^* (r)=t+r+2M ln|\frac{r}{2M}-1|$$
\begin{equation}\label{munu}
\mu= t-r^* (r)=t-r-2M ln|\frac{r}{2M}-1|
\end{equation}
As the photon moves inwards, the distance to the black hole decreases with time, this time is termed as advanced time. $(t, r)\rightarrow(\nu, r)$. So from the above Eq. (\ref{metintor1}), we can write as,
\begin{equation}\label{tormet2}
ds^2=\left(1-\frac{2M}{r}\right)d\mu d\nu-r^2 d\Omega
\end{equation}
From first part of Eq. (\ref{munu}), we have
\begin{equation}\label{metintor2}
e^{\frac{r^*-r}{2 M}}=\left(\frac{r}{2 M}-1\right)
\end{equation}
$$r^*=\frac{\nu -\mu }{2}$$
$$e^{-\frac{r}{2 M}} e^{\frac{\nu -\mu }{4 M}}=\left(\frac{r}{2 M}-1\right)d\mu d\nu$$
\begin{equation}\label{14}
2M e^{-\frac{r}{2 M}} e^{\frac{\nu -\mu }{4 M}}=\left(1-\frac{r}{2 M}\right)d\mu d\nu
\end{equation}
Let us define
$$e^{\frac{\nu}{4 M}}=V \qquad e^{\frac{-\mu }{4 M}}=U$$
$$ e^{\frac{\nu }{4 M}}d\nu=4 M dV \qquad e^{-\frac{\mu }{4 M}} d\mu=-4 M dU$$
So, from above Eq. (\ref{14}), we gave
$$\left(\frac{2 M}{r}\right)e^{-\frac{r}{2 M}} e^{\frac{\nu -\mu }{4 M}}d\mu  d\nu =-\left(\frac{32 M^3}{r}\right)e^{-\frac{r}{2 M}} dU dV$$
Hence
\begin{equation}
\left(1-\frac{r}{2 M}\right)d\mu d\nu=-\left(\frac{32 M^3}{r}\right)e^{-\frac{r}{2 M}} dU dV
\end{equation}
From the above metric in Eq. (\ref{tormet2}), we get
\begin{equation}\label{32M^2}
ds^2=-\left(\frac{32 M^3}{r}\right)e^{-\frac{r}{2 M}} dU dV-r^2 d\Omega
\end{equation}
For $V>0$, it can be extended to $-\infty <U< \infty$, which describes both the exterior and interior of a black hole. This means that by using tortoise coordinates one can remove the coordinate singularity. Now consider Eq. (\ref{metintor1})
$$dt=d\nu -dr^*\Rightarrow dt^2-dr^{*2}=d\nu ^2-2d\nu dr^*$$
$$dt^2-dr^{*2}=d\nu ^2-2 d\nu  dr\left(\frac{dr^*}{dr}\right)$$
we get
\begin{equation}
dt^2-dr^{*2}=d\nu^2 -2\left({1-\frac{r}{2 M}}\right)^{-1}d\nu  dr
\end{equation}
Use this equation in Eq. (\ref{metintor1}) and simplifying, we get
\begin{equation}
ds^2=-d\nu ^2 \left(1-\frac{2 M}{r}\right)-2 d\nu  dr-r^2 d\Omega ^2
\end{equation}
These coordinates $(\nu,r,\theta, \phi)$ are called outgoing Eddington Finkelstein coordinates. Similarly, for ingoing photons, the resulting metric can be written as
\begin{equation}
ds^2=-d\nu ^2 \left(1-\frac{2 M}{r}\right)+2 d\nu  dr-r^2 d\Omega ^2
\end{equation}
Generally, we can write as
\begin{equation}
ds^2=-d\nu ^2 \left(1-\frac{2 M}{r}\right)\pm 2 d\nu  dr-r^2 d\Omega ^2
\end{equation}
As these coordinates are also defined only on the horizon, but they are regular on the horizon and can be extended to the origin of a black hole. These two line elements are physically different from each other in every aspect. If we take radial null curves for which $d\theta=d\phi=0$, because $\theta$ and $\phi$ are constant and $ds^2=0$ then from Eq. (\ref{r*}) for ingoing geodesics, we can write as
$$d\nu =0=r^*+t=\text{constant}$$
and 
\begin{equation}
\frac{d\nu }{dr}=\left\{\left(0,2 \left(1-\frac{2 M}{r}\right)\right)\right\}
\end{equation}
From this solution, we see that when $r=2M$ falls towards the center $r=0$. Only massless particles $(M=0)$ move on the horizon. Now, if we use these conditions for the second line element in Eq. (\ref{munu}) the result is of the form
$$d\mu =0=r^*-t=\text{constant}$$
\begin{equation}
\frac{d\mu }{dr}=\left\{\left(0,-2 \left(1-\frac{2 M}{r}\right)\right)\right\}
\end{equation}
What is meant by these two results? For Eq. (\ref{metintor2}) everything goes away from the center at $r=2M$. This means that if the first part of Eq. (\ref{munu}) describes a black hole, then in contrast the second part shows a white hole, this issue is briefly discussed below in Fig. \ref{image-3.3}
\subsection{Kruskal Coordinates}

Now consider the Kruskal coordinate, which covers the whole space-time manifold of the maximally extended Schwarzschild solution and it is useful to find the coordinates which don't show the singularity beyond the event horizon. To do so the convenient way is to study the behavior of $r=r_s$ and choose the ingoing and outgoing radial null geodesics. Consider Eq. (\ref{metintor1}) and Eq. (\ref{munu}), we can write as 
\begin{equation}\label{metkrskl1}
ds^2=\left(1-\frac{r_s}{r}\right)dt^2+\left(1-\frac{r_s}{r}\right)^{-1}-r^2 d\Omega ^2
\end{equation}
For this equation the Kruskal coordinates as
$$T=\frac{1}{2}\mu +\nu )=\left(\frac{r}{r_s}-1\right)^{\frac{1}{2}}e^{\frac{r}{2 r_s}}sinh\left(\frac{t}{2 r_s}\right)$$
$$X=\frac{1}{2}(\mu -\nu )=\left(\frac{r}{r_s}-1\right)^{\frac{1}{2}}e^{\frac{r}{2 r_s}}cosh\left(\frac{t}{2 r_s}\right)$$
Using this, the metric along with the Eq.  (\ref{32M^2}) can be written as 
\begin{equation}\label{metTX}
ds^2= \frac{32r^3 _s}{r} e^{-\frac{r}{r_s}}\left ( -dT^2+dX^2 \right )+r^2 d\Omega ^2
\end{equation}
where the term r is defined from
$$\left( T^2-X^2\right) =\left(1-\frac{r}{r_s}\right)e^{-\frac{r}{r_s}}$$
These $(T, ~X,~\theta, \phi)$ coordinates are called Kruskal coordinates. They have many properties in flat and curved space-time
\begin{equation}\label{TXandtheta}
T^2=\pm X^2+constant \qquad T^2-X^2=constant \qquad and \qquad \frac{T}{X}=tanh \left(\frac{t}{2r_s}\right)
\end{equation}
This equation represents the straight line through the origin. For $t\rightarrow\infty$then it shows a null surface, where $T=\pm X$, this surface is similar to that $r\rightarrow r_s$ in Eq. (\ref{metkrskl1}). The region for these coordinates ranges between $-\infty<X<\infty$ and $T^2<X^2+1$. These coordinates are known as Kruskal coordinates and can be drawn as above.
\begin{figure}
\begin{center}
\includegraphics[width=0.6\textwidth]{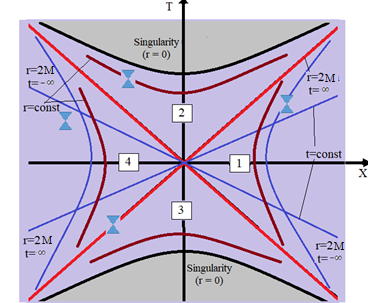}
\caption{A conformal diagram of Kruskal coordinates}
\label{image-3.3}
\end{center}
\end{figure}

Recall the Schwarzschild solution, which describes the simplest black hole. We used it as flat space-time, wherein the metric tensor, the time coordinate has a different sign from space coordinates, this corresponds to the fact that you can move in different directions in space, but only in one direction in one time and time always grows. Schwarzschild solution shows that inside the event horizon of a black hole signs for time and space-time coordinates are interchanged, therefore you can move freely in both directions with time, but you can move only in one direction towards the singularity, this is what makes it inevitable. Even light is also bound to do so. It can't move a millimeter away from the singularity. The best way to analyze different parts of a black hole is to work in Kruskal coordinates, which explains the Fig.  \ref{image-3.3}. In these coordinates, one can make four parts of the total space-time \cite{Carroll:2004st}.
\begin{itemize}
\item The space outside the black hole is in the right quadrant, which places equal r (radius) taking the form of hyperbolas going up and down. Here the event horizon looks like a straight diagonal line $(r=2M)$. This is the space in which our original coordinates are well-defined.
\item Following the future-directed null rays, we reach region $2$. Space inside the black hole is the upper quadrant, with constant $(r=2M)$ taking the form of hyperbolas going from left to right. Singularity is at the zero radius, which also looks like a hyperbola as above in region 2. Points with $t=const$. lie on straight lines crossing the center of the image. One good quality of these coordinates is that light cones always look the same here, light always follows diagonal lines as $45^0$. The future event horizon is the boundary of the region $2$.
\item Following the past null rays, we reach the region $3$. Which is opposite to the region $2$. This region is simply the time reverse of region $1$. This is the part of space from which the bodies can come to region $2$, but can't get back to Region $3$. This region can be thought of as a white hole. The past event horizon is the boundary of the region $3$.
\item The opposite region to our region is the region $4$. This region is connected to our region by a neck-like configuration called the warm hole (as discussed by Einstein Rosen Bridge). This is the mirror image of our region $1$.
\end{itemize}
All these parts are shown in fig. (\ref{image-3.3}) in detail above.
\section{Charged static black hole}
A Schwarzschild black hole is also known as a static charge-less black hole. Here in a charged static black hole, it is clear from the name, that any black hole has a charge. Along with charge, these black holes don't have the property of angular momentum. In general relativity the asymptotic final state of a gravitationally collapsed charged object coupled to electromagnetism is completely described by three quantities, they are mass $(M)$ and charge $(Q)$. All the other degree of freedom evaporates at the time of collapse. According to Wheeler this expression of the static black hole is known as No Hair Theorem \cite{Wald:1984rga}. This theorem is no longer valid, if the non-Abelian gauge field is not present, and the resulting black hole solutions are unstable see \cite{Novikov:1989sz}.

The solution for a spherically symmetric charged black hole can be obtained in the same way as that of a Schwarzschild black hole with function
\begin{equation}\label{RNBH}
f(r)=1-\frac{2 M}{r}+\frac{Q^2}{r^2}
\end{equation}
Where $Q$ is the charge of the black hole. So, from Eq. (\ref{schmetric}), one gets a metric for a static-charged black hole. It is the unique solution of the spherically symmetric and asymptotically flat space-time solution of the Einstein-Maxwell equation.
The only non-zero component of an electrically charged black hole is $T_{tr}=\frac{Q}{r^2}$, and for a magnetically charged black hole, the non-zero component is the magnetic field $F_{\theta\phi}=Qsin\theta$. We will discuss the electric charge case only. The location of the event horizon is the position where $g_{tt}\rightarrow0$, or we can write as 
$$f(r)=0$$
There are two solutions of this equation
\begin{equation}\label{RNBHradii}
r_{\pm}=M\pm \sqrt{M^2-Q^2}
\end{equation}
From this, we see that in a static-charged black hole, there are two horizons, the outer horizon (external horizon $r_+$) and the inner horizon (Cauchy horizon $r_-$). The conformal diagram is given below, where we will examine three cases ($M > Q $	$M=Q$ and $M>Q$) \cite{Wald:1984rga, Carroll:2004st, Gary}. Here we will discuss these cases one by one as below;
\begin{itemize}
\item{Case 1. We consider the first case as $Q<M$ so, from above Eq. (\ref{RNBH}), we can write as
\begin{equation}
f(r)=\frac{\left(r-r_+\right) \left(r-r_-\right)}{r^2}
\end{equation}
This equation shows, that the metric component is singular at $r\rightarrow r_\pm$and $r\rightarrow 0$, but, it also shows curvature singularity at $r\rightarrow0$. In this case, the singularity position is in space and can be avoided by some transformation of coordinates. Such as Eddington Finkelstein coordinates. The inner horizon is shown to be unstable at slight perturbation. Fig. (\ref{image-3.4}(a)) shows a maximally extended Reissner Nordstr$\ddot{o}$m black hole with $Q<M$. If we examine the figure, it shows the complete space-time. The infinity repeats in the vertical direction in proper intervals. There are two singularities, but both of these have opposite signs. It is claimed, that the electric field between these two singularities is always pointing from left to right, which shows that the left singularity has a charge $Q>0$ and that on right has a charge $Q < 0$. 
\begin{figure}
\begin{center}
\includegraphics[width=0.65\textwidth]{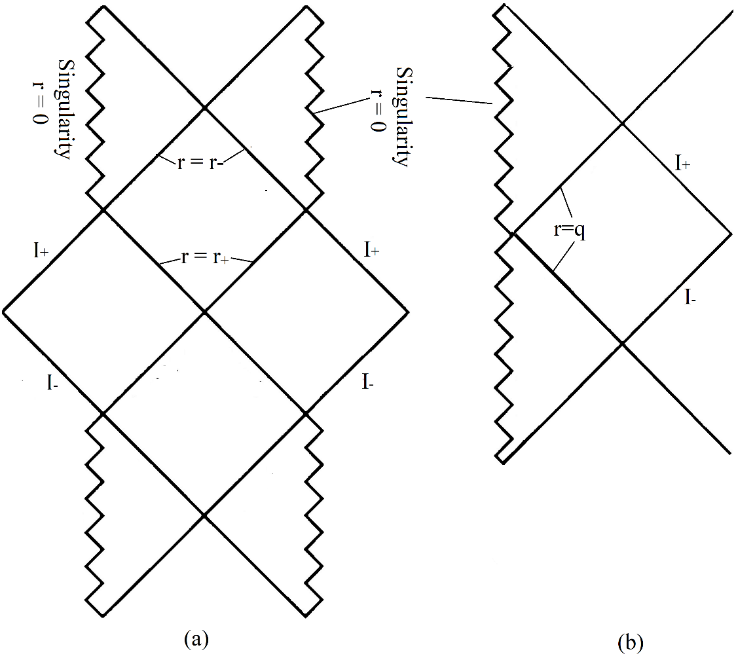}
\caption{Conformal diagram of maximally extended R-N black hole (a) $Q< M$ and (b). Extremal case $(M=Q$)}
\label{image-3.4}
\end{center}
\end{figure}}
\item{Case 2. The second case is $Q=M$ as the extremal limit, so from above Eq. (\ref{RNBH}), we can write as
\begin{equation}
[r]=1-\frac{2 M}{r}+\frac{Q^2}{r^2}=\left(1-\frac{Q}{r}\right)^2
\end{equation}
and the metric becomes
\begin{equation}
ds^2=\left(1-\frac{Q}{r}\right)^2dt^2+\left(1-\frac{Q}{r}\right)^{-2}dr^2+r^2d\Omega^2
\end{equation}
In this case, we have to assign a single charge. Let we take the charge is positive then the structure of the conformal diagram is shown in Fig. (\ref{RNBH}(b)). Similarly, one can take a negative charge for the opposite case.}

\item{Case 3. In this $q>M$ case there is no event horizon. As $r=0$ and $g_{tt}$ is negative, $r=0$ is a time-like singularity. The singularity in this case is called naked singularity i.e. an observer can see the singularity at all. This is not possible, because if so, then the coulomb repulsion will dominate the gravitational attraction and it will need an infinite amount of energy to compress the mass zero volume.}
\end{itemize}
\section{Rotating black hole}
The solution Einstein field equation for a rotating uncharged black hole is known as the Kerr solution. Astronomically, we know that stars are rotating and form a weak field approximating Einstein's field equation. We know the approximate form of the metric at long distances from a static isolated body having mass m and angular momentum $\rm J$ \cite{Carroll:2004st, Misner:1974qy}. The suitable coordinate system can be expressed as 
\begin{equation}
\begin{aligned}
V= \left(1-\frac{2 M}{r}+O\left(\frac{1}{r^2}\right)\right)dt^2- \left(\frac{4 Jsin^2 {\theta}}{r}+O\left(\frac{1}{r^2}\right)\right)dt d\phi \\ +\left(1-\frac{2 M}{r}+O\left(\frac{1}{r^2}\right)\right) \left(dr^2+r^2 d\Omega ^2\right)
\end{aligned}
\end{equation}
This coordinate system is valid for all solar systems. If a rotating star having angular momentum $\rm J$  undergoes a black hole during the collapsing process, then it is expected that it will have angular momentum to retain its initial process.  Kerr black hole is the one, that possesses the mass $M$ and angular momentum $\rm J$, but doesn't have charge i.e. $\rm Q=0$. So, Kerr black hole is characterized by two parameters i.e. mass and angular momentum. Unlike to Schwarzschild black hole, there exist two killing vectors $\partial_t$ and $\partial_\phi$. The Kerr solution is not spherically symmetric and static, but it has axis-symmetric property \cite{Visser:2007fj}. The Boyer Lindquist gives the line element for Kerr black hole coordinates $(t,r, \theta,\phi)$:
\begin{equation}
ds^2=\left(1-\frac{2 M r}{\rho }\right)dt^2 -\frac{4 a M \text{rsin}^2\theta}{\rho ^2}dt d\phi +\frac{\rho ^2}{\Delta }dr^2+\rho ^2 d\theta ^2+\left( a^2+r^2+\frac{2 a^2 M \text{rsin}^2\theta}{\rho ^2}\right)\theta  \sin ^2 d\phi ^2
\end{equation}

$$\Delta (r)=a^2-2 M r+r^2, \qquad \rho ^2(r, \theta )=a^2 \theta  \cos ^2+r^2-2, \qquad J=Ma$$
Where $a$ is the Kerr parameter, $J=Ma$ is the angular momentum. From the above line element, we can simply deduce the Schwarzschild black hole case by using $a=0$. The horizon is located at $\Delta(r)=0$. The external horizon and Cauchy horizon are
\begin{equation}
r_{\pm}=M\pm\sqrt{M^2-a^2}
\end{equation}
Similar to R-N black hole there are also two horizons, they are inner and outer event horizons. The space between the two horizons is called Ergo-sphere as shown in Fig.(\ref{image-3.5}) below
\begin{figure}
\begin{center}
\includegraphics[width=0.7\textwidth]{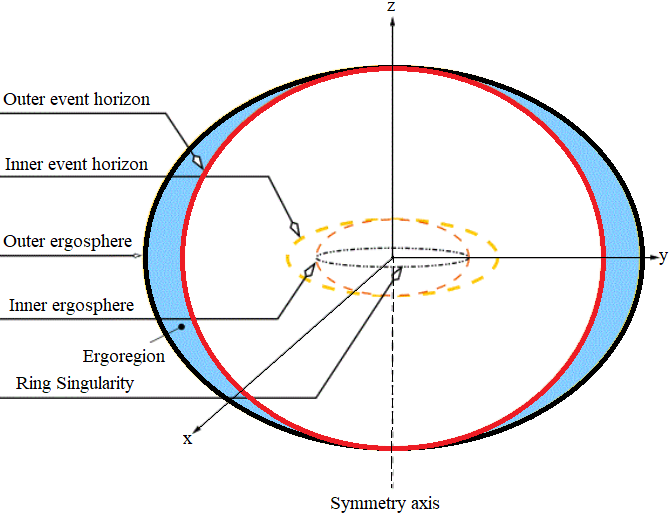}
\caption{A sketch of Kerr's black hole with different parts.}
\label{image-3.5}
\end{center}
\end{figure}
In the above metric equation, if $a\rightarrow 0$ then $r_\pm \rightarrow2M$, which shows that it has a coordinate singularity and coincides with Schwarzschild singularity. There also exists the curvature singularity corresponding to $\rho\rightarrow 0$. As in $R~- N$ black hole, there are also three cases for comparing the angular momentum and mass as follows
\begin{itemize}
\item $a<M$
\item $a=M$
\item $a>M$
\end{itemize}
Here the second case is also called the extreme case. For this case, we can write $J=M^2$. The first and third cases are unstable and have no physical significance. We can construct the laws of black hole thermodynamics using static or rotating black hole solutions \cite{Davies:1978mf, Hawking:1976de, Hawking:1974rv, Hawking:1982dh}. This will be discussed in the next chapter. In addition, we can extend the above discussion to the charged rotating black hole with related results, which is called the Kerr-Newman solution. The above discussion of black holes with mass $M$ can be summarized in the following table below.
\begin{figure}
\begin{center}
\includegraphics[width=0.7\textwidth]{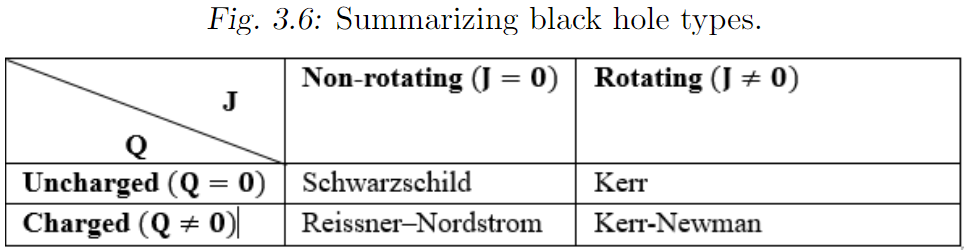}
\label{image-3.6}
\end{center}
\end{figure}

\section{Black hole thermodynamics}

After Einstein's field equation in 1915, Schwarzschild was the first to discover the vacuum solution known as the ''Schwarzschild metric''. Which consists of spherically symmetric space-time components. The distance from the center (where the mass of a collapsing body is compressed) to the boundary of the symmetric sphere is considered the Schwarzschild radius. The boundary of the sphere is termed as ''Event horizon'' the object was named as "Black hole".  Later in the seventies Jacob Bekenstein and Stephen Hawking proposed that black hole emits radiations during quantum mechanical processes near the horizon \cite{Hawking:1971tu, Hawking:1974rv, Li:2018bny}. They calculated the black hole temperature and entropy given below. 

\begin{equation}
T=\frac{\hbar}{8\pi G M} ,\quad  S=\frac{A}{4 \hbar G}
\end{equation}

They proposed that the emission of radiation is a quantum mechanical vacuum polarization which is quasi-static. It evaporates as time passes and eventually, the black hole evaporates. In the first stages, the question raised that from the second law of thermodynamics, we say that the entropy of a closed system never decreases, then what will be the entropy of our world, if a black hole swallows a hot body having certain entropy? We can't ignore the effect because it will contradict the law of thermodynamics. So, it is supposed that to retain the situation for consistency of the system's entropy, it needs to introduce the entropy of the black hole from one phase to the other during the swallowing of the object. In such a way, we can claim the decrease of entropy of the world due to its  transfer into the black hole. This means the second law of thermodynamics requires that a black hole has entropy, It can increase but never be decreased.

Hawking's area theorem was the first step in this regard for black hole thermodynamics, this theorem has a direct relation to the second law of thermodynamics. See the references \cite{Bekenstein:1973ur, Bardeen:1973gs, Hawking:1976de, Wald:1999vt}. Similarly, we can say that the radius of the black hole horizon is proportional to the black hole mass so when something falls into a black hole, it will be an increase in black hole mass, because nothing can come out from it. This supports the notion that a black hole has a non-decreasing area \cite{Gary, Natsuume:2014sfa}. So, the main reason for the black hole thermodynamics is its area and entropy. It is claimed that if the black hole is formed under a collapsed process, then initially it will be asymmetric (i.e. it will have no symmetry) and finally become symmetric to a static black hole. All the physical quantities evaporate during the collapsing process and only a few parameters can describe the black hole, like mass $M$, charge $Q$, and angular momentum $J$. This property of black is analogous to the thermodynamics.  

If we consider the implications of the area theorem in the context of black hole radiation, it becomes apparent that this theorem itself arises from classical general relativity. However, when we introduce quantum effects, such as those leading to Hawking radiation, the situation changes. This radiation suggests that black holes can emit particles, thereby implying a temperature associated with the black hole.  \cite{Hawking:1974sw, Hawking:1974rv, jacobson1996}. From this fact,  Bardeen accomplished the laws of black hole thermodynamics that are analogous to the laws of thermodynamics in our common world \cite{Bardeen:1973gs}. In the following section, we will explain the laws of black hole thermodynamics.

\subsection{Laws of black hole thermodynamics}

In $1973$ Bardeen et. al proposed that the mass of a black hole acts as the energy $(M=E)$ of a thermodynamic system. Its surface gravity $\kappa$ acts like temperature $T$ and the horizon area $A$ acts like entropy $S$ as that of a thermodynamic system. In classical general relativity, there are four laws of black hole thermodynamics. So, these laws are analogous to the law of thermodynamics \cite{Bardeen:1973gs, jacobson1996}. 

\subsubsection{Zeroth Law :}

The zeroth law states that for stationary black holes, the surface gravity is constant over the horizon. This law is valid only in spherical symmetric cases. The spherical symmetry in this contest means that the gravitational force is constant on the horizon. So, to obey the zeroth law, a black hole must be in equilibrium  temperature concerning its surroundings and gravity. The gravitational acceleration can be written as
\begin{equation}\label{acceleration}
a=\frac{M}{r_s ^2}
\end{equation}
The surface gravity (acceleration per unit mass on the horizon of a black hole is called the surface gravity) is
$$\kappa =a r_s=\frac{1}{4 M}$$
so the temperature $T$ of the black hole surface area is
\begin{equation}\label{temperature}
T=\frac{\kappa}{2\pi}
\end{equation}
 For simplicity, we used $c^2=G=1$. The analogous statement of zeroth law in thermodynamics is such that ''in equilibrium condition the temperature of the thermodynamic system is constant.''
\paragraph{\textbf{Theorem}: (Zeroth law of Black hole mechanics)}

\textit{ the surface gravity $\kappa$ is a constant on each consecutively connected component of the future event horizon for a stationary black hole space-time, which satisfies the dominant energy condition $(-T_\nu ^\mu V^\nu$ is future-directed timelike or null vector, where energy-momentum tensor $T_{\nu\mu}$ and future-directed timelike or null vectors $V^\nu$)}

\subsubsection{First Law :}

The first law of black hole thermodynamics is also called the law of energy conservation. This law states that any perturbation in the energy of a static black hole will relate to the change in area, charge, and angular momentum. Mathematically
\begin{equation}\label{firstlaw1}
dE=TdS+\phi dQ+\Omega dJ
\end{equation}
The entropy $S$ and temperature $T$ of the black hole are proportional to the area and surface gravity respectively. So, if we can write the energy in terms of mass one gets

\begin{equation}\label{firstlaw2}
dM=\frac{\kappa}{8\pi} dA+\phi dQ+\Omega dJ
\end{equation}\paragraph{\textbf{Theorem:}}
\textit{In a stationary asymptotically vacuum solution of Einstein equation with bifurcate killing horizon, for a small amount of matter carrying energy  $\delta M$, angular momentum $\delta J$ crosses $\Lambda_H$ and the black hole finally becomes stationary, then for Horizon $H_i=S_i{\bigcap} \Lambda_H$, (where  $H_i$ is the horizon and  $S_i$ is the surface.).They increase by an amount  $\delta{A_H}$ is given by Eq. (\ref{firstlaw2}).}
\subsubsection{Second Law :}

The entropy of a black hole always increases i.e.
\begin{equation}
dS\geq0 \Rightarrow\qquad dA\geq 0
\end{equation}

Bekenstein proposed that some multiple of black hole surface area (which can be measured in terms of squared Planks length) can also be considered as entropy. This statement is conjectured with the second law of black hole thermodynamics, also called the generalized second law (GSL). It can be written as
\begin{equation}
\delta  \left(\frac{A c^3}{G \hbar }+S\right)\geq 0
\end{equation}

From this, we can see that the sum of black hole entropy plus the entropy of the thermodynamic object (which falls into a black hole) is the total entropy and is always greater than zero. This law also explains that when something goes into a black hole, then we can't count for entropy.
\paragraph{\textbf{Theorem}}: (\textit{Second law of Black Hole Mechanics}) 
\textit{Hawking Area theorem \cite{jacobson1996, Hartman:2008b} Let $(M,g)$ be an symptomatically predictable space-time obeying the null energy condition states that $T_{\mu \nu} V^{\nu} V^{\mu}\geq 0$, with $V_\mu$ is a future-directed timelike or null vector. Let $S_1$ and $S_2$ are the Cauchy surfaces for $U$, such that $S_2\subset J^+ (S_1) $ then for Horizon $H_i=S_i{\bigcap} \Lambda_H$,}
$$Area(S_2)\geq Area(S_1)$$
\subsubsection{Third Law:}
The surface gravity of a black hole can't be reduced to zero by any procedure no matter how idealized. This was formulated by Bardeen, Carter, and Hawking  \cite{Bardeen:1973gs}. For a Kerr black hole the surface gravity $\kappa=0$, and if $\frac{J}{M^2}=1$, then it is possible to add mass to the black hole and decrease its angular momentum. Another way is to make a naked singularity i.e. when one gets negative mass. In actual situations, this case is not possible.

\section{Black hole evaporation}

According to Hawking's prediction in \cite{Hawking:1974sw}, a black hole emits Hawking radiations. He investigated that the emission of radiation is the result of particle and anti-particle pair creation in the vicinity of the black hole horizon. They have opposite momenta. In this incident, the particle with positive energy goes outward as a result black hole mass must be decreased at the rate at which the energy is radiated to infinity. For the maintenance of energy conservation, the particle with opposite momenta must act as back reaction \cite{Weber, Einstein}. This process is also known as the Hawking effect. It is claimed that the emission of Hawking radiation is analogous to the radiation from a black body. Then the rate of mass loss by the black hole is given by Stefan Boltzmann's law as:
\begin{equation}
\frac{dM}{d\nu}=-\sigma A T^4
\end{equation}
Where $\sigma$ is the Stefan Boltzmann constant. In the simplest case the Hawking temperature is $T\propto \frac{1}{M}$ and $A\propto m^2$, so we can write it as 
\begin{equation}
\frac{dM}{d\nu}=M^{-2}
\end{equation}

This shows that as the black hole loses mass, the temperature increases and at the final stage the black hole, it radiates all of its mass. The time of black hole mass evaporation is given by
\begin{equation}
\nu\approx M^3
\end{equation}
for which in cgs. system we can claim that 
\begin{equation}
\nu \approx 10^{74}\left(\frac{M}{M_0}\right)\sec.
\end{equation}

which shows the lifetime of a black hole to evaporate its total mass is more than the lifetime of the universe $(10^{74})$ \cite{Wald:1984rga}. For a star with a mass equal to the sun that collapsed into a black hole $M\rightarrow M_0$, its lifetime is approximately $10^{67}$ years. It is proposed in \cite{Stephani:2004ud}, that if any small black hole had formed in the early universe such that it would have been at the final stage of evaporation in recent times. Due to high temperature, a large amount of $\gamma$-rays will be emitted into the background, but since there are no such $\gamma$-rays. Hence, it seems that in our universe there are no such black holes. Recent analysis using the uncertainty principle reveals that during the evaporation the whole mass of the black hole doesn't evaporate; there is always some minimum remnant mass in plank scale order, which shows that black hole entropy never vanishes. 

Classically, in the last stages (when loses its mass down to the order of plank mass) of evaporation the effect on black holes cannot be possible to study. For such a description, one needs to wait for the theory of quantum gravity. The original theory of quantum gravity has yet not be discovered, but there are several ways that can lead us to the desired position. One of them is the consideration of a high-dimensional black hole, and the other is the existence of the Planck length $l_p$ and mass $M_p$.
	
\section{Discussions and conclusions}

In this chapter, we discussed briefly the concept of black holes. Starting from Einstein's theory of general relativity and field equations, we discussed different geometries as a result of solutions from the field equations. In classical physics, a black hole is a compact object with strong gravitational attraction, even though light also can't come out from it, i.e., a place of no return for anything in it, due to its strong gravitational field. A black hole is considered to be the final stage of a star's life after losing all of its energy. They are characterized by their boundary, called the event horizon. The existence of an event horizon is the reason for the incomplete study of the black hole interior by an outsider. In any black hole (static or rotating) an ingoing object will vanish at a 1-dimensional point in the center of a black hole, called a singularity. This was first proposed by Roger Penrose in $1965$. Due to insufficient information about the black hole interior, a singularity may be considered as the cutoff in the geometrical structure of space-time. Inside a black hole, the space-time is coupled to infinity. In $~1970$, Brandon Carter and Stephen Hawking proposed the No-Hair theorem, which shows a black hole can be characterized by mass $M$, charge $Q$, and angular Momentum J. Considering the black hole as a thermodynamic object, they also investigated the four laws of black hole thermodynamics. Which is also revised here. Hawking proposed that a black hole can emit radiation due to quantum effects.

\vspace{1.0cm}
\noindent\rule{16.5cm}{2.0pt}

\clearpage
\thispagestyle{empty}
\hfill
\clearpage
\newpage

\newpage
\chapter{The CR Volume for Black Holes and the Corresponding Entropy Variation: A Review}

In this chapter, we reviewed the work done on black hole interior volume, entropy, and evaporation. An insight into the basics for understanding the interior volume is presented. A general analogy to investigate the interior volume of a black hole, the associated quantum mode's entropy, and the evolution relation between the interior and exterior entropy is explained. Using this analogy, we predicted the future of information stored in a BH, its radiation, and evaporation. The results are noted in tables (\ref{tab:1}) and (\ref{tab:2}). To apply this analogy in BH space-time, we investigated the interior volume, entropy, and evaluation relation for different types of BHs. Finally, we also investigated the nature of BH radiation and the probability of particle emission during the evaporation process.
\section{Terminology} \label{terminology}

\subsection{Black hole interiors} \label{BH-int}
A Black Hole (BH) is a region of space-time with intense gravity such that nothing could escape from its surroundings \cite{Wald:1984rga, Weber, Carroll:2004st, Stephani:2004ud, Misner:1973prb}. According to Hawking's uniqueness theorem \cite{Hawking:1971vc}, all BHs belong to the Kerr Newman BH family that is characterized by its mass $m$, charge $q$, and angular momentum $J$ \cite{Mazur:2000pn, Saida:2011wj, Hawking:1976de}. Its horizon behaves like a trapped region where both the future-directed and null geodesics are orthogonal to it but converging due to strong gravity, i.e., the outgoing light is also dragged inward. The horizon is a cut-up region between its interior and the surrounding space. The study of the BH interior has challenging issues due to the interchanging space-time coordinates across the horizon \cite{Hawking:1994ss}.  The cosmic Censorship hypothesis work of Roger Penrose has provided insight into BH interiors \cite{Penrose:1999vj}. An important point of this work is that the inner horizon of a BH (charged, rotating, or both) should be unstable. Recently, a deep concern has been raised regarding these hypotheses in using mathematical relativity, semi-classical gravity, and numerical studies.  
\cite{Penrose:1971uk, Simpson:1973ua}.

\subsection{Hyper-surface}\label{hyper-surface}
Generally, a hyper-surface is the generalization of an ordinary $ 2-$dimensional surface embedded in $3-$dimensional space to an $(n-1)-$dimensional surface embedded in an $n-$dimensional space \cite{Christodoulou:2014yia, Christodoulou:2016tuua, Hsu:2007dr}. As the investigation of interior volume in flat space-time is not the same as in curved space-time, it is important to understand the meaning of hypersurface in both flat and curved space-time \cite{Misner:1973prb}.

\subsubsection{In flat space-time}\label{flat-spce-time}
Mathematically the interior volume $V$ bounded by a sphere $S$ of radius $R$ immersed in flat Minkowski space-time is $\frac{4}{3}\pi$ times its cubic radius \cite{Dolan:2013ft, Wang:2021llu, Johnson:2019wcq}. At the same time, its surface area is $\pi$ times to its squared radius. It is a $3-$dimensional space-like spherically symmetric hyper-surface $(say \sum)$ bounded by that sphere $S$ having volume $V$ \cite{Estabrook:1973ue, Cordero-Carrion:2001jpf}. It is the maximal volume determined by the largest $\sum$ in the interior of the sphere $S$. So, to define the interior volume bounded by a sphere $S$ in Minkowski space-time, one should choose the largest $\sum$ in the interior $S$. This  $\sum$ must satisfy two primary conditions:

    \begin{itemize}
        \item the simultaneity condition, and 
        \item It must be the largest space-like spherically symmetric hyper-surface bounded by sphere $S$.
    \end{itemize}
    
Both conditions could be equivalently treated in Minkowski's space-time \cite{dInverno:1992gxs, Weber, Carroll:2004st}. Consider a sphere $S$ with coordinates $R^2=x^2+y^2+z^2$ at $t=0$. A spherically symmetric hyper-surface bounded by a sphere $S$ can be defined as $t=t(r), ~ r \in[0, R]$, and $t(R)=0$ and the interior volume bounded by this hyper-surface is  
 \begin{equation}\label{V1}
 V=4\pi\int_0^R [r^4(1-[\partial_rt(R)]^2)]^\frac{1}{2} dr,
 \end{equation}
This equation shows the maximum volume bounded by the largest space-like hyper-surface at $t(R)=0$ is $V=\frac{4}{3}\pi R^3$ \cite{Padmanabhan:2002sha, Parikh:2005qs}. The same case can be understood by analyzing Fig. \ref{image-4.1}, where for an inertial observer, the simultaneity surfaces are straight lines in a $t-r$ plane. 

From the above two conditions, let us consider $R$ is the radius of the largest hyper-surface at time $t(R)=0$, then we can write the metric component as $g_{\mu v}=(0,~R)$, and the metric is $ds^2=dR^2$ so, its maximal volume is the same as $V=\frac{4}{3}\pi R^3$. Now, let us consider $R _i$ as another hypersurface at $t(R)=t_i$ with $i=1,2,3,4$, then the metric component $g_{\mu v}=(-t(R_i),~ R_i)$, the metric is $ds'^2=dR_i^2-dt(R_i)^2 $ that gives the volume $ V_i$. On comparison, we will get the volume as $V> V_i$, and hence $ds^2 > ds'^2$. Note that any spherically symmetric sphere can reside only one largest $\sum$ inside it.  The inertial frame will be defined for the $\sum$ bounding of the largest volume (also called the proper volume). Subsequently, from Eq. (\ref{V1}), we can say that any contribution to the time-like direction can reduce interior volume.
 
 \begin{figure}[H]
\begin{center}
\includegraphics[width=0.4
\textwidth]{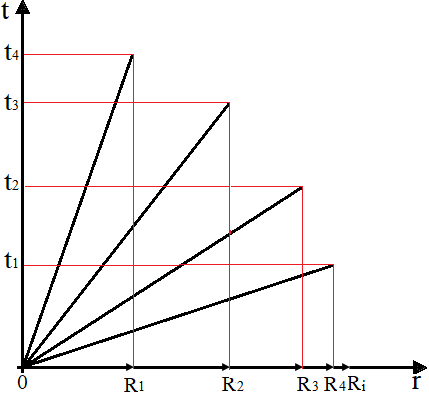}
\caption{Choosing the maximal hyper-surface in flat Minkowiski space-time with $t=0$ results in the maximal interior volume of the sphere $S$. This figure shows that any contribution to the time-like direction leads to a reduction of the interior volume.}
\label{image-4.1}
\end{center}
\end{figure}

\subsubsection{In curved space-time}\label{curved-space-time}
In curved space-time, the topic of BH interior volume is not as simple as that in Minkowiski space-time \cite{Carroll:2004st, Mann:2015luq, Parker:2009uva}. It demands different techniques to be adopted for choosing the largest space-like hypersurface.  Firstly, Christodoulou and C. Rovelli (CR) \cite{Christodoulou:2014yia, Christodoulou:2016tuua} investigated the maximal interior volume of BH after defining the largest hyper-surface in the interior of BH. According to their work, the largest hyper-surface bounded by an $n-$sphere can be explained by a Penrose diagram as shown in Fig. (\ref{image-4.2}). In this figure,  a space-like curve is drawn (from the horizon to the center of the collapsed object) and is divided into three parts labeled as $(1)$, $(2)$, and $(3)$. Section $(1)$ connects the hyper-surface to the horizon and is a null part of this hyper-surface. The long stretch part of the hyper-surface (i.e., section $(2)$) has a nearly constant radius (say $r_v$). According to CR work, the main contribution to the interior volume of a BH comes from section $(2)$, and section $(3)$ is part of the hyper-surface that connects the long stretched part to the center of the collapsing object i.e., $(~ r = 0)$. Let the volume bounded by each part be $V_1,~ V_2$, and $V_3$. By this terminology, $V_1=0$, and $V_3$ is the volume of the collapsed part. Since the space-time inside a collapsing object acquires a time-like killing vector field thus it has a finite volume. Hence, $V_3=constant$ (will give a finite contribution to the interior volume of the BH). Therefore, the total contribution of these two parts at the longest Eddington time $(v)$ is finite and can be ignored. This means that the total contribution to the BH's interior volume comes from $V_2$. The total volume bounded by the largest hyper-surface is $V$ which increases linearly with $v$ due to $V_2$. So, we can take part $(2)$ as the largest space-like hyper-surface with $r=r_v$. Hence, the interior volume bounded by the BH at $r=r_v$ will be its maximum interior volume $V$.

\begin{figure}
\begin{center}
\includegraphics[width=0.4\textwidth]{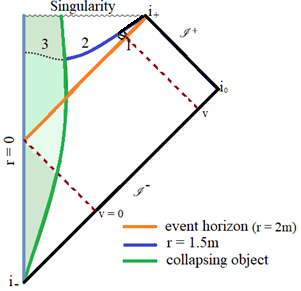}
\caption{(a) The largest hyper-surface (red line) drawn from the horizon to the center of a collapsed object. Part $1$ is the null part, Part $2$ is the stretched segment between Part $1$ and the boundary of the collapsed surface, and Part $3$ is the collapsed part with a finite volume. (b) A Penrose diagram of Kerr's geometry showing the largest hypersurface (red line) and its segments labeled by $1, ~2, ~\rm and ~3$.}
\label{image-4.2}
\end{center}
\end{figure}

\subsection{Advanced time or Eddington time (\textit{v}) and Space-time metric}\label{adv-time}
The metric on an arbitrary hyper-surface is generally written as \cite{Chandrasekhar:1985kt, Hawking:1973uf, Wald:1984rga, Weber, Carroll:2004st, Stephani:2004ud, Misner:1973prb} 
\begin{equation}
    ds^2=g_{\mu v}dx^{\mu}dx^{v},
\end{equation}
here $g_{\mu v}$ is the space-time metric. Due to its quadratic character, it is always easy to diagonalize by coordinate transformation. For example, the Schwarzschild metric case 
\begin{equation}
ds^2=-g_{00}dT^2+h_{ij}dx^idx^j=-\left(1-\frac{r_s}{r}\right)dt^2+\left(1-\frac{r_s}{r}\right)^{-1}dr^2+r^2\left(d\theta^2+sin^2\theta d\phi^2\right),
\end{equation}
There is an artificial curvature singularity at $r=r_s$ due to the bad selection of space-time coordinates. To avoid this singularity, one needs to choose the Eddington-Finkelstein coordinates where the time coordinate (Eddington time) is replaced as
\begin{equation}\label{Edd}
v:=t+r^*=t+ \int{\frac{dr}{f(r)}}=t+r+r_s ln(r-r_s), \qquad f(r)= 1-\frac{2M}{r},
\end{equation}
and the new Schwarzschild metric becomes
\begin{equation}\label{EDmet+T}
{ds}^2=-dT^2+\left(-f(r) \dot{v }^2+2 \dot{v } \dot{r}\right)d \lambda ^2+r^2d\Omega^2,
\end{equation}
here 
$$g_{TT}=-1,\quad g_{\lambda \lambda }=\left(-f(r)\dot{v }^2 +2 \dot{v } \dot{r}\right), \quad g_{\theta \theta }=r^2, \quad g_{\phi \phi }=  r^2 \sin ^2 \theta,$$

It is a time-dependent equation so the interior of the BH is not static, i.e., for any constant radial component, the hypersurface is variable at the given time $T$. Moreover, as these investigations are made for $v>>M$, and $r=\frac{3}{2}M$ near the largest hyper-surface as $T$ increases, the proper time between two neighboring hyper-surfaces must tend to be zero, and there will be no evolution for the BH. This means that the statistical quantities will not be affected by the time-dependent character of the metric. In our investigations, $T$ is approximately constant. That is why the space-like hypersurface leads to the CR volume. Next, we follow the statistical way to find the entropy of the quantum in the scalar field in Schwarzschild's BH. 
\begin{equation}
ds^2=(-f(r) \dot{v}^2+2\dot{v}\dot{r})d\lambda^2+r^2 (d\theta^2+sin^2\theta d\phi^2),
\end{equation}
These coordinates are acceptable for investigations in the interior of a BH. In the case of rotating BHs \cite{Wald:1984rga, Weber, Carroll:2004st, Wiltshire:2009zza, Narlikar:1986kr}, the Eddington form can be written as 
\begin{equation}\label{Kerrmet}
ds^2=-\frac{\Delta -a^2   \sin ^2\theta}{\rho ^2}dv^2+\rho ^2 d\theta ^2+2 dv dr+\frac{A^2  \sin ^2\theta}{\rho ^2}d\phi^2-2 a  \sin ^2 \theta dr d\phi -\frac{4 a   M r^2 sin\theta^2}{\rho ^2}dv d\phi,
\end{equation}
where $\Delta$, $\rho$, $A$, and $J$ have conventional meanings as in literature. Similarly, the metric in lower space-time dimensions (like BTZ BHs) is \cite{Banados:1992wn, Carlip:1995qv, Emparan:2020znc, Gukov:2003na}

\begin{equation}\label{E-Fmetric}
ds^2=-f(r) dt^2+\frac{dr^2}{f(r)}+r^2 (N^\phi (r) dt+d\phi)^2,
\end{equation}
here, $f(r)$ is the lapse function, and $N^\phi$ is the shift function defined as 
\begin{equation}
N^\phi (r)= -\frac{J}{2r^2},  \left (\left | J \right |\leq ml  \right ),
\end{equation} 
The Eddington form of this Eq. (\ref{E-Fmetric}) can be written as 
\begin{equation}\label{collapsedmass}
    ds^2=(-f(r) \dot{v}^2+2\dot{v}\dot{r})d\lambda^2+r^2 (N^\phi (r) dt+d\phi)^2,
\end{equation}
here, $l$ is the AdS radius of BH (related to cosmological constant $\Lambda=-\frac{1}{l^2}$), $\phi$ is the period ranging from $0$ to $2\pi$, $J$ is the angular momentum related to angular velocity $\Omega (r)$ and $m$ is the AdS mass. For $J=0$ (static BH), we get 
\begin{equation}\label{metric}
ds^2=-f(r) dt^2+\frac{dr^2}{f(r)}+4r^2 d\phi^2=(-f(r)\dot{v}^2+2 \dot{v} \dot{r}) d\lambda^2+4r^2 d\phi ^2,
\end{equation}
Some other geometries of space-times metrics in the Eddington coordinate can also be found in the literature e.g. \cite{Ong:2015tua, Ali:2019icq, Ali:2018sqk}. Note that the actual meaning of Eddington's time remains the same as the definition above (\ref{Edd}). 
\subsection{Black hole interior volume}\label{int-vol}

The main purpose of the BH study is to understand the nature of BH which is up to date a mystery in one way or another. Many attempts are made to probe its full structure and properties. Among these, the interior volume is a factor probed by many authors e.g., \cite{Grumiller:2005zk, DiNunno:2009cuq, Ballik:2010rx, Cvetic:2010jb, Gibbons:2012ac, Ballik:2013uia, Finch:2012vli, Iliesiu:2021ari, Chew:2020twk, Davidson:2010xe}. Such an attempt was made in 2015 by M. Christodoulou and C. Rovelli (CR) \cite{Christodoulou:2014yia} to solve the problem of BH interior volume by choosing the largest space-like hyper-surface in the interior of a spherically symmetric BH.  They considered the BH formed under the collapsed process (\ref{collapsedmass}) so, by using Eddington Finkelstein coordinates they defined the interior volume of BH.

Using CR's notion, this work was extended to the rotating BH in Ref. \cite{Bengtsson:2015zda}, to charged rotating BH \cite{Wang:2019ear, Haldar:2023pcv, Biro:2019rms}, RN BH case is discussed in \cite{Han:2018jnf, Wang:2018txl, Haldar:2019buj}, quasi-static spherically symmetric charged black hole \cite{Jiang:2020rxx}, the BTZ BH in Refs. \cite{Zhang:2019pzd, Ali:2020qkb}, its Lagrangian formalism are demonstrated in \cite{Maurya:2022vjd}, divergent volume \cite{Zhang:2020gbv}, non-commutative BH \cite{Zhang:2016sjy} and many others.

\subsection{Black hole exterior and interior Entropy}\label{int-ext-entropy}

In thermodynamics, the term entropy refers to the microscopic and macroscopic connections of a system e.g., the entropy of a gas is the microscopic heat transfer to the available number of micro-states for the gas molecules \cite{Jacobson:2005kr, Frolov:2018awz, Bekenstein:1973ur, Bekenstein:1972tm, Wald:1999vt, Wald:2002mon, Srednicki:1993im, Frampton:2008mw, Egan:2009yy, Bhaumik:2016sav}. Similarly, the BH entropy is a test for unifying gravitational and quantum mechanical theories \cite{tHooft:1984kcu}. According to Bekenstein and Hawking \cite{Bekenstein:1973ur, Hawking:1975vcx}, the BH is a thermodynamical object having entropy which is irreversibly related to its surface area. This entropy is called Horizon entropy or Hawking entropy denoted by $S_{BH}$   \cite{Bekenstein:1972tm, Bekenstein:1973ur, Almheiri:2020cfm, Strominger:1996sh}. Later, the four laws of BH thermodynamics are proposed in Ref. \cite{Bardeen:1973gs}. These investigations started a new dialogue of BH thermodynamics as a wide research area in BH's Physics. Several techniques in connection with the thermodynamics of BH were presented to explain its structure and properties. Due to the mysterious nature of  BH, the thermodynamics claim Hawking raised the questions of radiation and information loss which is termed an information paradox. It is suggested, that the entropy of the thermodynamics system is related to the information contained in the BH system \cite{Marolf:2017jkr}. So, by finding the correct entropy one can solve the problem of the information loss paradox. In search of these facts, Baocheng Zhang followed CR's investigations and found the entropy for the quantum modes of the scalar field. Both the interior volume and associated entropy are proportional to Eddington time \cite{Zhang:2015gda}. It means that the entropy variation in the interior of a BH may affect the statistical quantities. This is the main point to consider for understanding the BH interior. This point has been studied by several authors and found that the BH entropy variation can lead to understanding the interior of BHs with the evaporation and the information loss paradox \cite{Hawking:1974rv, Hawking:1996ny}.

\section{Introduction\label{introduction}}
A BH has a horizon making it a precise object compared to those we can see in our surroundings. The existence of the BH horizon is the main reason for the interchange of coordinates across the BH horizon \cite{Penrose:1969pc, Rindler:1956yx, Ng:1993jb} and creates the problem of the information paradox \cite{Hawking:1976ra}. After introducing the BH thermodynamics and its horizon the main question was "\textit{What is the size of a BH? How is the interior of BH and what is the future of an object entering BH's boundary?}" Amongst many attempts for these questions, Parikh discussed the interior volume of BH using stationary space-time \cite{Parikh:2005qs}. His proposed volume was independent of  time coordination. It means that his proposed volume for the interior of BH was constant in time. As for the information exchange from BH to its surroundings, the Bekenstein and Hawking entropy needs the BH's internal volume to be time-dependent (i.e., variable). The main point for understanding the BH interior is a space-time with a finite horizon area and infinite volume. When the radius goes to infinity the horizon must be constant. If this case is possible, one could construct the largest hypersurface that bounds the maximal interior volume inside the horizon. In Ref \cite{Grumiller:2005zk}, the BH volume was found proportional to BH's area. DiNunno and Matzner \cite{DiNunno:2009cuq} suggested that for the volume inside a BH, one needs to define particular three-space coordinates to evaluate the interior volume \cite{Caticha:2005qd}. These three-space coordinates may be explicitly time-dependent but have a limit of the integral to compute this volume. A similar $3d$ space coordinates can be constructed about a rotating BH (that must be axisymmetric rather than spherically symmetric). Any definition of time in a rotating BH space-time leads to the possibility of evaluating the volume bounded by the horizon. Ref. \cite{Cvetic:2010jb}, treating the cosmological constant (Pressure) as a dynamical variable in the first law of BH thermodynamics, and  the thermodynamics volume was calculated for static multi-charge solutions in four, five, and seven-dimensional gauged supergravities \cite{Ashtekar:1984zz}; rotating Kerr-AdS BHs in arbitrary dimensions \cite{Gibbons:2004uw}; and certain charged rotating BHs in four and five-dimensional gauged supergravities \cite{Myers:1986un} etc. For non-rotating BHs \cite{Tangherlini:1963bw},  the thermodynamical volume was claimed to be an integral of the scalar potential over the interior volume of BH. In contrast, for rotating BHs, the thermodynamics volume and the geometric volume differ from each other by a shift related to the angular momenta of the BH. Similarly, several other investigations are made to evaluate the interior volume of BH in such a way that it could satisfy the interior characteristics of a BH  \cite{Gibbons:2012ac, Ballik:2013uia, Kawai:2015uya}. 

Generally, the volume can be defined as 
\begin{equation}
V=\int_{S^n}{\sqrt{g}dr d\theta d\phi},
\end{equation}
where $g$ is the determinant of metric and $S^n$ represents an $n-$sphere. A $4-$dimensional flat BH can be regarded as an $n=2$ sphere. By determining $g$, Parikh \cite{Parikh:2005qs} found a constant volume of BH. Similarly \cite{DiNunno:2009cuq} found the volume of a stationary BH for $t=0$ as the null volume at the horizon. Due to the existence of BH's entropy and temperature \cite{Hawking:1975vcx, Bekenstein:1973ur, Bardeen:1973gs}, these investigations of BH interior volume using old techniques led the researcher to an unsatisfactory result about the BH's constant volume. In Ref. \cite{Christodoulou:2014yia, Christodoulou:2016tuua}, Christodoulou and Rovelli  considered Schawarzschild BH formed under a collapsed process with the largest hyper-surface bounding the maximal interior volume and found that the interior volume increases linearly with advanced time $(v)$. From their numerical analysis, the main contribution to the interior volume of BH was from the central part of the space-like hyper-surface at the radius $r=r_v$ as shown in Fig.(\ref{image-4.2}). They  showed that the interior volume of a space-like spherically symmetric hyper-surface in the interior of the BH is given by
\begin{equation}
V=\int ^v \int_{S^2} Max\left[-r^4 f(r)\right]^{\frac{1}{2}}d\theta d\phi dv,
\end{equation}
the factor $Max\left[-r^4 f(r)\right]^{\frac{1}{2}}$ is maximized for some value of $r=r_v$ to get
\begin{equation}
    V_{\sum}=4\pi \sqrt{-r_v^4 f(r_v)}v,
\end{equation}
here calculating $r_v$, one can easily get the interior volume of BH as a function of Eddington time $(v)$. Later, these investigations of Christodoulou and Rovelli are extended by many authors by using different BHs space-times with surprising results as noted in the table (\ref{tab:1}) below. In reference \cite{Yang:2018arj} the rate of mass change is taken into account, and it is found that volume increases toward evaporation. Whereas, in high-dimension cases, the increase in the rate of mass change decreases with the number of dimensions.

The proportionality relation between BH interior volume with $v$ distinguishes these investigations from those of earlier ones. This property could also change the statistical properties of quantum fields in the BH's interior and is expected to solve the problem of the BH information paradox as claimed by Parikh \cite{Parikh:2005qs}.  As the BH information is associated with its entropy, Baocheng Zhang first discussed the issue of the BH information paradox by using the quantum mode entropy of Schwarzschild's BH \cite{Zhang:2015gda}. 

Baocheng Zhang proposed the largest hyper-surface in the interior of BH and calculated the total number of quantum states $g(E)$ contained in the massless scalar field by using the Klein-Gordon equation \cite{Zhang:2015gda, Cordero-Carrion:2001jpf}

\begin{equation}
\frac{1}{\sqrt{-g}}{{\partial_ \mu }\left(\sqrt{-g} g^{\mu v } \partial_v \Phi \right)}=0,
\end{equation}

\begin{landscape}
\begin{table}[!th]
\caption{Results from Maximal hyper-surface and its bounded interior volume for different BHs.}
\label{tab:1}
\begin{tabular}{||c|p{3.5cm}|p{6cm}|p{3cm}|c|| }
\hline
 \hline
  S. No. &Black hole  & \qquad $r_v$ &The numerical values of $r_{v}$& $V_{\sum}$ \\ 
 \hline
 \hline
  1. & Schawarzschild \cite{Christodoulou:2014yia, Yang:2018arj, Zhang:2019abv, Wang:2020fgz}& \qquad $\frac{3}{2}m$  & 1.5 & $3 \pi v \sqrt{3}m^2$ \\  
  2. & RN\cite{Han:2018jnf, Wang:2018txl, Ali:2018sqk}&\qquad $\frac{1}{4}(m^2 +\sqrt{9m^2-8q^2})$ &1.4& $\frac{\pi v \sqrt{(\sqrt{9m^2-8q^2}+3m)^2(m\sqrt{9m^2-8q^2}+3m-4q^2)}}{2\sqrt{2}}$ \\ 
3. & Kerr \cite{Wang:2018dvo, Wang:2019ake, Wang:2019dpk}& To find the maximal space-like hyper-surface, an induced metric is required on the space-like hyper-surface in the interior of the Kerr BH in such a way that  $r$ coordinate takes its largest value \footnote{Consider a hyper-surface on the event horizon at a late advanced time $v$. Let us extend it from the event horizon with a constant radius $r$ in the interior of a BH. When it reaches the point where $v$ deviates give the maximal advance time and the hyper-surface at this point will bind the largest interior volume. The decomposed form of an arbitrary vector lying on a hyper-surface can be written
\begin{equation}
k=Zn^a+Z^a \qquad \Rightarrow \qquad n_a=-Z\nabla_a{r}
\end{equation}
where $Z$ and $Z^a$ are the Lapse and shift functions respectively, and $n^a$ is the co-vector with $\nabla_a{r}$ being the normal co-vector. For a space-like hyper-surface, the normal vector can  be written as 
\begin{equation}
n_a n^a=-1 \qquad \Rightarrow \qquad Z^2g^{ab}dr_a dr_b=-1,
 \end{equation}
Using the above equation, we can write as $Z^2=-g^{-rr}$. Now as the determinant of an induced metric on hyper-surface at constant radius $r$ is 
\begin{equation}
|h|=Z^{-2}(g)=g^{rr}g \qquad\Rightarrow \qquad |h|^2=-\Delta\rho^2 sin^2\theta,
\end{equation}
Now the volume for a rotating BH, the interior volume can be defined as 
\begin{equation}
V_{\sum}=\int^v\sqrt{|h|}dv d\theta d\phi=\int^v \sqrt{-\Delta}\rho sin\theta dv d\theta d\phi,
\end{equation}
This directly confirms the BH interior volume as in Ref. \cite{Bengtsson:2015zda}.}.& 1.402 &$2 \pi v \sqrt{-\Delta } \left(\sqrt{a^2+r^2}+\frac{r^2 \ln  \left(\sqrt{a^2+r^2}+a\right)}{(2 a) \left(\sqrt{a^2+r^2}-a\right)}\right)$ \\
4. & Kerr Newman \cite{Wang:2019ear, Ali:2021kdu}& Need the same procedure as that of Kerr BH&1.3005 &  $2 \pi v \sqrt{-\Delta' } \left(\sqrt{a^2+r^2}+\frac{r^2}{2 a}\ln \left(\frac{\sqrt{a^2+r^2}+a}{\sqrt{a^2+r^2}-a}\right) \right)
$ \\
5. & BTZ\cite{Zhang:2019pzd, Ali:2020olc}& \qquad $\sqrt{\frac{m}{2}}$ & 0.71 & \qquad $\frac{\pi v}{2} m$ \\
6. & Rotating BTZ \cite{Ali:2020qkb, Maurya:2022vjd}& \qquad $\sqrt{\frac{l^2 m (\sqrt{3X^2+1}+2)}{6}}$ & 0.45 & $\frac{\pi v}{3}(\sqrt{l^2 m^2 (-3X^2+2\sqrt{3X^2+1} +7)-9 J^2}$\\
7. & Charged  f(R) \cite{Ali:2019icq, Wen:2020thi}& \quad $\sqrt{6} \left(\sqrt{\frac{G}{R_0}-\frac{18 \sqrt{6} m}{b R_0 \sqrt{\frac{F}{R_0}}}}+\sqrt{\frac{F}{R_0}}\right)$ &- - - - - - -&$\frac{2}{3}\pi v \sqrt{\frac{2\sqrt{6} m}{b}A^3-6 q^2 A^2+\frac{A^6}{72 R_0}+A}$ \\
8. &  Neutral toral BH \cite{Ong:2015tua} & \qquad $\left (\frac{ m L^2}{\pi K^2}  \right )^{\frac{1}{3}}$&depends on the values of $n$, $K$ \& $L$& $4\pi m Lv$\\
 \hline
  \hline
\end{tabular}
\end{table}

\end{landscape} 
here, 

$$G=\frac{4 b \left(3 q^2 R_0-4\right)}{\sqrt[2]{Z}}-\frac{\sqrt[3]{Z}}{b}+16, \qquad F=-\frac{4 b \left(3 q^2 R_0-4\right)}{\sqrt[3]{Z}}+\frac{\sqrt[3]{Z}}{b}+8, \qquad  A=\sqrt{\frac{F}{R_0}}+\sqrt{\frac{Y}{R_0}},$$

$$Y=-\frac{18 m \sqrt{6 R_0}}{b \sqrt{-4 b \sqrt[-3]{Z} \left(3 q^2 R_0-4\right)+\frac{\sqrt[3]{Z}}{b}+8}}+4 b \sqrt[-3]{Z} \left(3 q^2 R_0-4\right)-\frac{\sqrt[3]{Z}}{b}+16,$$
$K$ is the compactification parameter and $L^2$ is the AdS length scale.

with $\phi$ as the scalar field for a respective BH, we get

\begin{equation}\label{Eqmot1}
    P^{\mu}P_{v}=g^{\mu v}P_{\mu}P_{v}=g^{00}E^2+h^{ij}P_iP_j, 
\end{equation}
$h^{ij} \quad (i, j =1,2, 3)$ is the $4-$dimensional induced inverse metric on the hyper-surface at a constant radius. It may be diagonal or non-diagonal but due to its quadratic property, it can be diagonalized by the coordinate transformation. So, we can get from Eq. (\ref{Eqmot1} )as 
\begin{equation}
E^2-\lambda^{ij}P_iP_j=0, \qquad \Rightarrow \qquad P_1^2=\frac{1}{\lambda^1}(E^2-\lambda^2P^2_2 -\lambda^3P^2_3),
\end{equation}

with $\lambda^{ij}$ as the diagonal elements (see the Appendix: \ref{A}). In $4d-$dimensional space-time the total number of quantum states are 
\begin{equation}\label{Qmode}
g(E)=\frac{1}{(2\pi)^3}\int dx_1 dx_2 dx_3 dP_1 dP_2 dP_3=\frac{1}{(2\pi)^3}\int dx_1 dx_2 dx_3 dP_2 dP_3 \sqrt{\frac{1}{\lambda^1}}\sqrt{E^2-\lambda^2P^2_2 -\lambda^3P^2_3},
\end{equation}
$$=\frac{E^3}{12\pi}V_{\sum},$$
In the case of lower dimensions, one can determine the number of quantum states as
\begin{equation}
    g(E)=\frac{1}{(2\pi)^2}\int dx_1 dx_2 dP_2\sqrt{\frac{1}{\lambda^1}}\sqrt{E^2-\lambda^2P^2_2},
\end{equation}
\begin{equation}\label{Qmodes}
g(E)=\frac{E^2}{8\pi}V_{\sum},
\end{equation}
here the integral formulae $\int\int{\sqrt{1-\frac{x^2}{a^2}-\frac{y^2}{b^2}}}dx=\frac{2\pi}{3}ab, \quad \& \quad \int_{0}^{a}{\sqrt{1-\frac{x^2}{a^2}}}dx=\frac{\pi}{4}a \quad$ are used for calculating the total number of quantum states. Next, defining the free energy $F(\beta)$ for $\beta=\frac{1}{T}$ is 
\begin{equation}\label{FE}
F(\beta)=\frac{1}{\beta}\int {ln(1-exp(-\beta E))}dg(E)=-\int{g(E)\frac{dE}{e^{\beta E}-1}},
\end{equation}
So, using the value of total quantum states one can easily determine the free energy either in $4-$dimension or $3-$dimensions. Hence, the quantum mode entropy of the massless scalar field is 
\begin{equation}\label{Ent1}
S_{\sum}=\beta ^2\frac{ \partial F(\beta)}{ \partial\beta}=\frac{\varpi}{\beta^2}\sqrt{-r_v ^2f(r_v)}v,
\end{equation}
here, $\varpi$ is some constant.  From this Eq. (\ref{Ent1}), the quantum mode entropy is also linearly proportional to $v$ i.e., any change in the quantum mode entropy could lead us to the change of the statistical quantities in the interior of the BH. 

We can calculate the variation of the statistical quantities in the interior BH using the relation of quantum mode entropy and advance time. For this, consider two assumptions that could lead us to find the evaluation relation between BH's interior and exterior entropy. These assumptions are

\begin{itemize}
\item \textbf{BHs radiation as black body radiations:} This assumption guarantees the BH temperature as seen by an observer from infinity is the same as event horizon temperature so, the Boltzmann law can be used for investigating the emission of radiation process \cite{landsberg:1989}. In $(n+1)$ dimension space-time of a symmetric BH, the Boltzmann law is 
\begin{equation}\label{dv}
\frac{dm}{dv}=-\sigma A T^{n+1} \Rightarrow dv=-\frac{\beta^{n+1} \gamma }{ A }dm,
\end{equation}
here, $n$ is the number of space dimensions, $A$ is  the event horizon's area, and $\beta$ is the inverse Hawking temperature of the BH.
\item Quasi-static process: The radiation emission is a quasi-static process. It means that the evaporation process is much slow i.e. $\frac{dm}{dv}<<1$ but Hawking's temperature continuously changes i.e., the thermal equilibrium between the scalar field and horizon of the BH is adiabatically preserved $(\Delta Q=0)$. This assumption guarantees the investigations of the variation of BH's radiation for an infinitely small  time interval ( quantum level.)
\end{itemize}

With these two assumptions, we can write the differential form of quantum mode entropy from Eq. (\ref{Ent1}) as
$$ \dot{S}_{CR}=\frac{\pi^2 \dot{V}_{CR}}{45\beta^{n+1}},$$
for a $4-$dimensional system $n=3$, so
\begin{equation}\label{diffEnt1}
    \dot{S}_{CR}=\frac{\pi^2 \dot{V}_{CR}}{45\beta^4},
\end{equation}
here, we need to find the differential volume for the respective BH and fit in this Eq. (\ref{diffEnt1})  to get the differential form of quantum mode entropy. As the interior volume of a BH is a function of advance-time so, one can easily take the differentiation of interior volume concerning advance-time and fit Eq. (\ref{dv}) in it to get the differential volume in terms of $\rm A, ~ \beta, ~\& ~M, ~Q, ~a$ (depending on the nature of BH geometry). To get the evaporation relation between the interior and exterior entropy of the BH, we need to compare the quantum mode entropy with Hawking entropy. It can be done in two ways: 
\begin{itemize}
    \item directly find the differential form of Hawking entropy for the respective BH i.e. 
\begin{equation}\label{diff BH}
    \dot{S}_{BH}=\frac{\dot{A}}{4\pi},
\end{equation} 
or
    \item use the first law of BH thermodynamics. For spherically symmetric rotating BHs the $1^{st}$ of BH thermodynamics is
\begin{equation}\label{1stlaw}
{dm}=\frac{d{S}_{BH}}{\beta},
\end{equation}
here $S_{BH}$ is Hawking's entropy of BH. In the case of rotating BHs, a spherically symmetric BH is considered thus, we can ignore exotic features of charge and angular momentum.
\end{itemize}

Now, either dividing Eq.  (\ref{diffEnt1}) by Eq. (\ref{diff BH}) or fitting Eq. (\ref{1stlaw}) in Eq. (\ref{diffEnt1}) one gets the evaluation relation of the two types of entropy. Some results are summarized in table (\ref{tab:2}). The negative sign in proportional relation represents the increase of quantum modes entropy advance or Eddington time whereas, Hawking's entropy decreases. Using these results, the effect of mass, charge, and angular momentum can also be elaborated from the curves (for detailed insight, see also the references given in the table (\ref{tab:2}));  

\begin{landscape}
\begin{table}[!th]
\caption{The interior entropy for different BHs}
\label{tab:2}
\begin{tabular}{||c|p{3.5cm}|p{8.5cm}|p{6cm} || }
\hline
 \hline
  S. No. &Black hole  
&Entropy $(S_{\sum})$&Relation b/w the variation of Interior Exterior Entropy  $\left(\frac{\dot{S}_{\sum}}{\dot{S}_{BH}}\right)$\\ 
 \hline
 \hline
1. & Schawarzschild \cite{Christodoulou:2014yia, Yang:2018arj, Zhang:2019abv, Wang:2020fgz}
& $\frac{3\sqrt{3}\gamma}{(90\times 8^4)\pi} A$ & $-\frac{\sqrt{3}\pi^2 \gamma}{240}$ \\  
2. & RN\cite{Han:2018jnf, Wang:2018txl, Ali:2018sqk}
& $\frac{v \left(m^2-q^2\right)^{3/2} \sqrt{m \left(\sqrt{9 m^2-8 q^2}+3 m\right)^2 \sqrt{\sqrt{9 m^2-8 q^2}+3 m-4 q^2}}}{720 \sqrt{2} \left(\sqrt{m^2-q^2}+m\right)^6}$ &$-\frac{\pi^2 \gamma}{360 \sqrt{2}}F(m,q)$  \\
3. & Kerr \cite{Wang:2018dvo, Wang:2019ake, Wang:2019dpk}&$\frac{\pi ^2  \left(m^2-a^2\right)^{3/2}}{360 \pi \left(\left\{\left(\sqrt{m^2-a^2}+m\right)^2+a^2\right\}\right)^3}V_{\sum}$ &$-\frac{\pi^2\gamma}{180}F\left(\frac{a(v)}{m(v)}\right)$  \\
4. &  Kerr Newman \cite{Wang:2019ear, Ali:2021kdu}&$\frac{1}{180}\frac{f_{max}(m,a,q)(m^2-a^2-q^2)^{\frac{3}{2}}}{(2m\sqrt{m^2-a^2-q^2}+2m^2-q^2)^3}v$ &$-\frac{\pi^2 \gamma}{90}F(m,a,q)$ \\
5. &BTZ\cite{Zhang:2019pzd, Ali:2020olc}
&$\frac{3\zeta(3)}{2\beta^2}\sqrt{-r_v^2 f(r_v)}$ &$-\frac{6\zeta(3) \gamma}{\pi}\frac{\sqrt{-r_v^2 f(r_v)}}{r_+}, \quad r_+=\sqrt{m}$ \\
6. & Rotating BTZ \cite{Ali:2020qkb, Maurya:2022vjd}
&$\frac{3\zeta(3)}{2\beta^2}\sqrt{-r_v^2 f(r_v)}v$ &$-\frac{4\gamma\zeta(3)}{3\pi}\frac{\sqrt{m(3X^2+\sqrt{3X^2+1}-1)}}{\sqrt{X+1}}$  \\
7. & Charged  f(R) \cite{Ali:2019icq, Wen:2020thi}
&$\frac{\pi^2 V_{\sum}}{45 \beta^3}=\frac{1}{4320}\alpha(m,q; b)$ & $-\frac{4\pi^2\gamma}{135}\gamma'(m,q;b)$ \\
8. & Neutral toral BH \cite{Ong:2015tua} &$-\frac{4 l m^2 v}{45 \pi  K^2 L^4}$ &$\frac{\gamma }{30}  \sqrt[3]{\frac{\pi ^{11} K^{10} m}{L}}$ \\
 \hline
  \hline
\end{tabular}
\end{table}

Here, we used $V_{\sum}$ as the volume of a respective BH as given in Table (\ref{tab:1}). The dot $(.)$ is used for differentiation concerning Eddington time $(v)$, $S_{BH}$ is the Bekenstein and Hawking entropy of the respective BH that is equal to one-quarter of its area. 
\begin{equation}
F(m, q)=\frac{\sqrt{(\sqrt{9m^2-8q^2}+3m)^2{(m\sqrt{9m^2-8q^2}+3m-4q^2)}}}{\sqrt{m^2-q^2}+m},
\end{equation}

\end{landscape}

\begin{equation}
F\left(\frac{a_o[v]}{m_o[v]}\right)=f_{max}\left(\frac{a_o[v]}{m_o[v]}\right)\left[1-\sqrt{1-\left(\frac{a_o[v]}{m_o[v]}\right)^2}\right]\left(\frac{a_o[v]}{m_o[v]}\right)^{-2},
\end{equation}

\begin{equation}
F(m, a,q)=\frac{f_{max}(m,a,q)(m^2-a^2-q^2)^{\frac{3}{2}}}{(2m\sqrt{m^2-a^2-q^2}+2m^2-q^2)^3},
\end{equation}

and 
\begin{equation}
f_{max}(m,a,q)=\sqrt{2mr-r^2-a^2-q^2}\left(\sqrt{r^2+a^2}+\frac{r^2}{2a}\frac{\sqrt{r^2+a^2}+a}{\sqrt{r^2+a^2}-a}\right)_{r=r_v},
\end{equation}


These investigations attract many authors due to their universality in explaining the BH's interiors with the Einstein theory of general relativity. The dynamical nature of these investigations with $v$  could be used for BH's interior volume to probe the interior information of BHs.

The structure of this chapter is such that in the section (\ref{terminology}), we present the meanings of important terms for the readers to understand the concept of the work done on the topic of the subject. The main literature on the BH interior volume, entropy, and its connection to the evaporation and information paradox is presented in sec. (\ref{introduction}). To explain work more simply, we present the review of interior volume, interior entropy, and their variation for Schawarzschild BH, Kerr BH, and BTZ BH to cover the $(3+1)$ as well as $(2+1)$ dimensional space-times in sec (\ref{Schw-BH}), (\ref{Kerr-BH}), and (\ref{BTZ-BH}). In sec. (\ref{ent-variation}), the entropy variation and evaporation relation for different BHs are being explained. In sec (\ref{natur-radiation}), we in-sighted the probability of emission and the nature of BH radiation. Finally, some remarks and discussion on the main point of the results are presented.
\section{A review of the BH interior Volume and Entropy variation}\label{vol-ent-review}

\subsection{Schwarzschild BH}\label{Schw-BH}

In Eddington Finkelstein coordinates $(~v,~r,~\theta,~\phi)$ the line element of Schwarzschild geometry is given in Eq. (\ref{collapsedmass}) e.g., see Refs. \cite{Christodoulou:2014yia, Zhang:2015, Wang:2018dvo, Ali:2021kdu}. For an uncharged spherically symmetric BH $f(r)=1-\frac{2m}{r}$ and the $2$-sphere form is $d\Omega^2=r^2(d\theta^2+sin^2 \theta d\phi)$ and the advance time is defined as $v=t+r^*$.
\begin{equation}
r^*=r+2mlog|r-2m|,
\end{equation}

here, the geometric units $G=c=\hbar=k_B=1$. The hyper-surface on the proposed sphere could be defined as the product of an affine parameter to the 2-sphere i.e., $\sum=\gamma\times S^2$,  where $\gamma\rightarrow(v(\lambda),r(\lambda))$ is the affine parameter. So, the Schawarzschild metric becomes
\begin{equation}\label{EDmet}
{ds}^2_{\sum}=\left(-f(r) \dot{v }^2+2 \dot{v } \dot{r}\right)d \lambda ^2+r^2d\Omega^2,
\end{equation}

We have considered only curved space-time so, the contribution to the volume will not be the same as $V=\frac{4}{3} \pi R^3$. Using Eq. (\ref{EDmet}) the interior volume of Schawarzschild BH is
$$V_{CR}=\int  \int d\lambda d\Omega \sqrt{r^4\left(-f(r)\dot{v }^2+2 \dot{v } \dot{r}\right)sin^2 \theta},$$
\begin{equation}\label{intvolgenfor}
=4 \pi\int{\text{d$\lambda $}}\sqrt{r^4\left(-f(r)\dot{v }^2+2 \dot{v } \dot{r}\right)},
\end{equation}

It shows that the proper length of a geodesic in auxiliary metric ($4\pi$-times) is the volume bounded by $\sum$. By maximizing this auxiliary metric, we need $\dot{r}=0$. So, we can find numerically the largest curve bounding the  maximal interior volume at $r=\frac{3}{2}M$. This equation also shows that finding the hyper-surface is similar to solving the geodesic equation to get the equation of motion (EOM) with Lagrangian. So, the auxiliary metric on the hypersurface becomes; 
\begin{equation}\label{effmetric}
{ds}^2_{eff}=r^4 \left(-f (r) \dot{v }^2+2 \dot{v } \dot{r}\right),
\end{equation}
This shows that calculating the volume of the hyper-surface $V_{\sum}$ is equivalent to calculating the volume from auxiliary metric with Lagrangian whose maximal value is

$$L\left(r,v, \dot{r}, \dot{v}\right)=1,$$
i.e.
\begin{equation}\label{Eq.1}
r^4\left(-f(r) \dot{v }^2+2 \dot{v } \dot{r}\right)=1,
\end{equation}
using this Eq. (\ref{intvolgenfor}), we gets 
\begin{equation}\label{VCR}
V_{CR}=4\pi \lambda_f,
\end{equation}

As the metric $\tilde{g}_{\alpha\beta}$   has a Killing vector i.e., $\zeta^\mu=(\partial_\mu)^\mu\propto (1, 0) $ and $\gamma$ is an affinely parameterized geodesic in auxiliary metric so, the inner product of Killing vector $\zeta^\mu$ with its tangent $\dot{x}^\alpha=(\dot{v}, \dot{r})$ will be conserved
\begin{equation}\label{Eq.2}
\zeta\times \dot{x}^\alpha=r^4\left(-f(r) \dot{v }^2+2 \dot{r}\right)=X,
\end{equation}
solving Eq. (\ref{Eq.1}) and (\ref{Eq.2}), we get
\begin{equation}\label{Eq.3}
\dot{r}=-r^{-4}\sqrt{A^2+r^4f(r)}, \qquad and  \qquad \dot{v}=\frac{1}{X+r^4\dot{r}},
\end{equation}

In the case of space-like geometry, $(-f(r) \dot{v }^2+2 \dot{v } \dot{r})>0$ so the Lagrangian $L>0$. since $r$ is positive and the Lagrangian will demolish at $r=0$, which is the endpoint of the geodesic. Hence from (\ref{Eq.3}), $\dot{r}$ becomes infinite. Thus, $\gamma$ is a space-like geodesic in $m_{aux}$. A well-suited parameterization is to take $\lambda$ as the proper length in auxiliary metric. So, the auxiliary metric given above becomes,
\begin{equation}\label{auxmet}
dS_{M_{aux}}=-\sqrt{-r^4f(r)}dv=Xdv,
\end{equation}

It can be easily seen that $X$ has to be negative for the geodesic to be space-like. Then, $\dot{v}$ and $\dot{r}$ are both negative and there are only positive terms in (\ref{Eq.1}). Integrating (\ref{Eq.2}), we get
\begin{equation}\label{VCR/4pi}
\frac{V_{\sum }}{4 \pi }=\lambda _f=\int _0 ^{2 M}\frac{r^4}{\sqrt{X^2+r^4f(r)}},
\end{equation}
Eq. (\ref{VCR/4pi}) shows the restriction could be imposed on $X$ as
$$X^2> -r_v^4f(r_v)>=0,$$
\begin{equation}
\Rightarrow X^2=\frac{27}{16}M^4=X_c ^2,
\end{equation}

This condition is obtained from the expression $(-f(r) \dot{v }^2+2 \dot{v } \dot{r})>0$ having the roots $r=0$ and $r=2m$ otherwise the position is maximum at $r_v=\frac{3}{2}m$. Since $(-f(r) \dot{v }^2+2 \dot{v } \dot{r})>0$ in the range $0<r<2m$. For every constant $r$, there is a solution or we can say $r$ is constant the surface is space-like geodesic of an auxiliary manifold. For a stationary (maximal) point of the volume given in Eq. (\ref{Eq.3}(b))
$$\frac{dv}{d\lambda}=\frac{1}{X}\Rightarrow d\lambda=Hdv,$$
Integrating, we get
\begin{equation}
\lambda_f=H(v_f-v),
\end{equation}
As $r =\rm constant$, the surface will have the maximal volume between two given values of advance time $v$ when $H$ is largest. This means that, $r=r_v$ which gives $H=H_c$. So, the volume will be the largest possible. These considerations provide the basis for the derivation of the asymptotic volume.
\begin{equation}\label{Acnu}
\lambda_f=H_c v,
\end{equation}
Using Eq. (\ref{Acnu}) in Eq. (\ref{VCR}), one could get the interior volume as
\begin{equation} \label{VCR1}
V_{CR}=-4\pi\sqrt{-r^4f(r)}v=4\pi H_c v,
\end{equation}
or we can write as
\begin{equation}\label{SCHVCR}
V_{CR}=3\sqrt{3}\pi m^2 v,
\end{equation}

Where $A=-\sqrt{-r^4f(r)}$, Which shows that the interior volume depends on advance time. This result can be extended to other cases by simply using metric Eq. (\ref{EDmet}) with lapse function $f(r)$ from the desired BH metric and finding the maximization of $H=H_c$, to get the asymptotic expression, using an analogy to the above equation, one gets the required BH interior volume.  Calculating the interior volume of charged and charged f(R) BH from Eq. (\ref{VCR1}). is convenient. For this, one needs to calculate the factor $\sqrt{-r^4f(r)}$ for $r=r_v$ from the lapse function of the respective BH metric as given in (\ref{tab:1}).

In Ref. \cite{Wang:2018dvo}, used the $a=0$ to degenerate the result of the Schawarzschild BH from the Kerr BH result and found the proportionality function for maximal hyper-surface as \cite{Wang:2018dvo}
\begin{equation}
    f\left(\frac{r}{m}, \frac{a}{m}\right)=f\left(\frac{r}{m}\right)_{max}=\frac{r}{m}\sqrt{2\frac{r}{m}-\left(\frac{r}{m}\right)^2},
\end{equation}
and the position of the maximal hypersurface is plotted in Fig. (\ref{image-3.4}).

\begin{figure}
\begin{center}
\includegraphics[width=0.55\textwidth]{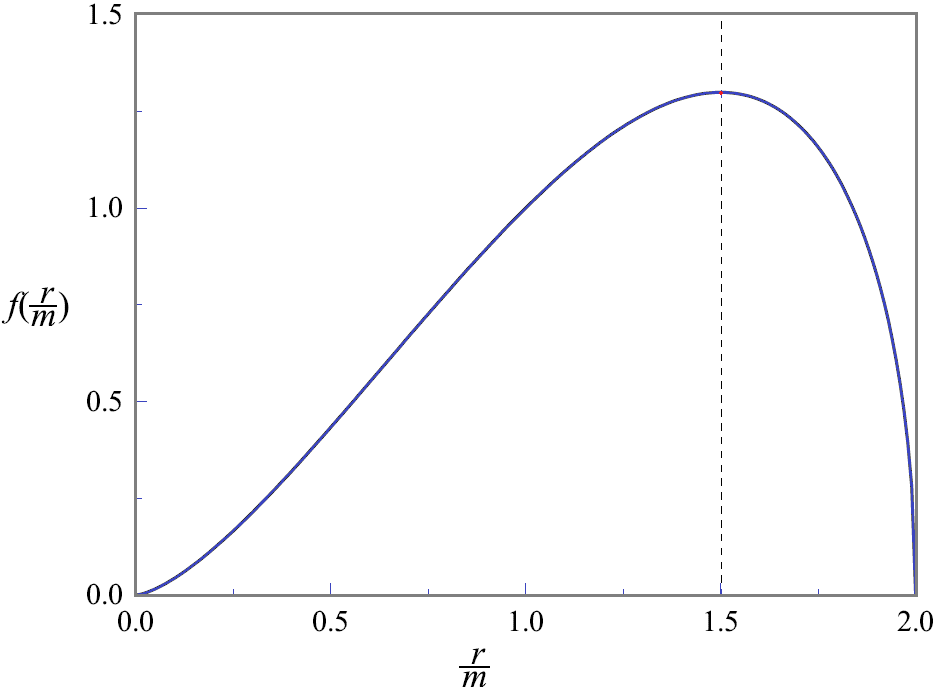}
\caption{The plot of proportionality function $f_{max}\left(\frac{r}{M}\right)$ vs $\frac{r}{M}$ for Schawarzschild BH showing the position of the maximal hypersurface is at $(\frac{r}{M})=\frac{3}{2}$ \cite{Wang:2018dvo}.}
\label{image-4.3}
\end{center}
\end{figure}
Baocheng Zhang \cite{Zhang:2015} considered a scalar field in the interior of Schwarzschild's BH and discussed its quantum modes entropy by determining the quantum states in the interior volume labeled by $(\lambda, \theta, \phi, p_{\lambda}, p_{\theta}, p_{\phi})$. So, the quantum mode corresponding to the cell of volume $(2\pi)^3$ in that phase space will have the total number of quantum modes given as
\begin{equation}
\frac{phase{\quad}space}{volume{\quad}of{\quad}the{\quad} cell}=\frac{d\lambda d\theta d\phi dp_{\lambda}dp_{\theta} dp_{\phi}}{(2\pi)^3},\qquad \hbar=1,
\end{equation}
For the total quantum modes, one needs to integrate the above expression as
\begin{equation}\label{freeEnergy}
g(E)=\frac{1}{(2 \pi )^3}\int{d\lambda d\theta d\phi dp_{\lambda}dp_{\theta} dp_{\phi}},
\end{equation}
This  gives the total number of quantum modes having energy less than $E$. As from the WKB approximation, the scalar field in the interior of a BH consisting of quantum modes is
\begin{equation}\label{Scalarfield}
\Phi =e^{\text{-iET}}e^{\text{iI}(\lambda, \theta, \phi)},
\end{equation}
Expanding and solving the above equation in components $(\lambda, \theta, \phi)$ and solving, gives the EOM (also see Appendix: \ref{A})
\begin{equation}\label{eqofenergy}
E^2-\frac{p_{\lambda }^2}{-f(r) \dot{v }^2 +2 \dot{v } \dot{r}}-\frac{p_{\theta }^2}{r^2}-\frac{p_{\phi }}{ r^2 \sin ^2 \theta }=0,
\end{equation}
here, we used
$$\frac{\partial I}{\partial \lambda }=p_{\lambda }, \qquad \frac{\partial I}{\partial \theta }=p_{\theta }, \qquad \frac{\partial I}{\partial \phi }=p_{\phi },$$
as eigenstates of the diagonal elements of the  metric. Solving Eq. (\ref{eqofenergy}) for $p_\lambda$, we get
\begin{equation}\label{P:lambda}
p_{\lambda }=\sqrt{-\dot{v }^2 f(r)+2 \dot{v } \dot{r}} \sqrt{E^2-\frac{p_{\theta }^2}{r^2}-\frac{p_{\phi }}{r^2 \sin ^2\theta}},
\end{equation}
Use Eq. (\ref{P:lambda}) in Eq. (\ref{freeEnergy}) and simplifying, we get 
$$g(E)=\int{d\theta  d\lambda  d\phi\sqrt{- f(r)\dot{v }^2+2 \dot{v } \dot{r}}\int{\sqrt{E^2-\frac{p_{\theta }^2}{r^2}-\frac{p_{\phi }}{r^2 \sin ^2\theta}}}dp_{\theta } dp_{\phi }},$$
$$=\frac{1}{(2 \pi )^3}\int d\theta  d\lambda  d\phi\sqrt{- f(r)\dot{v }^2+2 \dot{v } \dot{r}} \left (\frac{2 \pi }{3} E^2 r^2 sin^2 \theta  \right ),$$
$$=\frac{E^3}{(2 \pi )^3}\frac{2 \pi }{3}\int{d\lambda  \sqrt{r^4 \left(2 \dot{v } \dot{r}-\dot{v }^2 f(r)\right)}}{\int{\sin^2 \theta d\theta} \int{d\phi}},$$
$$\frac{E^3}{12 \pi ^2}\left ( 4\pi\int{d\lambda  \sqrt{r^4 \left(2 \dot{v } \dot{r}-\dot{v }^2 f(r)\right)}} \right ),$$
\begin{equation}\label{Qstate}
g(E)={\frac{E^3}{12 \pi ^2}{V_{CR}}},
\end{equation}

For the integration in the first step, we used the general formula $ \sqrt{1-\frac{x^2}{a^2}+\frac{y^2}{b^2}} dx dy=\frac{2 \pi }{3} a b$. As from Eq. (\ref{Qstate}),  $g(E)\propto V_{CR}$ so, it still has similarity with normal space-time. Note that the physical interpretation does not need to be the same because the volume in general relativity results from the curved space-time. The main reason for the physical difference can be considered as the volume is bound within the closed hyper-surface that is increasing with advanced time $v$. So, the number of quantum states inside the BH must also increase with time. This statement is crucial for interior volume and information storage in curved space-time. Considering this difference, we can calculate the free energy as
\begin{equation}\label{BfreeE}
F(\beta )=\frac{1}{\beta }\int {dg(E)} ln(1-e^{-\beta (E)}),  
\end{equation}
$$ F(\beta)=-\int{\frac{dg(E)}{-1+e^{-\beta(E)}}},$$
$$F(\beta)=-\frac{V_{\text{CR}}}{12 \pi ^2}\int{\frac{E^2 dE}{-1+e^{-\beta (E)}}},$$
Solving the integral, we get the result as
\begin{equation}\label{FreeEnergy}
F(\beta)=-\frac{\pi^2 V_{CR}}{180\beta^4},
\end{equation}
Finally, entropy is
\begin{equation}\label{Ent11}
S_{CR}=\beta^2 \frac{\partial F}{\partial \beta}=\frac{\pi^2 V_{CR}}{45\beta^3},
\end{equation}

Which is the entropy in the interior of the $V_{CR}$. Using the value of interior volume $V_{CR}$ from Eq. (\ref{SCHVCR}) and inverse temperature, we can get the quantum modes entropy in BH's interior. Since a BH has the property of emitting radiation, the emission is claimed to be quasi-static and increases with time due to variable temperature. Treating BH radiation as black body radiation can be defined by Stefan Boltzmann's law. Thus, the rate of mass loss from the Schwarzschild BH due to Hawking's radiation is
\begin{equation}\label{Boltzmannlaw}
\frac{dM}{dv}=-\frac{1}{\gamma M^2},
\end{equation}
This equation states that the time duration for the radiation to last from a BH is proportional to the triple power of mass $M$ i.e.,
$$v\approx\gamma M^3,$$
This also satisfies the condition of Ref. \cite{Christodoulou:2014yia} with $v>>M$.
Now, consider Schwarzschild BH inverse temperature as $\beta=\frac{1}{T}= 8\pi M$, then the interior entropy is 
\begin{equation}
S_{CR}=\frac{3 \sqrt{3} \gamma  M^2}{45\times 8^3}=\frac{3 \sqrt{3}\gamma A}{(45\times 8^4) \pi },
\end{equation}
here $ A=16\pi M^2$ is Schwarzschild BH's surface area  \cite{Bekenstein:1972tm, Zhang:2015}, as the radiations from the BH are in the Planks scale and the final evaporation stage has yet not been discovered, this means that the mass loss of the BH during the emission of radiation are so small. Hence, we can take $\frac{dM}{dv}\approx M$, which is inconsistent with the requirements of the CR volume because in such a case we can't take the growth in the BH's volume.  The above equation confirms  that the entropy of the quantum field in the CR volume is directly related to the horizon area and also the coefficient of $A$ is much smaller than $\frac{1}{4}$. This means it doesn't satisfy the first law of BH satisfied by the Bekenstein and Hawking relation. From this relation, we can say there is more information loss on the BH horizon. Here we can also raise the question of how to fit the above relation in the first law of BH thermodynamics. Let's compare the exterior and interior entropy that may have some entangled relation. This may justify these issues with the $1^{st}$ law of BH's thermodynamics.  So, considering the quasi-static emission of radiation, we can introduce the differential form of Eq. (\ref{Ent11}), and further, it can be represented in Hawking's entropy as

\begin{equation}\label{difEnt}
dS_{CR}=\frac{\pi^2 dV_{CR}}{45\beta^3}=-\frac{\sqrt{3}\pi^2\gamma}{30}mdm= -\frac{\sqrt{3}\pi\gamma}{240}(S_{BH}),
\end{equation}
This shows that interior quantum mode entropy is directly related to Hawking entropy. In this equation, the negative sign shows that as the Horizon entropy increases, the interior entropy decreases due to the loss of BH's information.

\subsection{Kerr BH}\label{Kerr-BH}
In the case of  Kerr BH \cite{Kerr:2007dk}, the interior and exterior horizons can be calculated from $\Delta=0$,
\begin{equation}\label{Kerrrad}
r_\pm=M\pm \sqrt{M^2-a^2},
\end{equation}
Due to the spinning property, the horizons of a Kerr BH may not be spherically symmetric and this character may affect the physical properties of Kerr's BH. As discussed in the Penrose diagram of CR's work, the interior volume is mainly the contribution of  the stretched part of the largest spherically symmetric $3d$ hyper-surface that can be calculated by either using the condition of vanishing curvature or by maximization of the factor $\sqrt{-r_v^4 f(r_v)}$. We also stated that this part of the hypersurface doesn't extend to singularity, so one can say that the spinning character doesn't affect the interior volume of a Kerr BH. Without loss of generality, for a Kerr BH, the  interior volume can be defined as
\begin{equation}
V_{CR}=\int{\sqrt{-\Delta}\rho sin\theta dv d\theta d\phi},
\end{equation}
that gives
\begin{equation}\label{KerrBHvol}
V_{\text{CR}}=2 \sqrt{-\Delta } \pi v \left(\sqrt{r^2 + a^2} + \frac{r^2}{2 a} log\left(\frac{\sqrt{a^2+r^2}+a}{\sqrt{a^2+r^2}-a}\right)\right),
\end{equation}

This result is also investigated in \cite{Bengtsson:2015zda}. At $r=r_s$, this equation gives the largest hyper-surface to calculate the maximal interior volume. Generally, this Eq. (\ref{KerrBHvol}) can be written as
\begin{equation}\label{Kerrvol}
V_{CR}=2\pi F(r,a)v =2\pi M^2 F_{max}\left(\frac{a}{M}\right)v,
\end{equation}
where 
$$F(r,a)=\sqrt{-\Delta } \left(\sqrt{r^2 + a^2} + \frac{r^2}{2 a} log\left(\frac{\sqrt{a^2+r^2}+a}{\sqrt{a^2+r^2}-a}\right)\right),$$
and 
$$F_{max}\left(\frac{r}{M}\right)=F\left(\frac{r_v}{M},\frac{a}{M}\right),$$

The position of the largest hyper-surface is numerically found at $r_v=1.402$ as shown in Fig. (\ref{image-2.4}).
\begin{figure}[ht]
\begin{center}
\includegraphics[width=0.53\textwidth]{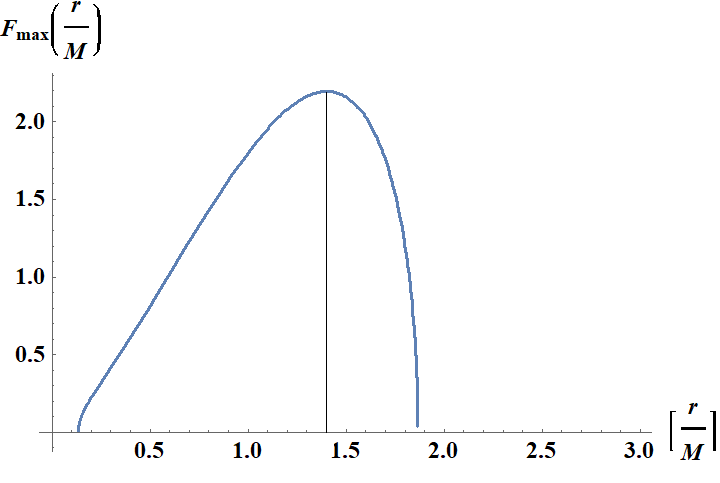}
\caption{The plot of $F_{max}\left(\frac{r}{M}\right)$ vs $\frac{r}{M}$ for a Kerr BH shows the proportional function has the  maximum value at $(\frac{r}{M})=1.402$ with $(\frac{a}{M})=0.2$}
\label{image-4.4}
\end{center}
\end{figure}

This relation can degenerate the Schwarzschild result by using $a=0$. Plotting the proportional relation, we can obtain the same maximal hypersurface at $r_v=1.5$ \cite{Wang:2018dvo}.

Following the Baocheng Zhang scenario \cite{Wang:2018dvo}, the entropy of static BHs (either charged or uncharged) is discussed in the above sections. We found that the entropy of the massless scalar field directly increases with $v$. So, this is the main character that affects the interior statistical quantities of the BH including its volume. To calculate the BH's entropy associated with quantum modes, we need to calculate the number of interior quantum states. Considering the metric for a maximal hyper-surface ( $r=r_v$), the volume is determined in Eq. (\ref{Kerrvol}). Next, considering the quantum modes of the massless scalar field in a Kerr BH, we obtain the EOM in general form as 
\begin{equation}
-E^2+h^{ab}P_a P_b=0,
\end{equation}

Where $h^{ab}$ is the auxiliary metric on the maximal hyper-surface. So, the total number of quantum states contained in the scalar field is given by Eq. (\ref{Scalarfield}). Where $V_{CR}$ is given in Eq. (\ref{KerrBHvol}). The free energy calculated is given in Eq. (\ref{FreeEnergy}) and the entropy in Eq. (\ref{Ent1}). Next, using two conditions. The BH emission rate is a quasi-static process (it needs the emission rate to satisfy the condition $v>>M$) and BH's radiation is black body radiation. The latter condition guarantees the horizon temperature as Hawking temperature given by,
\begin{equation}\label{insvtemp}
\beta =\frac{1}{T}=\frac{4\pi \left(r_{\pm }+a^2\right)}{r_+-r_-},
\end{equation}

here, $r_\pm$ is given in Eq. (\ref{Kerrrad}), the first condition guarantees that the temperature of the scalar field and the horizon will be in equilibrium if the process is considered at the quantum level. So, the temperature of the scalar field will be considered equal to the temperature of the horizon, and the rate of mass loss can be seen by the Stefan Boltzmann law in Eq. (\ref{Boltzmannlaw}). Using the values of $\beta$ and $A$, we can get Eq. (\ref{Boltzmannlaw}) as
\begin{equation}\label{BoltzmannlawKerr}
\frac{dM}{dv }=\frac{\left(r_+-r_-\right){}^4}{32 \gamma \pi ^3 M^3 r_+^3}=\frac{\left(M^2-a^2\right)^2}{32 \gamma \pi ^3 M^3 \left(\sqrt{M^2-a^2}+M\right)^3},
\end{equation}
Where $\gamma$ is a positive constant, it depends on quantum modes coupled with gravity. Its value doesn't affect our discussion. As the volume of BH changes with advanced time hence, using the above equation, we can introduce the volume in terms of differential form Eq. (\ref{Kerrvol}) as
\begin{equation}\label{KerrvolBoltzmannlaw}
dV_{CR}=2 \pi M^2   F\left(\frac{r}{M},\frac{a}{M}\right)dv,
\end{equation}
In addition to Eq. (\ref{BoltzmannlawKerr}), and Eq. (\ref{KerrvolBoltzmannlaw}) gives
\begin{equation}
\dot{V}_{CR}=-64 \gamma \pi ^4 M^4 F_{\max }\left(\frac{r_{v }}{M},\frac{a}{M}\right)\frac{\left(\sqrt{M^2-a^2}+M\right)^3}{\left(M^2-a^2\right)^2}\dot{M},
\end{equation}
Finally, the differential entropy is
\begin{equation}
\dot{S}_{\text{CR}}=-\frac{\pi ^2}{45 \beta }\dot{V}_{CR}=-\frac{8 \pi^3  \gamma  M^2}{45} F_{\max }\left(\frac{r_{v }}{M},\frac{a}{M}\right)\frac{\left(\sqrt{M^2-a^2}+M\right)^3}{\sqrt{M^2-a^2} \left(\sqrt{M^2-a^2}+a^2+M\right)^3} \dot{M},
\end{equation}

Now, to get the evolution relation between the exterior and interior entropy, we need the variation of Bekenstein and Hawking entropy for Kerr BH  \cite{Bekenstein:1972tm, Bekenstein:1973ur, Bardeen:1973gs}. As the Hawking entropy is
$$S_{BH}=\frac{A}{4}=\pi(r^2 _+ -a^2),$$
Where $A=4\pi (r_+^2+a^2 )$ is the area of the Kerr BH. Its differential form is 
\begin{equation}\label{DRNBHent}
\dot{S}_{BH}=\frac{2\pi(1+\sqrt{ M^2- a^2})^2}{\sqrt{ M^2- a^2}}M\dot{M},
\end{equation}
Here $\dot{M}<0$ so, the proportional relation between the two types of entropy can be written as 
\begin{equation}\label{KerrENT}
\dot{S}_{CR}=-\frac{\pi ^2}{180}\gamma f\left(\frac{r_{v }}{M},\frac{a}{M}\right)\dot{S}_{BH},
\end{equation}
here,
$$f\left(\frac{r_{v }}{M},\frac{a}{M}\right)=f_{\max }\left(\frac{a}{M}\right)=F_{\max }\left(\frac{r_{v }}{M},\frac{a}{M}\right)\left (1-\sqrt{1-\left(\frac{a}{M}\right)^2}  \right )\left ( \frac{a}{M} \right )^{-2},$$

Alternatively, this result can be obtained by determining the volume bounded by the largest hyper-surface in the interior of Kerr BH and fitting its value in the general form of entropy Eq. (\ref{Ent11}). Next, using the two assumptions to determine the differential form and finally by comparison with Bekenstein and Hawking entropy, one gets the same result as Eq. (\ref{KerrENT}). The plot of this function $f_{max} (\frac{a}{M})$ vs $(\frac{a}{M})$ is given above. The proportionality function is plotted against the spin parameter in Fig. (\ref{image-4.5}). As the value of the spin parameter increases, the value of the proportionality function decreases.
\begin{figure}
\begin{center}
\includegraphics[width=0.5\textwidth]{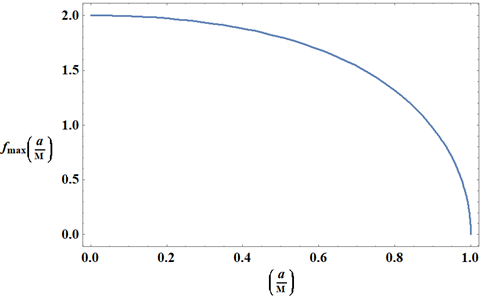}
\caption{The plot of $f_{max} (\frac{a}{M})$ vs $(\frac{a}{M})$ for a Kerr BH\cite{Wang:2018dvo}.}
\label{image-4.5}
\end{center}
\end{figure} 
\subsection{BTZ BH}\label{BTZ-BH}
BTZ BH is a $(2+1)-$dimensional solution of Einstein-Maxwell equations having negative cosmological constant and constant negative curvature \cite{Banados:1992wn, Banados:1992gq, Birmingham:2001dt}. In Ref. \cite{Carlip:1995qv}, it is stated that BTZ BH shares many classical and quantum properties
with the $(3+1)-$ dimensional BH systems. The metric of $(2+1)$-dimensional rotating BTZ BH is defined in Eq. (\ref{metric}), with the lapse function and angular shift are
\begin{equation}
f(r)=-m^2+\frac{r^2}{l^2}+\frac{J^2}{4 r^2},
\end{equation}  
here, we can define $\Lambda=-\frac{1}{l^2}$ as the cosmological constant, $j, m, l$ are the azimuthal angular momentum, AdS mass and AdS radius and $N^\phi (r)$ is the shift function corresponding to angular velocity $\Omega (r)$ and $\phi$ is the period in the range of $0<\phi<2\pi$. For the symmetry the metric of rotating BTZ BHs needs an azimuthal symmetry so that the angular momentum is conserved under a coordinate transformation. The mass $m$ and Hawking entropy $S_{BH}$ of BTZ rotating BH at the horizon are defined as \cite{Banados:1992wn, Carlip:1995qv}
$$m= \frac{r^2}{l^2}+\frac{J^2}{4 r^2}, \quad S_{BH}=2\pi r_+,$$
At the horizon
\begin{equation}\label{radius}
f(r)=0 \qquad \Rightarrow \qquad r_\pm =\sqrt{\frac{l^2 m}{2}  \left(1\pm X\right)},\quad with \quad X=\sqrt{1-\left(\frac{J}{l m}\right)^2},
\end{equation}
 According to this Eq. (\ref{radius}), the coordinates singularity is at  $\frac{J}{l m}=1$. Using the lapse function, one can easily define the surface gravity

$$\kappa=\frac{1}{2}\frac{\partial f(r_+)}{\partial r}=\frac{\sqrt{2}mX}{\sqrt{l^2m(1+X)}} ,$$
and the horizon temperature
\begin{equation}\label{BTZTemp}
T=\frac{mX}{\pi\sqrt{2l^2m(1+X)}},
\end{equation}
The dragging coordinate transformation could help to avoid the dragging effect of the BH so, one should introduce dragging coordinate transformation as

\begin{equation}\label{angularvelocity}
d\phi=-N^{\phi}dt=\frac{J}{2r_+ ^2}dt=\Omega dt,
\end{equation}
This definition of angular velocity is only satisfied in the case of $(2+1)-$dimensional space-time. So, we can investigate the total number of quantum states in the interior of the proposed scalar field. For defining the hyper-surface, the following form of the Eddington Finkelstein coordinates can be used

\begin{equation}\label{Eddmet}
ds^2=(-f(r)\dot{v}^2+2 \dot{v} \dot{r}) d\lambda^2+r^2(N^{\phi}dt+d\phi)^2,
\end{equation}
In the case of  axially symmetric BH, there will be negligible effects on the BH horizon, thus the deformative forces will be less effective in the interior of the at $r=r_v$ BH. Summarizing the results, we can write the general form of the BH interior volume as \cite{Ali:2020qkb} 
\begin{equation}
    V_{CR}=\int _0 ^{2\pi} d\phi \int \sqrt{r^2(-f(r)\dot{v}^2+2 \dot{v} \dot{r})}d\lambda=2\pi v \sqrt{-r_v ^2f(r_v)},
\end{equation}
 By the maximization of the factor $\sqrt{-r_v ^2 f(r_v)}$ one can get the maximal hyper-surface
 \begin{equation}
     r_c=\frac{\sqrt{l^2 m \left(\sqrt{3 X^2+1}+2\right)}}{\sqrt{6}},
 \end{equation}
and hence, the interior volume obtained as 
\begin{equation}
    V_{CR} =\frac{v\pi}{3} \sqrt{l^2 m^2 \left(3 X^2+\sqrt{3 X^2+1}+7\right)-9J^2},
\end{equation}
with a numerical position at $r_v=0.45$ as shown in Fig. (\ref{image-4.6}). 

\begin{figure}[ht]
\begin{center}
\includegraphics[width=0.65\textwidth]{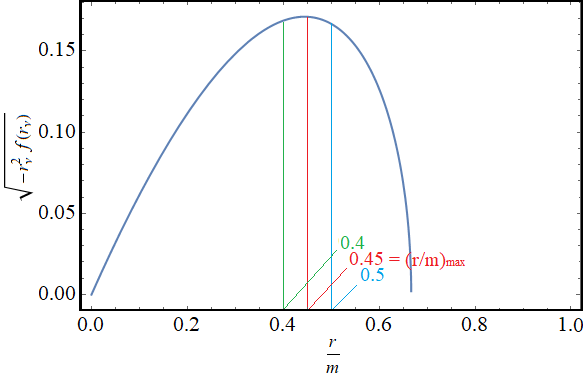}
\caption{The plot of maximization factor $\sqrt{-r_v ^2f(r_v)}$  vs $\frac{r}{m}$ for the rotating BTZ BH shows the maximal hyper-surface located at $0.45m$ with \quad $J=0.5$\cite{Ali:2020qkb}}.
\label{image-4.6}
\end{center}
\end{figure}

\section{Entropy Variation and Evaporation}\label{ent-variation}

Using Eq. (\ref{Ent1}), for the rotating BTZ BH the interior entropy of BTZ BH is obtained as
\begin{equation}\label{BTZEnt}
S_{\sum}=\frac{3\zeta (3)}{2\beta^2}\sqrt{-r_v ^2f(r_v)}v,
\end{equation}
From this Eq. (\ref{Ent1}) the entropy of BTZ BH is also proportional to $v$ and this feature may affect the statistical quantities in its interior. To see this, we considered two assumptions to get an evaluation relation between quantum modes entropy and Bekenstein-Hawking entropy. These assumptions are:

\begin{itemize}
\item {BH radiation as black body radiations: So, in $(2+1)$ dimension space-time of rotating BTZ BH, the Boltzmann law is 
\begin{equation}
\frac{dm}{dv}=-\sigma A T^3 \Rightarrow dv=-\frac{\beta^3 \gamma }{ A }dm,
\end{equation}
here $A=\pi  l \sqrt{2 m (X+1)}$ and $\beta$  are the area and the inverse temperature at the event horizon of BTZ BH respectively.}
\item {The BH radiation emission is a quasi-static process (so slow) i.e. $\frac{dm}{dv}<<1 \Rightarrow m<<v$ but Hawking's temperature varies continuously.}
\end{itemize}

So, considering these assumptions with Stefan Boltzmann's law by using the values of $A$ and $\beta$, the differential form of quantum mode entropy is obtained as
\begin{equation}\label{Ent3}
dS_{\sum}=-\frac{3\zeta(3)\gamma \sqrt{-r_v ^2f(r_v)}}{2} \left(\frac{\beta}{ A }\right)dm,
\end{equation}
For a spherically symmetric rotating BH the first law of BH thermodynamics is \cite{Bardeen:1973gs}
\begin{equation}\label{1stlawJ}
{dm}=\frac{d{S}_{BH}}{\beta}+\Omega_{H} dJ,
\end{equation}
here $S_{BH}$ is the Hawking entropy also called horizon entropy. As a conserved quantity, the distortion of angular momentum at the horizon will be small, and at $r=r_v$ it will be negligibly small. So, for onward discussion, the effects of angular momentum will be ignored so, Eq. (\ref{1stlawJ}) becomes
\begin{equation}\label{1^{st}_law}
{dm}=\frac{d{S}_{BH}}{\beta},
\end{equation}

Using Eq.  (\ref{1^{st}_law}) in Eq.   (\ref{Ent3}), the relation between the interior and exterior entropy of a BH is obtained as
\begin{equation}
dS_{\sum}=-\frac{3\zeta(3)\gamma \sqrt{-r_v ^2f(r_v)}}{2} \left(\frac{dS_{BH}}{ A }\right),
\end{equation}
here, $\zeta$ is the zeta function. This equation is a direct relationship between the two types of entropy. Since the BH's interior entropy is directly related to the interior volume so, by maximizing $\sqrt{-r_v ^2f(r_v)}$, one can maximize the interior entropy. On the other hand at $r=r_v$, BH will have a constant area $A$. Both area and entropy are mass-dependent. This means that this relation between the interior and exterior entropy of a BH is a function of $m$, that is
\begin{equation}\label{proprela}
 dS_{\sum}=-\frac{4 \gamma \zeta (3)}{3 \pi }F(m)d{S}_{\text{BH}},
\end{equation}
here the proportionality function $F(m)$ is 
\begin{equation}\label{Proprel}
F(m)=\frac{\sqrt{m \left(3 X^2+\sqrt{3 X^2+1}-1\right)}}{\sqrt{X+1}},
\end{equation}
The proportional function vs. mass of the BH is plotted in Fig. (\ref{image-4.7}). This plot represents the power function of some variables between $0$ and $1$. We can see from this curve that the BTZ BH mass $(m)$ grows from $1.0$, and the slope of the curve also increases as the proportionality function increases. At the start point, the BTZ BH mass seems constant with some increase in evolution function. After that, the mass gradually increases, and gaining some mass limits the evolution relation grows with an increase in BH mass uniformly without any deviation, also see Ref. \cite{Ali:2020olc}.

 \begin{figure}
\begin{center}
\includegraphics[width=0.5\textwidth]{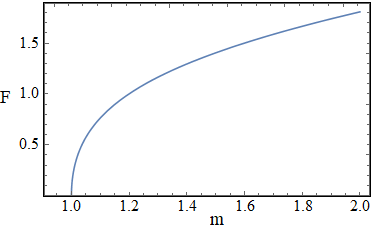}
\caption{A plot of $F(m )$ vs. mass $m$ of rotating BTZ BH \cite{Ali:2020qkb}.}
\label{image-4.7}
\end{center}
\end{figure}
\begin{itemize}
\item The BH emission rate is quasi-static i.e. $\frac{dm}{dv}\ll 1$: This means that the evaporation rate is slow enough and the horizon temperature can be considered as degenerated. This assumption enables us to consider the BH emission rate at the quantum level \cite{Hossenfelder:2012jw, Li:2002xb}. This also guarantees the preservation of the thermal equilibrium between the interior scalar field of BH and the event horizon in such an adiabatic process.\end{itemize}

\section{Nature and Probability of Hawking Radiation with Hawking Entropy} \label{natur-radiation}

As in the above discussion, we treated an entangled relation between the interior and exterior entropy for understanding the concept of BH evaporation so, it will be good to understand the nature of radiation emitting from the BH horizon. This relation between BH evaporation and entropy could be the basis for understanding the puzzle of the information loss paradox which is expected to give a detailed theory of quantum gravity \cite{Hawking:2005kf}.

A BH emits Hawking radiation and evaporates. In practice, observing them directly from an astrophysical BH is quite difficult. Many ideas are presented to understand the nature and features of these radiations \cite{Wald:1975kc, Hawking:1974rv, Hawking:1975vcx, Brout:1995wp, Michel:2014zsa, Robertson:2012ku, Unruh:1976db, MunozdeNova:2018fxv, Shi:2021nkx} but none of them could explain the proper mechanism for the emission of Hawking radiation.  
According to Hawking, these radiations are one of the quantum features of BH which can be understood by quantum tunneling across the BH horizon and retrieved by many authors e.g., \cite{Wald:1999vt, Vanzo:2011wq, Deng:2016qua}. Considering these tactics for the BH radiation, one can understand the BH interior.  As the entropy of a BH is directly related to its horizon area, what happens to its entropy after losing energy?  Once the black hole starts evaporating and the outgoing Hawking radiation escapes the BH horizon, the quantum mode entropy of this region will no longer be zero due to the entanglement between the interior and exterior quantum modes. This entropy continues to grow as the BH evaporates, see the review \cite{Almheiri:2020cfm}. In our discussion, we considered the BH emission rate as a quasi-static process, where the emission process is slow enough and the BH has a variable temperature. As long as the temperature increases, entropy increases, and hence evaporation.


One can find the nature of radiations created during the creation-annihilation process of matter \cite{Parikh:1999mf, Kraus:1994fh, Iso:2006ut, Sakalli:2014sea}.  Let us consider the KG equation
\begin{equation}
\left(\partial _{\mu }\left(\sqrt{-g}g^{\text{$\mu $v}}\partial _v\right)+m\right)\psi =0,
\end{equation}
where $m$ is the mass of the particle created. In the case of $4-$dimensional space-time coordinates, expanding the above equation gives

$$\frac{\partial}{\partial t}\left(\sqrt{-g}g^{\text{tt}}\frac{\partial\psi}{\partial t}\right) +\partial _r\left(\sqrt{-g}g^{\text{rr}}\partial _r\right)+\partial _{\phi }\left(\sqrt{-g}g^{\phi \phi }\partial _{\phi }\right)-\frac{\sqrt{-g} m^2 \psi }{\hbar ^2}=0,$$
\begin{equation}
\frac{\partial }{\partial t}\left(r g^{\text{tt}} \frac{\partial \psi }{\partial t}\right)+\frac{\partial }{\partial r}\left(\text{rg}^{\text{rr}} \frac{\partial \psi }{\partial r}\right)+\frac{\partial }{\partial \phi }\left(r g^{\phi \phi } \frac{\partial \psi }{\partial \phi }\right)-\frac{m^2 r \psi }{\hbar ^2}=0,
\end{equation}

here $\psi =e^{-i E t}e^{{iI}(r,\phi )}$ so, we get
\begin{equation}
\frac{1}{f(r)}\left(\frac{\partial I}{\partial t}\right)^2-f(r) \left(\frac{\partial I}{\partial r}\right)^2-\frac{1}{r^2}\left(\frac{\partial I}{\partial \phi }\right)^2-m^2=0,
\end{equation}
Applying the method of separating variables with classical action $I(t, r, \phi )=-E t+W(r)+L\phi +c$. So, from the above equation, we get
\begin{equation}\label{W}
f(r)\left(r \frac{\partial W(r)}{\partial r}\right)^2=\frac{E^2}{f(r)}-\frac{L^2}{r^2}-m^2=0 \Rightarrow W(r)=\pm \int \frac{\sqrt{E^2-f(r) \left(\left(\frac{L}{r}\right)^2+m^2\right)}}{f(r)} dr,
\end{equation}

Here, the $\pm$ signs point to the scalar particles moving away or toward the event horizon (emission/absorption). As $f(r)=0\Rightarrow r=r_h$, one can solve the above equation by using the residue theorem and expanding $f(r_h)$ by the Taylor series, we get 
\begin{equation}
f(r)=f(r)+\left(r_h-r\right) f'(r)+O\left(\left(r_h-r\right)^2\right)\approx \left(r_h-r\right) f'(r),
\end{equation}
So, from above Eq. (\ref{W}), we get the distribution of particles as 
\begin{equation}
W(r)=\pm \frac{\widetilde{E}}{f'(r)}\int \frac{1}{\left(r_h-r\right) } dr,
\end{equation}
here $\widetilde{E}$ is the modified energy. Solving the Residue problem, we get
\begin{equation}
W(r)=\pm \frac{i\pi \widetilde{E}}{f'(r)},
\end{equation}
the emission and absorption probabilities of particles entering and leaving the horizon is 
\begin{equation}\label{absorption}
\Gamma_{+}=e^{-\frac{2}{\hbar}ImI}=e^{-\frac{2}{\hbar}(ImW_+ +Imc)},
\end{equation}
\begin{equation}\label{emission}
\Gamma_{-}=e^{-\frac{2}{\hbar}ImI}=e^{-\frac{2}{\hbar}(ImW_- +Imc)},
\end{equation}
An object in the vicinity of the BH horizon will be swallowed by it. So, the absorption probability in Eq. (\ref{absorption}) could be normalized to unity. We can do this by considering $Imc=-ImW_-$. We also know that $ImW_+=-ImW_-$. So, from the above two Eqs. (\ref{absorption}) and (\ref{emission}), we get
\begin{equation}
\Gamma_{+}=e^{-\frac{4}{\hbar}ImW_+}=e^{-\frac{4\pi}{f'(r)} \frac{\widetilde{E}}{\hbar}},
\end{equation}
It means that during the tunneling process of particles from the BH horizon, one can't distinguish the particles. Using the value of $f'(r)$, we have
\begin{equation}\label{bolt}
\Gamma_+=e^{-\beta \frac{\widetilde{E}}{\hbar}}=e^{-\beta \omega},
\end{equation}
Here, we used $\widetilde{E}=\hbar \omega$ and $\beta$ is the inverse Hawking temperature of BH. The above Eq. (\ref{bolt}) is the Boltzmann distribution formula. Here the leading order term $e^{-\beta \omega}$ is the Boltzmann factor for emitted radiation. From this Eq. (\ref{bolt}), we can easily obtain the horizon temperature and entropy \cite{Li:2020qqa, Sakalli:2015raa}.
\section{Summary}
The study of the BH interior could reveal many facts that need to be understood about the BH. After the CR work \cite{Christodoulou:2014yia} in 2015, the notion of BH interior volume got a great attraction from many authors and great work has been done on this topic. Starting from BH's interior volume, BH's entropy, evaporation, and the information paradox issue are discussed with groundbreaking results. Many valuable results have been found so far.  This chapter reviews all related works on BH's interior volume, entropy, evaporation, and a possible solution to the information paradox by using BHs in different spacetime dimensions.

From Parikh's work (the black hole volume remains constant over time) to the perspective of Christodoulou and Rovelli (the volume of a black hole varies with time), we employed a unique method to determine the volume of the black hole interior. The concept of BH's interior volume seems different from that bound by a sphere in flat $3d$ spacetime. The interior volume bound by a $2-$sphere immersed in curved space-time is the volume of the largest space-like spherically symmetric hyper-surface bounded by two-sphere $S$. This means that the BH bounds a maximum volume equal to the volume bounded by the largest $3d$ hyper-surface and is found proportional to $v$. Following their investigations, the quantum mode entropy in the scalar field is investigated directly proportional to $v$ by Baocheng Zhang \cite{Zhang:2015gda}. From these time-dependent relations of BH interior volume and entropy with $v$, it is greatly possible to affect the statistical quantities with a small change in interior volume or entropy in the interior of BH. It is a great step to follow up for probing the idea of information paradox. After investigating the interior volume and entropy, we determined an entanglement relation between the interior and exterior entropy to understand the change in BH thermodynamical quantities and the evaporation status with these changes. For this purpose, two assumptions of BH radiation as black body radiation and the emission process of radiation as a quasi-static were introduced. The $1^{st}$ assumption led us to use the Boltzmann law and the $2^{nd}$ assumption guaranteed us to use the differential form investigation up to the quantum level for an infinitely small interval of time. Using these assumptions, we got the differential form of the interior and exterior entropy. By comparing the two types of entropy, we obtained proportional relationships. It shows some important features consistent with Hawking's investigation of BH evaporation. By applying this technique in different BH space-times, we get an extended confirmation of our results as given in the tables (\ref{tab:1}) and (\ref{tab:2}). An exemplary review is also made for Schawarzschild, Kerr, and BTZ type BH space-times.

Using the assumption of quasi-static emission, we also investigated the emission probability and nature of BH radiation that satisfies the Boltzmann distribution. The largest number of quantum states $\left(N_{BH}\propto \frac{1}{\Gamma_+}\right)$, that a BH could reside is related to $S_{BH}$ as 
 \begin{equation}
     N_{BH}=e^{S_{BH}}
 \end{equation}
But during the evaporation process, the Bekenstein and Hawking entropy gradually reduces as also clear from Eq. (\ref{Qmode}) and the number of quantum states grows with the expansion of BH's volume, thus the interior entropy of the scalar field also increases. due to this reason, at the final stage of evaporation (stop point) the number of quantum states in the BH's interior is much more than the exterior quantum modes of BH. Hence this investigation confirms the result of \cite{Christodoulou:2014yia} and one can claim that the large interior volume will have more space to store information even if the horizon shrinks to a small size (Remnant BH) \cite{Chen:2014jwq}. This is the main point of our investigation to claim a large interior of the BH to store information.

Finally, we can suggest that work is needed on these topics to more deeply understand BH physics and solve the issue of the information loss paradox. In this regard, we could consider the modified gravity theories in our future work. 

\vspace{1.0cm}
\noindent\rule{16.5cm}{2.0pt}

\clearpage
\thispagestyle{empty}
\hfill
\clearpage
\newpage

\newpage
\chapter{Configuration entropy and thermodynamics phase transition of black hole in f(R) gravity}

Starting from a d$-$dimensional black hole (BH) in $f(R)$ gravity, we analyzed the effect of modified gravity on critical point parameters, the difference in number densities, and configuration entropy during the BH phase transition phenomenon. From our investigations, consistent results with charged AdS BH are obtained that is holographic dual of van der Waal's fluid and hence the BH in modified gravity. The thermodynamic pressure, temperature, and free energy are affected by $f(R)$ gravity. The difference in the number densities of molecules (small and large BHs) and configuration entropy are investigated as a function of reduced temperature $(\tilde{\tau})$. The difference in the number densities of BH molecules in $f(R)$ gravity decreases with the increase in $\tilde{\tau}$, whereas, $S_{con}$ increases monotonically and becomes a concave function with the increase in space-time dimensions. The relation between the difference in the number density of BH molecules and space-time dimensions $(d)$ decreases with the increase of $d$. Finally, using our results, the laws of BH Physics are also discussed.
\section{Introduction} \label{sec: intro}

Einstein's theory of General Relativity (GR) gives a unified description of space-time. According to this theory, the curvature of space-time is directly related to the energy-momentum tensor. This theory predicts the existence of BH as the solution to Einstein's Equation of GR \cite{Wald:1984rga, Stephani:2004ud}. Later, it was expected that a BH has entropy associated with its surface area, and the temperature is related to surface gravity \cite{Bekenstein:1972tm, Hawking:1974sw}. Due to their thermal property, BH can radiate and evaporate under the creation, and annihilation phenomena near the event horizon. Hawking showed that an AdS-Schwarzschild BH can be thermodynamically stable and manifest the first-order phase transition due to a negative free energy at high temperatures \cite{Hawking:1982dh}. This study of BH in anti-de Sitter (AdS) space was later refreshed in the AdS/CFT correspondence, comprising a relationship between string theory in a 5-dimensional AdS space and Yang-Mills theories on the AdS conformal boundary. Subsequently, the Hawking–Page phase transition was understood in the gauge theory set-up as a phase transition from confinement to de-confinement \cite{Witten:1998qj, Witten:1998zw, Sotiriou:2008rp}. 

$f(R)$ gravity is an extension of Einstein's GR that received increased attention due to combined motivation from high-energy physics, cosmology, and astrophysics e.g., \cite{Capozziello:2011et, Shaikh:2023wsc, Nojiri:2008nt, Stelle:1976gc, Utiyama:1962sn, CANTATA:2021ktz} and a huge literature can be found. Using $f(R)$ theories, intensive research could give a wider understanding of the consequences and limitations related to cosmological applications. Generally,  metric (or second-order) formalism, Palatini (or first-order) formalism, and metric-affine formalism are used to discuss $f(R)$ theory \cite{CANTATA:2021ktz, Faraoni:2008mf, Sotiriou:2008rp}.  Compared to other theories, various considerations are important before an $f (R)$ theory is accepted as a potentially viable candidate for the underlying theory of gravity e.g., Cosmological dynamics, free from instabilities and ghost solutions, correct Newtonian and post-Newtonian limits, cosmological perturbations compatible with the data from the Cosmic Microwave Background (CMB) and large-scale structure survey, and Well-posed Cauchy problem discussed by many authors for detail see  \cite{delaCruz-Dombriz:2006kob, Dunsby:2010wg, Nojiri:2021mxf, Nojiri:2019dio, Khoury:2003aq, Sokolowski:2008kf, Appleby:2009uf, deMartino:2015zsa, Song:2006ej, Martinelli:2021hir, Capozziello:2011gw, Lanahan-Tremblay:2007sxd}. In the strong-field regime and on cosmological scales, the study of gravitational waves from binary black hole mergers \cite{Schmidt:2006jt} and a binary neutron star merger \cite{Connaughton:2016umz, LIGOScientific:2017vwq, LIGOScientific:2016aoc, Ali:2023zva} along with its optical counterpart showed constrain on deviation from GR with high accuracy and a stronger constraint could be anticipated from near-future observations. Thus, it is important to identify theories of modified gravity that intrinsically possess the same solutions as in GR among different theories. Some differences between various gravity theories can be found in linearized gravity by analyzing gravitational wave polarization via the interferometric response functions \cite{Motohashi:2018wdq, Corda:2009re, LIGOScientific:2017ync, LIGOScientific:2017zic}. This needs a wide range of analysis for single-/multi-field scalar-tensor theories of modified gravity in the presence of matter component, spacetime geometry including a cosmological background as well as spacetime around black holes and neutron stars \cite{DeLaurentis:2016jfs}, and references therein.

BH phase transition is a fascinating phenomenon after the Hawking-Page prediction about the BH and a standard thermodynamics system. The study of charged AdS BH and critical phenomenon unveiled the first-order phase transition of Van der Wall (VdW) like system (that obeys the liquid-gas phase transition) \cite{Chamblin:1999tk, Chamblin:1999hg, Kubiznak:2012wp, Gunasekaran:2012dq}. Following this analogy, several authors have revealed the most fascinating and significant physics in BH thermodynamics, e.g. \cite{Altamirano:2013ane, Wei:2014hba, Frassino:2014pha, Altamirano:2013uqa, Belhaj:2012bg, Hendi:2012um, Altamirano:2014tva, Dolan:2014jva}. The VdW phase transition and $P-v$ criticality of AdS BH in general framework are discussed in \cite{Kubiznak:2012wp, Bhattacharya:2017hfj, Majhi:2016txt} and also extended to massive gravity \cite{Ghodrati:2020mtx, Chougule:2018cny} where the logarithmic corrections for entropy are found necessary for holographic correction of VdW fluid. \cite{Upadhyay:2019hyw}. The equation of state for VdW fluid is read as,
\begin{equation}\label{vdW-eq}
    (P+\frac{\alpha}{v^2})(v-\beta)=k_BT\Rightarrow Pv^3-(k_BT+\beta P)v^2+\alpha v-\alpha \beta=0,
\end{equation}
here, $v=\frac{V}{N}$ is the specific volume, $\alpha$ and $\beta$ are constants, and $k_B$ is the Boltzmann constant. Later, we will use an analogous technique to obtain the equation of state for BH in $f(R)$ gravity with its VdW fluid-like nature.

From the holographic principle \cite{Bousso:2002ju, Abdalla:2001as}, the information content of a region of space can be encoded on its boundary rather than within its volume. In the case of black holes, the holographic principle suggests that the entropy and other properties of a black hole can be described in terms of degrees of freedom residing on the surface of the black hole's event horizon. In the AdS/CFT correspondence \cite{Ramallo:2013bua, Hubeny:2014bla}, a gravitational theory in Anti-de Sitter (AdS) space is dual to a quantum field theory without gravity living on the boundary of that space. This duality allows for the study of black hole entropy and other gravitational phenomena through the lens of a well-understood quantum field theory. In this framework, the configurational entropy of a black hole is associated with the number of possible microscopic configurations of the quantum fields living near the event horizon. The details of these configurations determine the black hole's entropy and are encoded in the boundary theory.

Configuration entropy (information entropy) is related to the position of its constituent particles rather than their velocity or momentum. It is the best tool for investigating the stability of a physical system under consideration. It provides an extraordinary tool to analyze compact astrophysical objects. Stability bound for stellar distributions was first obtained in the context of the configurational entropy by \cite{Gleiser:2013mga, Gleiser:2015rwa, Casadio:2016aum}. In reference \cite{Casadio:2016aum} a relevant procedure is provided to scrutinize the stability of Bose-Einstein condensation for long-wavelength gravitons in the quantum portrait of a BH. In these investigations, the critical stability region of stellar configurations has been consistently defined in complementary distinct paradigms by matching with observational data. Hence, the critical stellar densities were shown to match the Chandrasekhar limit associated with critical points of the configuration entropy.

In this chapter, we discussed the critical phenomenon and entropy of microscopic states for BH in $f(R)$ gravity following the work of \cite{Lee:2017ero}. The key idea is to generalize the critical phenomenon and investigate the configuration entropy in $f(R)$ gravity. After getting these results, we used a $ 4-$dimensional space-time to manipulate the effects of $f(R)$ gravity on configuration entropy and other parameters compared to the AdS BHs \cite{Kubiznak:2012wp, Sadeghi:2016dvc}. The $P-v$ criticality of BH in $f(R)$ gravity is the holographic duality of a VdW fluid and satisfies the first-order phase transition. The organization of this chapter is such that in the next section (\ref{sec1}), we will discuss the basic concept of configuration entropy and its main differences from the Bekenstein-Hawking entropy. In section (\ref{sec2}), we will follow a general approach to investigate relations for the co-existence curve, the difference in molecular number densities of molecules, and the configuration entropy of a $d-$dimensional BH in an $f(R)$ gravity by using either thermodynamic relations as well as reduced parameters space. In section (\ref{sec3}), we have applied the general approach to a $4d-$ space-time to evolve the difference in number density and configuration entropy $(S_{con})$ in the context of molecule number densities $(n's)$, which in turn are a function of $\tilde{\tau}$. Using these general investigations, we obtained several interesting results. These results are consistent with those of charged AdS BH \cite{Kubiznak:2012wp} and hence show the dual nature of BH in $f(R)$ gravity with the VdW fluid. We also calculate the relation between the difference in number density and dimensional coordinates for vanishing reduced pressure (i.e. at its maximum value). In section (\ref{sec4}), some remarks and discussions are added and the final section (\ref{sec5}) is devoted to the results of this work.

\section{The basic concept of Configuration Entropy}\label{sec1}

Configuration entropy for an unknown number of states $(N)$ is defined by Shannon relation $\log _2 N=k_B \log N$, where $k_B$ is the Boltzmann constant \cite{shannon:1948}. In a thermodynamic process, the variation of configuration entropy is equivalent to the change of the system's microscopic entropy so, the Shannon entropy is proportional to Boltzmann-Gibbs entropy. Whereas, in statistical thermodynamics, it is related to micro-states $(w)$ distributed at a given energy states of the system as,

\begin{equation}\label{GBent1}
S_{BG}=k_B \log w,
\end{equation}
here, $k_{\beta}$ is the Boltzmann constant. The probability of particle configuration over the micro-state distribution has the form $w=\frac{1}{p_i}=\frac{N_1 ! N_2 !}{N_o !}$. For each molecule, there is an equal probability in the given distribution so,
\begin{equation}\label{GBent2}
S_{BG}=-k_B\sum_{i=1} ^w p_i \log p_i,
\end{equation}
here $\sum p_i=1$ is the total probability of a particle in the given distribution. Let, $N_o$ be the total of $N_1$ and $N_2$ types molecules then the configuration entropy can also be written as
\begin{equation}\label{GBent3}
S_{BG}=k_B log \left(\frac{N_o !}{N_1 ! N_2 !}\right),
\end{equation}
Here the effective molecules number density $n$ of molecules can be written as \cite{Gunasekaran:2012dq, Kubiznak:2012wp}
\begin{equation}\label{noDen}
n=\frac{N_o}{V}=\frac{1}{4 l_p ^2 r_+},
\end{equation}
where $V$ is the thermodynamic volume and $l_p$ is the Plank's length $\left(l_p ^2=\frac{G \hbar}{c^3}\right)$. Using probability distribution, the total configuration entropy of the system can be calculated as,
\begin{equation}\label{GBent4}
S_{con}=-\left(n_1 \log \left(\frac{n_1}{n_1+n_2}\right)+n_2 \log \left(\frac{n_2}{n_1+n_2}\right)\right),
\end{equation}
with geometric units, $G=\hbar=k_{\beta}=c=1$ are used. Here, we have used Stirling's approximation with effective molecule number densities $(n_1, n_2)$. In connection with BH thermodynamics, configuration entropy is discussed by many authors. Our work used the most common formula between the entropy and the micro-states distribution in Eq. (\ref{GBent1}) to drive Eq. (\ref{GBent4}). The special property of this equation is the Boltzmann distribution i.e. the distribution of energy between identical but distinguishable particles or we can say to distinguish the state of the particles by their energies. The main perspective of Boltzmann distribution is from the information entropy that we will discuss in the case of BH physics. Note that the number density of small and large molecules is related to the small and large fields as discussed in the literature \cite{Wei:2015iwa} and to make parallelism of our work with the concept of thermodynamics phase transition. By investigating molecular number densities $n_1$ and $n_2$ of molecules, one can find the configuration entropy black hole associated with $f(R)$ gravity. 

The configuration entropy in black holes is a topic of interest for many researchers and is related to understanding the microscopic origin of black hole entropy. It is also known as information entropy is related to the position and orientation of the constituent particles. The Bekenstein-Hawking entropy of BH is proportional to its surface area showing the BH as a thermodynamic system. It provides a macroscopic description of BH but does not provide insight into the underlying microscopic degrees of freedom responsible for the entropy. This could be done by studying the configuration entropy. Hence, understanding the microscopic origin of black hole entropy, including the role of configurational entropy, is an active area of study in theoretical physics and is connected to the broader quest for a theory of quantum gravity. The configurational entropy is related to the location of micro-states of the constituent particles or observables (existing in the BH) where they could be packed together in the maximum possible number of ways. The Hawking entropy counts for the area bounded by the black hole’s horizon instead of the discrete field. Generally, the configuration entropy is the disorderliness of the arrangement of atoms or micro-states whereas, Hawking's entropy is related to the horizon temperature governing the profound relation with laws of black hole thermodynamics. A more detailed difference between the two types of entropy can also be found in the literature \cite{Shah:2023vxf, Pessoa:2022fwo, Casadio:2022pla, Braga:2016wzx, Correa:2015vka, Barreto:2022ohl, daRocha:2021jzn}. 

\section{A General Framework to Configuration Entropy} \label{sec2}

Consider the line element of a $d-$dimensional BH in an $f(R)$ gravity as
\begin{equation}\label{mettric}
ds^2=-f(r)dt^2+f(r)^{-1}dr^2+r^2\Omega_{d-2},
\end{equation}
here, the lapse function takes the form
$$f(r)=1-\frac{2m}{r^{(d-3)}}+\frac{q^2}{br^{2(d-3)}}-\frac{R_0r^2}{12}, $$
here $b=1+f'(R)$. It represents an asymptotically AdS solution if $R_0=-\frac{12}{l^2}=4\Lambda$, $l$ is the AdS length. A negative cosmological constant $(\Lambda = -\frac{3}{l^2})$ has interesting features in AdS studying the BH thermodynamics. Thus, it has a huge advantage in holography and AdS/CFT correspondence. The critical point in charged AdS BH shows the Van der Waal's fluid-like nature and it is found that charged AdS BH is a holographic dual of VdW fluid \cite{Niu:2011tb, Kubiznak:2012wp, Sadeghi:2016dvc}. The asymptotically AdS BH admits a gauge duality description with dual thermal field theory in the presence of $\Lambda$, which leads to the famous Hawking-Page phase transition \cite{Hawking:1982dh} for more detail also see \cite{Niu:2011tb, Cai:2007wz, Cai:2007vv, Eune:2013qs}.  The parameters $m$ and $q$ correspond to the ADM mass and electric charge of BH,
\begin{equation}\label{masscharge}
m=H=\frac{16\pi M}{(d-2)\Omega'_{d-2}} \qquad q=\frac{8{\pi} Q}{\Omega'_{(d-2)}\sqrt{2(d-2)(d-3)}}\qquad
\Omega'_{d-2}=\frac{2\pi ^{\frac{d-1}{2}}}{\Gamma[\frac{(d-1)}{2}]},
\end{equation}

From Ref. \cite{Chen:2013ce}, the relation between scalar curvature and thermodynamic pressure can be read as $R_0=-\frac{64 \pi P}{(d-2)b}$ so, the horizon temperature $T$ of a $d$-dimensional BH in an $f(R$) gravity is
\begin{equation}\label{temp}
T=\frac{\partial_r f(r)}{4\pi}=\frac{1}{12\pi r_+}\left (3(d-3)\left ( 1-\frac{q^2 r_+ ^{6-2d}}{b} \right )+\frac{16(d-1)\pi r_+ ^2 P}{(d-2)b} \right ),
\end{equation}
and the entropy is given by 
\begin{equation}\label{Ther-ent}
S=\frac{\Omega '_{d-2} r_+ ^{d-2}b }{4}
\end{equation}
where $A$ is the area of the BH horizon. From Eq. (\ref{temp}), the thermodynamic pressure can be written as
\begin{equation}\label{pressure}
P=\frac{3b}{d-1}\left (\frac{(d-2)}{4 r_+ }T-\frac{(d-2)(d-3)}{16\pi r_+ ^2}\left ( 1-\frac{q^2 r_+ ^{6-2d}}{b}\right )\right ),
\end{equation}
Its conjugate thermodynamic volume \cite{Cvetic:2010jb}
\begin{equation}
V=\frac{2\pi ^{\frac{d-1}{2}}b r_+ ^{d-1}}{\Gamma(\frac{d-1}{2})}
\end{equation}

Using the specific volume of the BH as $\nu=\frac{4l_p ^2 r_+}{(d-2)}$, Eq. (\ref{pressure}) becomes
\begin{equation}\label{P2}
P=\frac{3b}{(d-1)}\left (\frac{T}{\nu}-\frac{(d-3)}{\pi (d-2){\nu}^2}+\frac{4^{2d-5}(d-3)q^2}{4\pi b (d-2)^{2d-5}{\nu}^{2d-4}}\right ),
\end{equation}
This equation is analogous to the VdW equation of state and one can claim to investigate the critical point. The Gibbs free energy $(G)$ from the Legendre transformation of enthalpy is
\begin{equation}\label{FE2}
G=H-TS=\frac{(d-2)^{d-3}\pi^{\frac{d-1}{2}}}{2^{2d-3} \pi \Gamma{(\frac{d-1}{2}})}\left (b\nu^{d-3}-\frac{(d-2)\pi P \nu^{d-1}}{d-1}+\frac{4^{2(d-3)}(2d-5)q^2}{(d-2)^{2(d-3)}\nu^{d-3}} \right ),
\end{equation}
Note, here $l_p ^2=1$ is used from geometric units. The value of Gibbs free energy may be positive or negative corresponding to the stability or instability of BH respectively. The $P-v$ diagram in $4-$dimensional space-time is shown in Fig. (\ref{image-5.1}).
\begin{figure}
\begin{center}
\includegraphics[width=0.58\textwidth]{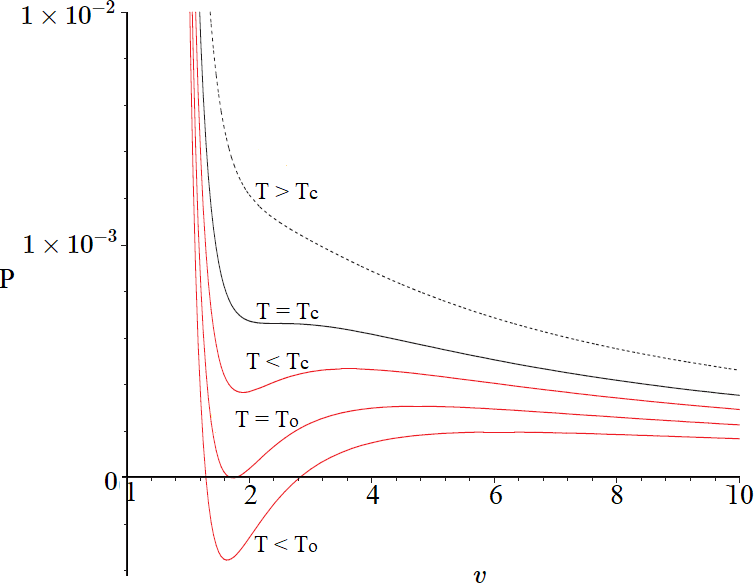}
\caption{$P-v$ diagram for a $4-$dimensional charged BH in $f(R)$ gravity. [From top to bottom]. The upper top dashed line is for "ideal gas". The black curve is the critical isotherm $T=T_c$, the lower red line is two-phase states at $T<T_{c}$ at $q=1$ and $f'(R)=0.1$.}
\label{image-5.1}
\end{center}
\end{figure}
The critical point must satisfy the conditions
\begin{equation}\label{CC}
\partial_\nu P_{r_h=r_\nu, T=T_c}=0 \qquad \partial_\nu ^2 P_{r_h=r_\nu, T=T_c}=0,
\end{equation}
which leads us to calculate the critical point parameters both in general and for $d=4$ are
\begin{equation}\label{SVCP}
\nu_c =\frac{4}{d-2}\left(\frac{(d-2)(2d-5)q^2}{b}\right)^{\frac{1}{2(d-3)}} = 2 \sqrt{6}q b^{-\frac{1}{2}},
\end{equation}
\begin{equation}\label{CPP}
P_c=\frac{3b(d-3)^2}{(d-2)^2(d-1)\pi \nu_c ^2}=\frac{1}{48\pi}\left(\frac{b}{q}\right)^2,
\end{equation}
and
\begin{equation}\label{CPT}
T_c=\frac{4(d-3)^2}{(d-2)(2d-5)\pi \nu_c}=\frac{\sqrt{b}}{3 \sqrt{6}q\pi},
\end{equation}
here the temperature of the critical point is dependent on modified gravity parameter $b$ as it is a function of $v_c$. Using the values of $v_c$ and $P_c$ in Eq. (\ref{FE2}), the free energy at the critical point becomes
\begin{equation}\label{CPFE}
G_c=\frac{\pi ^{\frac{d-3}{2}} \left(d^3-6 d^2+21 d-28\right)}{8 (d-1)^2 \Gamma \left(\frac{d-1}{2}\right)}\sqrt{\frac{b (2 d-5) q^2}{d-2}}=\sqrt{\frac{2}{3}}q \sqrt{b},
\end{equation}
 Eq. (\ref{SVCP})-(\ref{CPFE}) are the thermodynamic parameters of the critical point, where the phase of molecules is identical but can't be identified at vanishing latent heat \cite{Mo:2016sel, Chen:2013ce}. These equations of critical point show that the values of critical temperature and pressure increase while the critical volume decreases with increasing modified gravity factor $b$. For phase transition, the modified gravity factor $b>1$. It means, all the critical values of $P_c$, $T_c$, $G_c$, and $v_c$ must be positive. From this, the universal inflection point on the critical line (also called the critical ratio) can be written as,
\begin{equation}\label{Pt.inter}
\rho_c=\frac{P_c\nu_c}{T_c}=\frac{3b(2d-5)}{4(d-2)(d-1)}=\frac{3 }{8}b,
\end{equation}
It is a valid universal constant predicted for an ideal gas. It shows the same relationship as that of VdW fluid. In the case of $4-$dimensional charged AdS BH $\rho_c=\frac{3}{8}$ \cite{Kubiznak:2012wp}. This result is surprising because it proves the coincidence of BH in modified gravity with the VdW fluid. In the case of $f(R)$ BH, the value of $\rho$ increases monotonically with the increase in $f'(r)$. 

One can consider the co-existence curve as a borderline between the two regions. On both sides of the critical curve, the number densities $n_1$ and $n_2$ are different as represented by the subscripts $1$ and $2$. It is found that during phase transition, the specific volume and number density suffer a sudden change from a small black hole (SBH) to a large black hole (LBH). For the microscopic structure of the thermodynamics system having discontinuous changes during the phase transition where the non-vanishing latent heat of each BH is determined by the formula \cite{Wei:2015iwa},
\begin{equation}\label{LHeat}
L=\frac{T\Delta S}{N}=T\Delta\nu\partial_T P=T\left(\frac{1}{n_1}-\frac{1}{n_2}\right)\partial_T P,
\end{equation}
At the co-existence line $T=T_c, \quad n_1=n_2 \quad \Rightarrow \quad L=0 \quad $ so, one can't distinguish between particles from SBH and LBH during the phase transition. 

Now to calculate the critical exponent, let us define the reduced parameters. Consider the relations
\begin{equation}\label{Red.Para}
\tilde{p}=\frac{P}{P_c}\qquad \tilde{\tau}=\frac{T}{T_c}\qquad \tilde{\nu}=\frac{\nu}{\nu_c}\qquad\tilde{G}=\frac{G}{G_c},
\end{equation}
As the critical point parameters depend on the modified gravity parameter, this will affect the critical exponents. Considering this fact, the equation of state can be rewritten in terms of reduced parameters as
\begin{equation}\label{Red.P}
\tilde{p}=\frac{4\tilde{\tau}(d-2)}{(2d-5)\tilde{\nu}}-\frac{(d-2)}{(d-3)\tilde{\nu}^2}+\frac{1}{(d-3)(2d-5)\tilde{\nu}^{2d-4}},
\end{equation}
or we can also write as,
\begin{equation}\label{Red.T}
\tilde{\tau}=\frac{1}{4} \left(\frac{(2 d-5) \tilde{\nu } \tilde{p}}{d-2}+\frac{2 d-5}{(d-3) \tilde{\nu }}-\frac{1}{(d-2) (d-3) \tilde{\nu }^{2 d-5}}\right),
\end{equation}
and the reduced Gibbs free energy from Eq. (\ref{FE2}) can be obtained as,
\begin{equation}\label{Red.FE}
\tilde{G}=\frac{\sqrt{b} \pi ^{\frac{d-1}{2}} q\sqrt{(d-2) (2 d-5)}}{8 \pi \Gamma \left(\frac{d-1}{2}\right)}\left ( \tilde{\nu }^{d-3}+\frac{1}{(d-2) \tilde{\nu }^{d-3}}-\frac{3 (d-3)^2 \tilde{p} \tilde{\nu }^{d-1}}{(d-2) (d-1)^2} \right),
\end{equation}
The reduced parameters in Eq. (\ref{Red.P}) and (\ref{Red.T}) are independent of factor $b$, except the reduced Gibbs free energy $\tilde{G}$, depending on it. The first order phase transition occurs between SBH and LBH along the co-existence curve except for the critical point $\tilde{\tau}=T_c$, where the two states have the same Gibbs free energies and temperatures i.e. $\tilde{G}_1=\tilde{G}_2, \qquad \tilde{\tau_1}=\tilde{\tau_2}$.  The second-order phase transition occurs when the molecules cross the critical point. Consequently, we get:
\begin{equation}\label{A1}
\left ( \tilde{\nu}_1^{d-3}-\tilde{\nu _2}{}^{d-3} \right )-\frac{3 (d-3)^2 \tilde{p}}{(d-2) (d-1)^2}\left ( \tilde{\nu _1}{}^{d-1}-\tilde{\nu _2}{}^{d-1} \right )-\frac{1}{d-2}\left ( \frac{\tilde{\nu _1}{}^{d-3}-\tilde{\nu _2}{}^{d-3}}{\left(\tilde{\nu _1} \tilde{\nu _2}\right){}^{d-3}} \right )=0,
\end{equation}
\begin{equation}\label{A2}
\frac{(2 d-5) \tilde{p}}{d-2}\left ( \tilde{\nu }_1-\tilde{\nu }_2 \right )-\frac{2 d-5}{d-3}\left ( \frac{\tilde{\nu }_1-\tilde{\nu }_2}{\tilde{\nu }_1 \tilde{\nu }_2} \right )-\frac{1}{(d-2) (d-3)}\left ( \frac{1}{\tilde{\nu _1}{}^{2 d-5}}-\frac{1}{\tilde{\nu _2}{}^{2 d-5}} \right )=0,
\end{equation}
Similarly, the average temperature of SBH and LBH at the co-existence curve in reduced parameters can be written as:
\begin{equation}\label{A3}
 \tilde{\tau }=\frac{1}{8}\left(\frac{(2 d-5) \tilde{p}}{d-2}\left ( \tilde{\nu _1}+\tilde{\nu _2} \right )+\frac{2 d-5}{d-3}\left ( \frac{\tilde{\nu _1}+\tilde{\nu _2}}{\tilde{\nu _2} \tilde{\nu _1}} \right )-\frac{1}{(d-2) (d-3)}\left ( \frac{\tilde{\nu _1}{}^{2 d-5}+\tilde{\nu _2}{}^{2 d-5}}{\left(\tilde{\nu _1} \tilde{\nu _2}\right){}^{2 d-5}} \right )\right ),
\end{equation}
Eq. (\ref{A1})-(\ref{A3}) are the general solutions for reduced parameters of a $d-$dimensional BH in $f(R)$ gravity in terms of an effective volume. The main purpose of these solutions is to find a relation between reduced pressure and temperature at the co-existence state, molecule number densities, and configuration entropy of BH in $f(R)$ gravity \cite{Lee:2017ero}. In the next section, we will consider the $4-$dimensional case of the above results.

\section{Approach to 4d f(R) gravity framework \label{sec3}}
Here we will analyze our general framework for $4-$dimensional $f(R)$ AdS space-time. Let's take $\tilde{\nu _1}+\tilde{\nu _2}=x$ and $\tilde{\nu _1} \tilde{\nu _2}=y$, then Eq. (\ref{A1}) - (\ref{A3}) can be written as.
\begin{equation}\label{A12}
y \tilde{p} \left(x^2-y\right)-6 y+3=0,
\end{equation}
\begin{equation}\label{A22}
3 y^3 \tilde{p}+x^2-6 y^2-y=0,
\end{equation}
\begin{equation}\label{A32}
16 y^3 \tilde{\tau }-3 x y^3 \tilde{p}+x \left(x^2-3 y\right)-6 x y^2=0,
\end{equation}
Solving these equations, we get the reduced pressure as,
\begin{equation}\label{RPnRT}
\tilde{p}=\frac{2^{4/3} \tilde{\tau }^2 \left(\sqrt{\tilde{\tau }^2-2}-\tilde{\tau }\right)^{2/3}}{\left(\left(\sqrt{\tilde{\tau }^2-2}-\tilde{\tau }\right)^{2/3}+\sqrt[3]{2}\right)^2},
\end{equation}
This equation shows that the reduced pressure is a function of reduced temperature, independent of factors like charge, Ricci scalar $R_0$, and modified gravity factor $b$. The co-existence curve can also be found from Maxwell's equal area law. It is simply a curve in the $\tilde{p}-\tilde{\tau}$ plane where the Gibbs free energy and the temperature at $r=r_l$ (for LBH) and $r=r_s$ (for SBH) coincide. The slope of the co-existence curve in $\tilde{p}-\tilde{\tau}$ plane is depicted in (\ref{image-5.2}). The slope of this curve can be explained by the Clapeyron equation which will be discussed later. This curve is similar to that of the Van der Wall's fluid. The transition occurs at temperature $T<\tilde{\tau}_c$. 
\begin{figure}
\begin{center}
\includegraphics[width=0.6\textwidth]{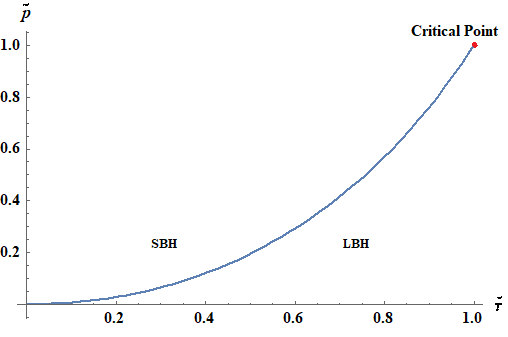}
\caption{The plot of $\tilde{p}$ vs $\tilde{\tau}$ for $d=4$ of the BH in $f(R)$ gravity. The co-existence curve (blue line) and critical point (red dot on the top of the co-existence curve) are also shown for $(\tilde{p}, \tilde{\tau})$-plane for $f'(R)=0.8$.}
\label{image-5.2}
\end{center}
\end{figure}

Introducing the difference in molecule number densities by using $n=\frac{1}{\nu}$, we get
\begin{equation}\label{NDD1}
\frac{n_1-n_2}{n_c}=\frac{\sqrt{(\tilde{\nu _1}+\tilde{\nu _2})^2-4{\tilde{\nu _1}\tilde{\nu _2}}}}{\tilde{\nu _1}\tilde{\nu _2}}=\sqrt{6-6\sqrt{\tilde{p}}},
\end{equation}
which is independent of factor $b$ and the curve is similar to the curve of charged AdS BH. So, we can claim that BH in $f(R)$ gravity obeys the VdW phase transition. Using Eq. (\ref{RPnRT}), the above Eq. (\ref{NDD1}) can also be written as:
\begin{equation}\label{NDD2}
\frac{n_1-n_2}{n_c}=\sqrt{6-\frac{6 \left(2^{2/3} \tilde{\tau} \sqrt[3]{\sqrt{\tilde{\tau} ^2-2}-\tilde{\tau} }\right)}{\left(\sqrt{\tilde{\tau} ^2-2}-\tilde{\tau} \right)^{2/3}+\sqrt[3]{2}}},
\end{equation}
The $\frac{n_1-n_2}{n_c}$ is also a function of reduced temperature, independent of factor $b$. Its plot is shown in Fig. (\ref{image-5.3}), which shows that $\frac{n_1-n_2}{n_c}$ decreases with the increase of $\tilde{\tau}$ and approaches zero at the critical point. This means that the difference in number density is maximum at $\tilde{\tau}=0$. 
\begin{figure}
\begin{center}
\includegraphics[width=0.6\textwidth]{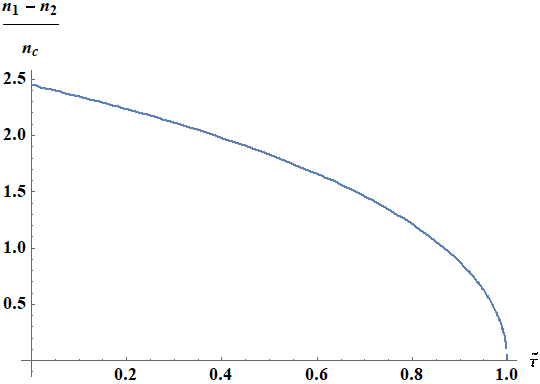}
\caption{The plot of $\frac{n_1-n_2}{n_c}$ vs. $\tilde{\tau }$ for $d=4$ for the BH in $f(R)$ gravity. The difference in molecule number densities decreases with the increase of reduced temperature and its maximum value is at $\tilde{\tau}=0$.}
\label{image-5.3}
\end{center}
\end{figure}
The individual molecule number densities $n_1$ and $n_2$ can be determined as:
\begin{equation}\label{NDDTZ}-
\begin{split}
n_{(1,2)}=\frac{\sqrt{(\tilde{\nu _1}+\tilde{\nu _2})^2-4{\tilde{\nu _1}\tilde{\nu _2}}}\pm(\tilde{\nu _1}+\tilde{\nu _2})}{\tilde{\nu _1}\tilde{\nu _2}}=\frac{\sqrt{x^2-4y}\pm x}{2y}\\
=\frac{\left (\sqrt{3-\sqrt{\tilde{p}}}\pm \sqrt{3-3\sqrt{\tilde{p}}} \right )\sqrt{\tilde{p}}}{\sqrt{2}},
\end{split},
\end{equation}
here the minus sign $(-)$ is taken for small molecules before phase transition and the $(+)$ sign for corresponding large molecules after phase transition. Using the value of $\tilde{p}$ from Eq. (\ref{RPnRT}), we get:
\begin{equation}\label{n1n2}
\begin{split}
n_{(1,2)}=\frac{1}{\sqrt{2}}\left ( \sqrt{3-\frac{ 2^{2/3} \tilde{\tau} \sqrt[3]{\sqrt{\tilde{\tau} ^2-2}-\tilde{\tau}}}{\left(\sqrt{\tilde{\tau}^2-2}-\tilde{\tau}\right)^{2/3}+\sqrt[3]{2}}} \pm \sqrt{3-\frac{3\times 2^{2/3}\tilde{\tau}\sqrt[3]{\sqrt{\tilde{\tau}^2-2}-\tilde{\tau}}}{\left(\sqrt{\tilde{\tau}^2-2}-\tilde{\tau} \right)^{2/3}+\sqrt[3]{2}}} \right )\times \\ {\frac{ 2^{2/3} \tilde{\tau} \sqrt[3]{\sqrt{\tilde{\tau} ^2-2}-\tilde{\tau}}}{\left(\sqrt{\tilde{\tau}^2-2}-\tilde{\tau}\right)^{2/3}+\sqrt[3]{2}}},
\end{split}
\end{equation}
The distribution probability of small and large molecules depends on $n_{(1,2)}$, which can be expressed as: 
\begin{equation}
p_1=\left(-\frac{n_1}{n_2}\right)^{n_1} \left(1-\frac{n_1}{n_2}\right)^{n_2} \left(\frac{n_1}{n_2}-1\right)^{-n_1},
\end{equation}

Note that this equation is obtained using Stirling's approximation for the factorial equation as $N_i != N_i^{N_i} e^{-N_i}$ (\textit{Note:} Taking the $\log$ of both sides, we can get Sterling's approximation in its general form). Using the values of $n_{(1,2)}$, one can get the $p_1$ as a function of reduced temperature. The $S_{con}$ can be calculated from Eq. (\ref{GBent2}) or Eq. (\ref{GBent3}). Here, we use the values of number density in Eq. (\ref{GBent4}) so, the $S_{con}$ can be written as:
\begin{equation}\label{conf-Ent}
\begin{split}
S_{con}=\sqrt{\frac{\tilde{p}}{2}}\left(\sqrt{3-3 \sqrt{\tilde{p}}}-\sqrt{3-\sqrt{\tilde{p}}}\right)\log \left(\frac{\sqrt{3-3 \sqrt{\tilde{p}}}}{2 \sqrt{3-\sqrt{\tilde{p}}}}-1\right)-\\ \sqrt{\frac{\tilde{p}}{2}} \left(\sqrt{3-\sqrt{\tilde{p}}}+\sqrt{3-3 \sqrt{\tilde{p}}}\right) \log \left(1+\frac{\sqrt{3-3 \sqrt{\tilde{p}}}}{2 \sqrt{3-\sqrt{\tilde{p}}}}\right),
\end{split}
\end{equation}

Using the value of $\tilde{p}$ from Eq. (\ref{RPnRT}), we can get the configurational entropy for $d=4$ as a function of reduced temperature, and the plot of $S_{con}$ vs $\tilde{\tau}$ is shown in Fig. (\ref{image-5.4}). Similarly, by calculating $\tilde{p}$ for higher dimensions, the configuration entropy can be calculated from the above equation. From the above plot, it is clear, that the configurational entropy increases monotonically with reduced temperature and becomes a concave function of reduced temperature with an increase in space-time dimensions as shown in Fig. (\ref{image-5.5}).
\begin{figure}
\begin{center}
\includegraphics[width=0.6\textwidth]{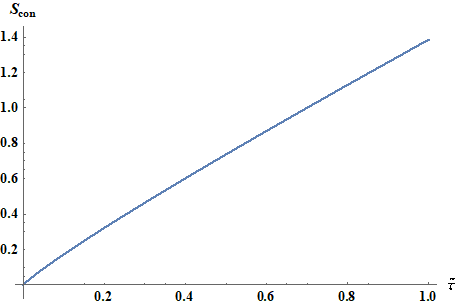}
\caption{The plot of $S_{con}$ vs. $\tilde{\tau}$ for $d = 4$}
\label{image-5.4}
\end{center}
\end{figure}
In higher-order dimensional cases like $d=5, 6,...$, the solution of these equations becomes more complicated. So, we considered a simple numerical approach to avoid complexity in its solution. The resultant plot for $d=5$ is shown in Fig. (\ref{image-5.5}), here the $S_{con}$ increases with reduced temperature, and at the final stage, $S_{con}$ becomes a concave function of reduced temperature. For a d-dimensional $f(R)$ BH, the $S_{con}$ is also dependent on $\tilde{\tau}$ and independent of factor $b$ as evident from Eq. (\ref{conf-Ent}).
\begin{figure}
\begin{center}
\includegraphics[width=0.6\textwidth]{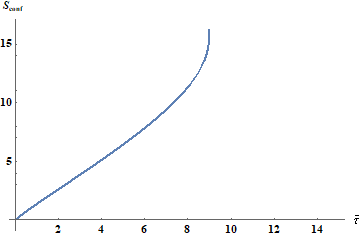}
\caption{The plot of $S_{con}$ vs. $\tilde{\tau}$ for $d = 5$.}
\label{image-5.5}
\end{center}
\end{figure}
At $\tilde{\tau}=0 \Rightarrow \tilde{p}=0$, from Eq. (\ref{A2}) and (\ref{A3}), the maximum difference in molecule number densities becomes:
\begin{equation}\label{DD-d}
\frac{n_1-n_2}{n_c}=((d-2) (2 d-5))^{\frac{1}{2 (d-3)}},
\end{equation}
\begin{figure}
\begin{center}
\includegraphics[width=0.6\textwidth]{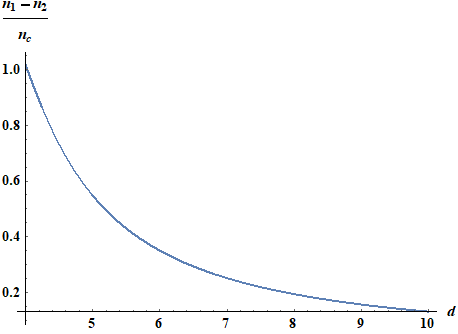}
\caption{The plot of $\frac{n_1-n_2}{n_c}$ vs $d$ of the BH in $f(R)$ gravity.}
\label{image-5.6}
\end{center}
\end{figure}
Using the value of reduced pressure, the difference in molecule number density becomes a function of reduced temperature, and its plot in Fig. (\ref{image-5.6}) shows that the difference in molecular number density decreases with an increase in space-time dimensions. 
\section{Validation of Physics Laws}
As we have seen the modified gravity factor $b$ greatly affects the parameters. Let us now consider thermodynamic laws to validate our results as in Ref. \cite{Wei:2019gur}.
\subsection{The first law of black hole thermodynamics} 
We have defined the reduced parameters in Eq. (\ref{Red.Para}) where the critical point parameters are already obtained. So, we will try to validate the laws of BH physics. As we know the first law of BH thermodynamics is
\begin{equation}
dM=TdS+VdP,
\end{equation}
Since we have introduced the critical point parameter in Eq. (\ref{CC}) to (\ref{CPT}). So, to get these results like ordinary thermodynamics, we need to treat the BH mass as enthalpy rather than the total energy of the BH. So, using Eqs. (\ref{Red.Para}) the first law of BH thermodynamics and free energy in reduced parameters can be written as 
\begin{equation}\label{1stlawa}
d\tilde{H}=c_1\tilde{\tau} d\tilde{S}+c_2 \tilde{v} d\tilde{p},
\end{equation}
\begin{equation}\label{1stlawb}
d\tilde{G}=-c_3\tilde{S} d\tilde{T}+c_4 \tilde{v} d\tilde{p},
\end{equation}

with critical ratios $c_1$, $c_2$, $c_3$, and $c_4$ are 
\begin{equation}\label{CRs}
c_1= \frac{T_c S_c}{H_c}, \qquad c_2= \frac{P_c V_c}{H_c}, \qquad c_3= \frac{T_c S_c}{G_c} \qquad c_4= \frac{P_c V_c}{G_c},
\end{equation}
The reduced parameters are already discussed in section \ref{sec2}. So, to calculate the Eq. (\ref{1stlawa}), and (\ref{1stlawb}), we need to consider the reduced parameter along with $c_1$, $c_2$, $c_3$, and $c_4$. By doing so, we calculate the critical entropy from Eq. (\ref{Ther-ent}) by using the critical volume from Eq. (\ref{SVCP}) as 
\begin{equation}
S_c=24 \pi q^2,
\end{equation}
and enthalpy for the proposed BH from Eq. (\ref{CPFE}) as 
\begin{equation}
H_c=5 \sqrt{\frac{2b}{3}} q,
\end{equation}and the critical volume is 
\begin{equation}
V_c=24 \sqrt{6} \pi \left(\frac{q^2}{b}\right)^{3/2},
\end{equation}

So now, we can calculate the numerical values of critical ratios as 

\begin{equation}
c_1 = 0.8, \qquad c_2=0.3, \qquad c_3=4.0, \qquad c_4=1.5,
\end{equation}So, using these values, we can write the first law of BH thermodynamics in reduced form for the BH in $f(R)$ gravity.
\subsection{Clapeyron Equation}
We have plotted the $\tilde{p}-\tilde{\tau}$ diagram in Fig. (\ref{image-5.3}) whose slope of this curve can be described by the Clapeyron equation. In reduced form, the Clapeyron equation can be obtained from Eq. (\ref{1stlawb}) only at the phase transition point where the free energy for both the small and large BH is vanishing i.e. $\tilde{G_1}=\tilde{G_2}=0$. So,
\begin{equation}
\frac{\partial \tilde{P}}{{\partial\tilde{T}}}=\frac{1}{\rho_c} \frac{\Delta\tilde{S}}{{\partial\tilde{V}} },
\end{equation}
here $\rho_c=\frac{c_3}{c_4}$. This equation is also discussed in Ref. \cite{Wei:2014qwa}. The vanishing Gibbs free energy condition is only valid at the co-existence curve, the Clapeyron equation can only be defined at this curve.
\subsection{Latent Heat}
The latent heat is an interesting thermodynamic quantity of BH to determine its stability or instability. In Eq. (\ref{LHeat}), we have defined the latent heat so, in reduced parameters the latent heat can be written as 
\begin{equation}
\tilde{L}=\frac{L}{T_c S_c}=\tilde{T}\Delta \tilde{S},
\end{equation}
This is the amount of heat that must be removed/added to the existing state of a BH to manifest the phase transition phenomenon. At the co-existence state, we can not distinguish particles so we can take the variation of free energy and enthalpy will be zero and, from Eqs. (\ref{1stlawa}) and \ref{1stlawb} we can write as 
\begin{equation}
\tilde{T}\Delta\tilde{S}=-\tilde{S}\Delta\tilde{T}
\end{equation}
here $c_1c_4=c_2c_3$. The Clapeyron equation is also holding at this stage. Using these results, we can validate Maxwell's equal area law and heat Capacity as in Ref. \cite{Wei:2014qwa}.

\section{Remarks and Discussion\label{sec4}}

The pioneering work of the Big Bang theory has been proved by the supernova cosmology project and high Z search team that led researchers to predict the forces of gravity and the expansion of the universe as the main dynamical factors. These forces can easily be considered by using the concept of modified gravity. Such as $f(R)$ gravity approaches can be introduced by adding higher powers of scalar curvature $R$, the Riemann and Ricci tensors, or their derivatives in the Lagrangian formulation. The Love-lock gravity, bran world cosmology, scalar-tensor theory, or $f(R)$ theory can be adopted. Among these, the modified gravity theory, $f(R)$ theory is proved to mimic the cosmological history from inflation to the actual accelerating expansion era \cite{Nojiri:2003ft, Atazadeh:2008zj}. So, it is interesting and meaningful to extend the BH in $f(R)$ gravity from cosmology to holographic dual along with their corrections \cite{Balasubramanian:2001nb}. These modifications could provide us with some different features of the BH compared to Einstein's gravity. 

The VdW equation of state is obtained from the horizon temperature (\ref{temp}) that satisfies the conditions (\ref{CC}). So, we can conclude the existence of $P-v$ criticality that provides an opportunity to make a relationship between the $f(R)$ gravity and holographic theory through the VdW behavior. The $P-v$ diagram of BH in $f(R)$ gravity coincides with that of charged AdS BH, and hence, we can claim that BH in $f(R)$ gravity is a holographic dual of VdW fluid. Using the critical point parameters the universal value of the inflection point is found dependent on $b$. Using $b=1$, this degenerates the inflection point of VdW fluid similar to charged AdS BH and thus proves BH in $f(R)$ gravity as the holographic dual of VdW fluid.

A d-dimensional BH in $f(R)$ gravity is considered to investigate the effect of factor $(b)$ on phase transition, co-existence curves, differences in molecule number densities, and configuration entropy. We obtained analytical expressions for the temperature, pressure, specific volume, and Gibbs free energy. Using critical point parameters, an expression for the inflection point is obtained depending on factor $b$. Critical point parameters showed the increasing values of critical temperature and pressure, while the critical volume decreased with the increase of factor $b$ i.e. $f'(R)$. The condition $b>1$ is considered for the phase transition of $f(R)$ BH and the inflection point was found consistent with the universal number as $3/8$ for $b=1$ i.e. when $f'(R)=0$ is the same as that of charged AdS BH. It shows a monotonic increase with the positive change in $f'(R)$ value. We also investigated consistent results by using the reduced phase space parameters. The critical point parameters and their inflection point are found without compromising their global generality. Several important results related to the thermodynamics of BH in $f(R)$ gravity are obtained. These results are also comparable to these $4-$dimensional charged AdS BHs. 

This study provides a beautiful insight for understanding the basic physics related to configurational entropy (information entropy), phase transition, co-existence curves, laws of BH physics, and the underlying microscopic freedom of states of BH in $f(R)$ gravity. It also provides us with important features of BHs that differ from those obtained in Einstein’s theory. It is a nice step for researchers to probe the universe with GR theory even the Einstein field equation in $f(R)$ gravity where the GR theory becomes more complicated with the presence of matter \cite{Moon:2011hq, Braga:2016wzx}. The investigation of configuration entropy is a useful tool to investigate the stability and (or) the relative dominance of states for diverse physical systems. This work also provides a comparison with charged AdS BHs and one can analyze the differences between them by investigating the effect of factor $b$. Considering the small and large BH phase transition one can easily discuss the analogy of a BH with VdW’s liquid gas phase transition system \cite{Ovgun:2017bgx, Kubiznak:2012wp, Khan:2020zpz}. Especially at the critical point, where the small and large molecules are indistinguishable hence, the number of micro-states $W$ will be at its peak. It is important to explain that we developed the molecule number densities (of small and large BHs) associated with the contribution from the horizon to the mass of the BH. As an artifact to find a well-defined configurational entropy, this molecule number density is defined in the inner region of a BH horizon \cite{Wei:2015iwa}. This study can also be extended to discuss the thermodynamic fluctuations and quantum corrections of BH physics which is a subject of interest to many researchers \cite{Upadhyay:2018bqy, Upadhyay:2018gee, Khan:2020zpz}

\section{Results\label{sec5}}

In this chapter, we considered BH in $f(R)$ gravity to discuss thermodynamics phase transition and configuration entropy in extended phase space. We analyzed the effect of modified gravity factor $b$ on thermodynamic parameters like critical point parameters, molecule number densities, and configuration entropy. The thermodynamics pressure, temperature, and free energy for $d-$dimensional and $d=4-$dimensional BH in $f(R)$ gravity are calculated as a function of factor $b$. Using the critical point conditions at the BH horizon, we calculated critical point parameters $(v_c, P_c, T_c, G_c)$ similar to VdW's system at the vanishing of $f'(R)$ value. As the value of $f'(R)$ increases, the inflection point also increases monotonically. A $P-v$ diagram is plotted for $d=4-$dimensional and at $f'(R)=0.1$ as shown in Fig. (\ref{image-5.1}) that coincides with that of charged AdS BH and proves the BH in $f(R)$ gravity as a holographic dual of VdW fluid. Using transformation of generalized Eqs. (\ref{A1}) to (\ref{A3}) from and their $4-$dimension form in Eq. (\ref{A12}), (\ref{A22}) and (\ref{A32}) in reduced phase-space. Solving these equations, reduced pressure is obtained as a function of reduced temperature. As we see in Eq. (\ref{Red.T}) the reduced temperature is independent of other factors like $q$, $R_o$, and $b$. The main reason is that the reduced form is a ratio of thermodynamics quantity to its critical value at the phase transition point. So in this coincidence, the reduced pressure also satisfies these properties. Similarly, the difference in molecule number density and configuration entropy is also derived as a function of reduced temperature independent of factor $b$. 

The plot of $\tilde{p}-\tilde{\tau}$ is shown in Fig. (\ref{image-5.2}). This plot shows that as long as the reduced temperature increases, the reduced pressure also increases, as a result, the volume would decrease. This curve also shows that the $1^{st}$ order phase transition between SBH/LBH starts at some specific temperature $T<T_c$, whereas the $2^{nd}$ order phase transition occurs when the SBH/LBH crosses the co-existence curve. The plot of the difference in number density vs. reduced temperature is depicted in Fig. (\ref{image-5.3}), which shows the difference in molecule number densities of SBH/LBH decreases with the increase in reduced temperature i.e. its value is maximum for $\tilde{\tau}=0$. The individual molecule number densities of SBH and LBH as a function of $\tilde{\tau}$ are evaluated in Eq. (\ref{n1n2}). Using these results in Eq. (\ref{GBent4}), we calculated the $S_{con}$ given in Eq. (\ref{conf-Ent}) as a monotonic function of reduced temperature and its plot is also shown in Fig. (\ref{image-5.4}). We also extended these analyses to $d=5$. it is found that as long as the space-time dimensions increase, the $(S_{con}-\tilde{\tau})$ plot becomes a more and more concave function of reduced temperature as shown in Fig. (\ref{image-5.5}). The maximum value of $\frac{n_1-n_2}{n_c}$ is found for $\tau=0$, which decreases with an increase in space-time dimensions $d$ as shown in Fig. (\ref{image-5.6}).

Moreover, we also discussed the validation of our results by using the laws of Physics to get their interpretation. The results are comparable with charged AdS space-time but independent of parameter $b$. The first law of BH thermodynamics is defined with four critical parameters $c_1$, $c_2$, $c_3$, and $c_4$. Using these critical values, enthalpy, entropy, volume, pressure, and free energy, $c_1$, $c_2$, $c_3$, and $c_4$ are calculated. Next, using the co-existence curves, we derived the Clapeyron equation and the Latent heat in their reduced parameters. These results show a beautiful sketch of BH Physics at the co-existence and phase transition stages.

\noindent\rule{16.5cm}{2.0pt}

\clearpage
\thispagestyle{empty}
\hfill
\clearpage
\newpage

\newpage

\section{Appendices}
{\subsection{Appendix A.}\label{A}
Consider a hyper-surface at a late advanced time $v$. Let us extend it from the event horizon with a constant radius $r$ in the interior of a BH. When it reaches the point where $v$ deviates give the maximal advance time and the hyper-surface at this point will bind the largest interior volume. The decomposed form of an arbitrary vector lying on a hyper-surface can be written
\begin{equation}
k=Zn^a+Z^a \qquad \Rightarrow \qquad n_a=-Z\nabla_a{r}
\end{equation}
where $Z$ and $Z^a$ are the Lapse and shift functions respectively, and $n^a$ is the co-vector with $\nabla_a{r}$ being the normal co-vector. For a space-like hyper-surface, the normal vector can  be written as 
\begin{equation}
n_a n^a=-1 \qquad \Rightarrow \qquad Z^2g^{ab}dr_a dr_b=-1,
 \end{equation}
Using the above equation, we can write as $Z^2=-g^{-rr}$. Now as the determinant of an induced metric on hyper-surface at constant radius $r$ is 
\begin{equation}
|h|=Z^{-2}(g)=g^{rr}g \qquad\Rightarrow \qquad |h|^2=-\Delta\rho^2 sin^2\theta,
\end{equation}
Now the volume for a rotating BH, the interior volume can be defined as 
\begin{equation}
V_{\sum}=\int^v\sqrt{|h|}dv d\theta d\phi=\int^v \sqrt{-\Delta}\rho sin\theta dv d\theta d\phi,
\end{equation}
This directly confirms the BH interior volume as in Ref. \\cite{Bengtsson:2015zda}.

\subsection{Appendix B.}\label{B}
 The Klein-Gordon equation in curved space-time is
\begin{equation}
    \frac{1}{\sqrt{-g}}{{\partial_ \mu }\left(\sqrt{-g} g^{\mu v } \partial_v \Phi \right)}=0,
\end{equation}
Expanding this equation in $(T,\lambda, \theta, \phi)$ and using
$$\partial\Phi_v=\partial_ T  \Phi+\partial_{\lambda} \Phi+\partial_ {\theta} \Phi+\partial_ \phi \Phi,$$
First solving $(\partial_ T  \Phi, \partial_{\lambda} \Phi, \partial_ {\theta} \Phi, \partial_ \phi \Phi)$, we get
\begin{equation}
    \begin{aligned}
        \partial_ T  \Phi=\partial_ T (e^{\text{-iET}}e^{\text{iI}(\lambda ,\theta ,\phi )})=-i E\Phi,\\ \partial_{\lambda} \Phi=\partial_{\lambda} (e^{\text{iET}}e^{\text{iI}(\lambda ,\theta ,\phi )})=i I\partial_{\lambda} \Phi,\\ \partial_ {\theta} \Phi=\partial_ {\theta} (e^{\text{iET}}e^{\text{iI}(\lambda ,\theta ,\phi )})=i I\partial_ {\theta} \Phi, \\ \partial_ \phi \Phi)=\partial_ \phi (e^{\text{iET}}e^{\text{iI}(\lambda ,\theta ,\phi )})=i I \partial_ \phi \Phi,
    \end{aligned}
\end{equation}

So, from above we can write the scalar field as 
$$\partial_v \Phi=i (E^2+\partial_{\lambda} I+\partial_ {\theta} I+\partial_ \phi I)\Phi,$$
The first part of the above general equation can be written as 
\begin{equation*}
    \begin{aligned}
        \frac{1}{\sqrt{-g}}{{\partial_ T }\left(\sqrt{-g} g^{T v } \partial_v \Phi \right)} \\
=\frac{1}{\sqrt{-g}}{{\partial_ T }\left ( \sqrt{-g}\left ( g^{TT} +g^{T\lambda}+g^{T\theta}+g^{T\phi}\right )\partial_v \Phi  \right )},
    \end{aligned}
\end{equation*}
\begin{small}
    \begin{equation*}
    \begin{aligned}
        \frac{1}{\sqrt{-g}}{{\partial_ T }\left(\sqrt{-g} g^{T v } \partial_v \Phi \right)}\\ =\frac{1}{\sqrt{-g}}{{\partial_ T }\left ( \sqrt{-g}\left ( g^{TT} +g^{T\lambda}+g^{T\theta}+g^{T\phi}\right )\times i (E^2+\partial_{\lambda} I+\partial_ {\theta} I+\partial_ \phi I)\Phi \right)},
    \end{aligned}
\end{equation*}
\begin{equation*}
\begin{aligned}
    \frac{1}{\sqrt{-g}}{{\partial_ T }\left(\sqrt{-g} g^{T v } \partial_v \Phi \right)} \\ =\frac{1}{\sqrt{-g}}{{\partial_ T }\left ( \sqrt{-g}\left ( g^{TT} +g^{T\lambda}+g^{T\theta}+g^{T\phi}\right ) \times i (E^2+\partial_{\lambda} I+\partial_ {\theta} I+\partial_ \phi I)\Phi \right)},
\end{aligned}
\end{equation*}
\end{small}
where for the metric Eq. (\ref{EDmet+T}), the coordinates are $g^{TT}=-1, g^{T\lambda}=g^{T\theta}=g^{T\phi}=0$

\begin{equation}
    \begin{aligned}
        \frac{1}{\sqrt{-g}}{{\partial_ T }\left(\sqrt{-g} g^{T v } \partial_v \Phi \right)} \\ =\frac{1}{\sqrt{-g}}{{\partial_ T }\left ( \sqrt{-g}\left ( -1 +0+0+0\right )\times i (E^2+\partial_{\lambda} I+\partial_ {\theta} I+\partial_ \phi I)\Phi \right)},
    \end{aligned}
\end{equation}
or we can write as 
$$\frac{1}{\sqrt{-g}}{{\partial_ T }\left(\sqrt{-g} g^{T v } \partial_v \Phi \right)}=- E^2\Phi,$$
Similarly, the other parts of the general equation are

\begin{small}
    \begin{equation}
    \begin{aligned}
        \frac{1}{\sqrt{-g}}{{\partial_ \lambda }\left(\sqrt{-g} g^{\lambda  v } \partial_v \Phi \right)} \\ =\frac{1}{\sqrt{-g}}{{\partial_ \lambda }\left ( \sqrt{-g}\left ( g^{\lambda T} +g^{\lambda\lambda}+g^{\lambda\theta}+g^{\lambda\phi}\right )\times i (E^2+\partial_{\lambda} I+\partial_ {\theta} I+\partial_ \phi I)\Phi \right)},
    \end{aligned}
\end{equation}
\end{small}
Here $g^{\lambda T}=0, g^{\lambda\lambda}=\frac{1}{\left(-f(r) \dot{v }^2+2 \dot{v } \dot{r}\right)} ,g^{\lambda\theta}=0, g^{\lambda\phi}=0$ so,
$$\frac{1}{\sqrt{-g}}{{\partial_ \lambda }\left(\sqrt{-g} g^{\lambda  v } \partial_v \Phi \right)}=\frac{1}{\left(-f(r) \dot{v }^2+2 \dot{v } \dot{r}\right)} \partial_ \lambda ^2 I \Phi,$$

\begin{small}
\begin{equation}
    \begin{aligned}
        \frac{1}{\sqrt{-g}}{{\partial_ \theta }\left(\sqrt{-g} g^{\theta  v } \partial_v \Phi \right)} \\ =\frac{1}{\sqrt{-g}}{{\partial_ \theta }\left ( \sqrt{-g}\left ( g^{\theta T} +g^{\theta\lambda}+g^{\theta\theta}+g^{\theta\phi}\right )\times i (E^2+\partial_{\lambda} I+\partial_ {\theta} I+\partial_ \phi I)\Phi \right)}
    \end{aligned}
\end{equation}
\end{small}

Here $$g^{\theta T}=0, g^{\theta\lambda}=0 ,g^{\theta\theta}=\frac{1}{r^2}, g^{\theta\phi}=0$$ so,
$$=\frac{1}{\sqrt{-g}}{{\partial_ \theta }\left ( \sqrt{-g}\left ( 0 +0+\frac{1}{r^2}+0\right )\times i (E^2+\partial_{\lambda} I+\partial_ {\theta} I+\partial_ \phi I)\Phi \right)}$$ 
$$\frac{1}{\sqrt{-g}}{{\partial_ \theta }\left(\sqrt{-g} g^{\theta  v } \partial_v \Phi \right)}=\frac{1}{r^2}\partial^2 _ {\theta}I\Phi$$

and 

\begin{small}
\begin{equation}
    \begin{aligned}
       \frac{1}{\sqrt{-g}}{{\partial_ \phi }\left(\sqrt{-g} g^{\phi  v } \partial_v \Phi \right)} \\ =\frac{1}{\sqrt{-g}}{{\partial_ \phi }\left ( \sqrt{-g}\left ( g^{\phi T} +g^{\phi\lambda}+g^{\phi\theta}+g^{\phi\phi}\right )\times i (E^2+\partial_{\lambda} I+\partial_ {\theta} I+\partial_ \phi I)\Phi \right)},
    \end{aligned}
\end{equation}
\end{small}

here $g^{\phi T}=0, g^{\phi\lambda}=0 ,g^{\phi\theta}=0, g^{\phi\phi}=\frac{1}{r^2sin^2 \theta}$ so,

\begin{small}
    \begin{equation}
    \begin{aligned}
        =\frac{1}{\sqrt{-g}}{{\partial_ \phi }\left ( \sqrt{-g}\left ( 0 +0+0+\frac{1}{r^2 sin^2 \theta}\right )\times i (E^2+\partial_{\lambda} I+\partial_ {\theta} I+\partial_ \phi I)\Phi \right)},
    \end{aligned}
\end{equation}

\end{small}

$$\Rightarrow \frac{1}{\sqrt{-g}}{{\partial_ \phi }\left(\sqrt{-g} g^{\phi v } \partial_v \Phi \right)}=\frac{1}{r^2 sin^2 \theta}\partial^2 _ {\phi}I\Phi,$$
using the above results, the general solution of the Klein-Gordon equation in curved space-time for $(T,\lambda, \theta, \phi)$ can be written as 
$$- E^2\Phi+\frac{1}{\left(-f(r) \dot{v }^2+2 \dot{v } \dot{r}\right)} \partial_ \lambda ^2 I \Phi+\frac{1}{r^2}\partial^2 _ {\theta}I\Phi+\frac{1}{r^2 sin^2 \theta}\partial^2 _ {\phi}I\Phi=0,$$
using $\partial_ \lambda ^2 I=p_\lambda ^2, \partial^2 _ {\theta}I=p_\theta ^2, \partial^2 _ {\phi}I=p_\phi ^2$, we gets 

\begin{equation}
    -E^2+\frac{1}{\left(-f(r) \dot{v }^2+2 \dot{v } \dot{r}\right)} p_\lambda ^2 +\frac{1}{r^2}p_\theta ^2+\frac{1}{r^2 sin^2 \theta}p_\phi ^2=0,
\end{equation}

Which is the Eq. (\ref{eqofenergy}). One can easily deduce these investigations for the lower dimensional BH and obtain the EOM.

\vspace{1.0cm}
\noindent\rule{16.5cm}{2.0pt}
\clearpage
\thispagestyle{empty}
\hfill

\noindent\rule{16.5cm}{2.0pt}


\clearpage
\newpage

\begin{thebibliography}{300}
\addcontentsline{toc}{section}{References}


\bibitem{Klebesadel:1973iq}
R.~W.~Klebesadel, I.~B.~Strong and R.~A.~Olson,
Astrophys. J. Lett. \textbf{182}, L85-L88 (1973).
doi:10.1086/181225.

\bibitem{Fishman:1995st}
G.~J.~Fishman and C.~A.~Meegan,
Ann. Rev. Astron. Astrophys. \textbf{33}, 415-458 (1995).
doi:10.1146/annurev.aa.33.090195.002215.

\bibitem{vanParadijs:1997wr}
J.~van Paradijs, P.~J.~Groot, T.~Galama, C.~Kouveliotou, R.~G.~Strom, J.~Telting, R.~G.~M.~Rutten, G.~J.~Fishman, C.~A.~Meegan and M.~Pettini, \textit{et al.}
Nature \textbf{386}, 686-689 (1997).
doi:10.1038/386686a0.

\bibitem{Costa:1997obd}
E.~Costa, F.~Frontera, J.~Heise, M.~Feroci, J.~J.~M.~in 't Zand, F.~Fiore, M.~N.~Cinti, D.~D.~Fiume, L.~Nicastro and M.~Orlandini, \textit{et al.}
Nature \textbf{387}, 783-785 (1997).
doi:10.1038/42885.
[arXiv:astro-ph/9706065 [astro-ph]].

\bibitem{Frail:1997qf}
D.~A.~Frail, S.~R.~Kulkarni, S.~R.~Nicastro, M.~Feroci and G.~B.~Taylor,
Nature \textbf{389}, 261 (1997).
doi:10.1038/38451.

\bibitem{Heng:2008nr}
K.~Heng, D.~Lazzati, R.~Perna, P.~Garnavich, A.~Noriega-Crespo, D.~Bersier, T.~Matheson and M.~Pahre,
Astrophys. J. \textbf{681}, 1116 (2008).
doi:10.1086/588279.
[arXiv:0803.2879 [astro-ph]].

\bibitem{Zhang:2018ond}
B.~Zhang,
Cambridge University Press, 2018,
ISBN 978-1-107-02761-9.

\bibitem{Kumar:2014upa}
P.~Kumar and B.~Zhang,
Phys. Rept. \textbf{561}, 1-109 (2014).
doi:10.1016/j.physrep.2014.09.008.
[arXiv:1410.0679 [astro-ph.HE]].

\bibitem{Wijers:1997xu}
R.~A.~M.~J.~Wijers, M.~J.~Rees and P.~Meszaros,
Mon. Not. Roy. Astron. Soc. \textbf{288}, L51-L56 (1997).
doi:10.1093/mnras/288.4.L51.
[arXiv:astro-ph/9704153 [astro-ph]].

\bibitem{Woosley:1993wj}
S.~E.~Woosley,
Astrophys. J. \textbf{405}, 273 (1993).
doi:10.1086/172359.

\bibitem{Woosley:2006fn}
S.~E.~Woosley and J.~S.~Bloom,
Ann. Rev. Astron. Astrophys. \textbf{44}, 507-556 (2006).
doi:10.1146/annurev.astro.43.072103.150558.
[arXiv:astro-ph/0609142 [astro-ph]].

\bibitem{Perna:2018wft}
R.~Perna, D.~Lazzati and M.~Cantiello,
Astrophys. J. \textbf{859}, no.1, 48 (2018).
doi:10.3847/1538-4357/aabcc1.
[arXiv:1803.04983 [astro-ph.HE]].

\bibitem{Dagoneau:2020mbg}
N.~Dagoneau, S.~Schanne, J.~L.~Atteia, D.~G\"otz and B.~Cordier,
Exper. Astron. \textbf{50}, no.1, 91-123 (2020).
doi:10.1007/s10686-020-09665-w.
[arXiv:2005.12560 [astro-ph.HE]].

\bibitem{MacFadyen:1998vz}
A.~MacFadyen and S.~E.~Woosley,
Astrophys. J. \textbf{524}, 262 (1999).
doi:10.1086/307790.
[arXiv:astro-ph/9810274 [astro-ph]].

\bibitem{Popham:1998ab}
R.~Popham, S.~E.~Woosley and C.~Fryer,
Astrophys. J. \textbf{518}, 356-374 (1999).
doi:10.1086/307259.
[arXiv:astro-ph/9807028 [astro-ph]].

\bibitem{Beloborodov:2008nx}
A.~M.~Beloborodov,
AIP Conf. Proc. \textbf{1054}, 51 (2008)
doi:10.1063/1.3002509
[arXiv:0810.2690 [astro-ph]].

\bibitem{Hanami:1997yy}
Hitoshi ~ Hanami
ApJ, 687, \textbf{491} 2(1997).
doi; 10.1086/304964.

\bibitem{Liu:2016olx}
T.~Liu, G.~E.~Romero, M.~L.~Liu and A.~Li,
Astrophys. J. \textbf{826}, no.1, 82 (2016).
doi:10.3847/0004-637X/826/1/82.
[arXiv:1602.06907 [astro-ph.HE]].

\bibitem{Janiuk:2010xs}
A.~Janiuk, Y.~F.~Yuan, R.~Perna and T.~Di Matteo,
IAU Symp. \textbf{275}, 349 (2011).
doi:10.1017/S1743921310016388.
[arXiv:1010.0903 [astro-ph.HE]].

\bibitem{MacFadyen:2004ab}
 Andrew ~ MacFadyen
Science, \textbf{303}, Issue 5654, pp. 45-46, (2004).
doi: 10.1126/science.1091764.

\bibitem{Liu:2017rwh}
T.~Liu, C.~Y.~Song, B.~Zhang, W.~M.~Gu and A.~Heger,
Astrophys. J. \textbf{852}, no.1, 20 (2018)
doi:10.3847/1538-4357/aa9e4f
[arXiv:1710.00141 [astro-ph.HE]].

\bibitem{Evans:2008wp}
P.~A.~Evans, A.~P.~Beardmore, K.~L.~Page, J.~P.~Osborne, P.~T.~O'Brien, R.~Willingale, R.~L.~C.~Starling, D.~N.~Burrows, O.~Godet and L.~Vetere, \textit{et al.}
Mon. Not. Roy. Astron. Soc. \textbf{397}, 1177 (2009).
doi:10.1111/j.1365-2966.2009.14913.x.
[arXiv:0812.3662 [astro-ph]].

\bibitem{Duncan:1992hi}
R.~C.~Duncan and C.~Thompson,
Astrophys. J. Lett. \textbf{392}, L9 (1992)
doi:10.1086/186413



\bibitem{Eichler:1989ve}
D.~Eichler, M.~Livio, T.~Piran and D.~N.~Schramm,
Nature \textbf{340}, 126-128 (1989).
doi:10.1038/340126a0.

\bibitem{Narayan:1992iy}
R.~Narayan, B.~Paczynski and T.~Piran,
Astrophys. J. Lett. \textbf{395}, L83-L86 (1992).
doi:10.1086/186493.
[arXiv:astro-ph/9204001 [astro-ph]].

\bibitem{DAvanzo:2015kdp}
P.~D'Avanzo,
JHEAp \textbf{7}, 73-80 (2015)
doi:10.1016/j.jheap.2015.07.002

\bibitem{Kouveliotou:1993yx}
C.~Kouveliotou, C.~A.~Meegan, G.~J.~Fishman, N.~P.~Bhyat, M.~S.~Briggs, T.~M.~Koshut, W.~S.~Paciesas and G.~N.~Pendleton,
Astrophys. J. Lett. \textbf{413}, L101-104 (1993).
doi:10.1086/186969.

\bibitem{Wei:2022ygk}
Y.~F.~Wei and T.~Liu,
Universe \textbf{8}, no.10, 529 (2022)
doi:10.3390/universe8100529

\bibitem{Liu:2017kga}
T.~Liu, W.~M.~Gu and B.~Zhang,
New Astron. Rev. \textbf{79}, 1-25 (2017)
doi:10.1016/j.newar.2017.07.001
[arXiv:1705.05516 [astro-ph.HE]].

\bibitem{Yuan:2014gma}
F.~Yuan and R.~Narayan,
Ann. Rev. Astron. Astrophys. \textbf{52}, 529-588 (2014)
doi:10.1146/annurev-astro-082812-141003
[arXiv:1401.0586 [astro-ph.HE]].

\bibitem{Liu:2022cph}
B.~F.~Liu and E.~Qiao,
iScience, \textbf{25}, Issue 1, 103544, pp. 1 -16, (2022).
doi;  https://doi.org/10.1016/j.isci.2021.103544.
[arXiv:2201.06198 [astro-ph.HE]].

\bibitem{Narayan:2008bv}
R.~Narayan and J.~E.~McClintock,
New Astron. Rev. \textbf{51}, 733-751 (2008)
doi:10.1016/j.newar.2008.03.002
[arXiv:0803.0322 [astro-ph]].

\bibitem{Blandford:1977ds}
R.~D.~Blandford and R.~L.~Znajek,
Mon. Not. Roy. Astron. Soc. \textbf{179}, 433-456 (1977).
doi:10.1093/mnras/179.3.433.

\bibitem{Bell:1996ad}
D.~ Lynden ~ Bell,
Mon. Not. R. Astron. Soc., \textbf{279}, Issue 2, pp.  389-401, (1996).
doi;  https://doi.org/10.1093/mnras/279.2.389.

\bibitem{LIGOScientific:2017vwq}
B.~P.~Abbott \textit{et al.} [LIGO Scientific and Virgo],
Phys. Rev. Lett. \textbf{119}, no.16, 161101 (2017)
doi:10.1103/PhysRevLett.119.161101
[arXiv:1710.05832 [gr-qc]].

\bibitem{Savchenko:2017ffs}
V.~Savchenko, C.~Ferrigno, E.~Kuulkers, A.~Bazzano, E.~Bozzo, S.~Brandt, J.~Chenevez, T.~J.~L.~Courvoisier, R.~Diehl and A.~Domingo, \textit{et al.}
Astrophys. J. Lett. \textbf{848}, no.2, L15 (2017)
doi:10.3847/2041-8213/aa8f94
[arXiv:1710.05449 [astro-ph.HE]].

\bibitem{Tan:2017hqa}
W.~W.~Tan, X.~L.~Fan and F.~Y.~Wang,
Mon. Not. Roy. Astron. Soc. \textbf{475}, no.1, 1331-1339 (2018)
doi:10.1093/mnras/stx3242
[arXiv:1712.04641 [astro-ph.HE]].

\bibitem{LIGOScientific:2016aoc}
B.~P.~Abbott \textit{et al.} [LIGO Scientific and Virgo],
Phys. Rev. Lett. \textbf{116}, no.6, 061102 (2016)
doi:10.1103/PhysRevLett.116.061102
[arXiv:1602.03837 [gr-qc]].

\bibitem{Janiuk:2016qpe}
A.~Janiuk, M.~Bejger, S.~Charzy\'nski and P.~Sukova,
New Astron. \textbf{51}, 7-14 (2017)
doi:10.1016/j.newast.2016.08.002
[arXiv:1604.07132 [astro-ph.HE]].

\bibitem{Connaughton:2016umz}
V.~Connaughton, E.~Burns, A.~Goldstein, M.~S.~Briggs, B.~B.~Zhang, C.~M.~Hui, P.~Jenke, J.~Racusin, C.~A.~Wilson-Hodge and P.~N.~Bhat, \textit{et al.}
Astrophys. J. Lett. \textbf{826}, no.1, L6 (2016)
doi:10.3847/2041-8205/826/1/L6
[arXiv:1602.03920 [astro-ph.HE]].

\bibitem{Khan:2018ejm}
A.~Khan, V.~Paschalidis, M.~Ruiz and S.~L.~Shapiro,
Phys. Rev. D \textbf{97}, no.4, 044036 (2018)
doi:10.1103/PhysRevD.97.044036
[arXiv:1801.02624 [astro-ph.HE]].

\bibitem{Janiuk:2017tbq}
A.~Janiuk, M.~Bejger, P.~Sukova and S.~Charzynski,
Galaxies \textbf{5}, no.1, 15 (2017)
doi:10.3390/galaxies5010015
[arXiv:1701.07753 [astro-ph.HE]].

\bibitem{Balbus:1998az}
S.~A.~Balbus and J.~F.~Hawley,
RevModPhys. \textbf{70}, 1 - 53 (1998).
doi:10.1103/RevModPhys.70.1.

\bibitem{Velikhov:1959xx}
E. ~P. ~Velikhov
J. Exptl. Theoret. Phys., \textbf{36}, no. 9, 1398–1404, (1959).

\bibitem{Balbus:1991ay}
S.~A.~Balbus and J.~F.~Hawley,
Astrophys. J. \textbf{376}, 214-233 (1991).
doi:10.1086/170270.

\bibitem{Proga:2006ir}
D.~Proga and B.~Zhang,
Mon. Not. Roy. Astron. Soc. \textbf{370}, L61-L65 (2006)
doi:10.1111/j.1745-3933.2006.00189.x
[arXiv:astro-ph/0601272 [astro-ph]].

\bibitem{Liu:2012ek}
T.~Liu, E.~W.~Liang, W.~M.~Gu, S.~J.~Hou, W.~H.~Lei, L.~Lin, Z.~G.~Dai and S.~N.~Zhang,
Astrophys. J. \textbf{760}, 63 (2012)
doi:10.1088/0004-637X/760/1/63
[arXiv:1209.4522 [astro-ph.HE]].

\bibitem{Nattila:2021qag}
J.~N\"attil\"a and A.~M.~Beloborodov,
Phys. Rev. Lett. \textbf{128}, no.7, 075101 (2022).
doi:10.1103/PhysRevLett.128.075101.
[arXiv:2111.15578 [astro-ph.HE]].

\bibitem{MacDonald1982}
D.~MacDonald and K.~S.~Thorne,
Mon. Not. Roy. Astron. Soc. \textbf{198}, 345-383 (1982).
doi: https://doi.org/10.1093/mnras/198.2.345.

\bibitem{Pugliese:2011xn}
D.~Pugliese, H.~Quevedo and R.~Ruffini,
Phys. Rev. D \textbf{84}, 044030 (2011)
doi:10.1103/PhysRevD.84.044030
[arXiv:1105.2959 [gr-qc]].

\bibitem{Shapiro:1983du}
S.~L.~Shapiro and S.~A.~Teukolsky,
pp. 645, (1983).
ISBN: 9780471873167.

\bibitem{Aliev:2013jqz}
A.~N.~Aliev, G.~D.~Esmer and P.~Talazan,
Class. Quant. Grav. \textbf{30}, 045010 (2013).
doi:10.1088/0264-9381/30/4/045010.
[arXiv:1205.2838 [gr-qc]].

\bibitem{Bardeen:1973gs} 
  J.~M.~Bardeen, B.~Carter and S.~W.~Hawking,
  Commun.\ Math.\ Phys.\  {\bf 31}, 161 (1973).
  doi:10.1007/BF01645742

\bibitem{Bardeen:1972fi}
J.~M.~Bardeen, W.~H.~Press and S.~A.~Teukolsky,
Astrophys. J. \textbf{178}, 347 (1972).
doi:10.1086/151796.

\bibitem{Novikov:1989}
Igor D.~Novikov, Valery P.~Frolov
Springer Netherlands, (1989).
doi; https://doi.org/10.1007/978-94-011-5139-9.

\bibitem{Kato:2008xy}
S. ~Kato, J. ~Fukue, S.~Mineshige
``Black-Hole Accretion Disks: Towards a New Paradigm,''
Kyoto University Press, p. 549 (2008).

\bibitem{Carroll:2004st} 
  S.~M.~Carroll,
  ``Spacetime and geometry: An introduction to general relativity,''
  San Francisco, USA: Addison-Wesley, 513 p, (2004).

\bibitem{Okazaki:1987st} 
 A. Okazaki, S. Kato, J. Fukue,
Astronomical Society of Japan, Publications (ISSN 0004-6264), \textbf{39}, no. 3, p. 457-473, (1987).

\bibitem{Chandrasekhar:1985kt}
S.~Chandrasekhar,
``The mathematical theory of black holes,''
ISBN: 9780198503705, https://inspirehep.net/literature/224457

\bibitem{Wald:1984rga} 
  R.~M.~Wald,
  ``General Relativity,''
  doi:10.7208/chicago/9780226870373.001.0001, (1984).

\bibitem{Misner:1973prb}
C.~W.~Misner, K.~S.~Thorne and J.~A.~Wheeler,
W. H. Freeman, (1973).
ISBN 978-0-7167-0344-0, 978-0-691-17779-3.

\bibitem{LIGOScientific:2016vbw}
B.~P.~Abbott \textit{et al.} [LIGO Scientific and Virgo],
Phys. Rev. D \textbf{93}, no.12, 122003 (2016).
doi:10.1103/PhysRevD.93.122003.
[arXiv:1602.03839 [gr-qc]].

\bibitem{Lee:1999se}
H.~K.~Lee, R.~A.~M.~J.~Wijers and G.~E.~Brown,
Phys. Rept. \textbf{325}, 83-114 (2000).
doi:10.1016/S0370-1573(99)00084-8.
[arXiv:astro-ph/9906213 [astro-ph]].

\bibitem{Meszaros:2006rc}
P.~Meszaros,
Rept. Prog. Phys. \textbf{69}, 2259-2322 (2006)
doi:10.1088/0034-4885/69/8/R01
[arXiv:astro-ph/0605208 [astro-ph]].

\bibitem{Piran:2004ba}
T.~Piran,
Rev. Mod. Phys. \textbf{76}, 1143-1210 (2004)
doi:10.1103/RevModPhys.76.1143
[arXiv:astro-ph/0405503 [astro-ph]].

\bibitem{Woosley1993}
S.~E.~Woosley,
Astrophys. J. \textbf{405}, 273 (1993).
doi:10.1086/172359.

\bibitem{Kouveliotou1993}
C.~Kouveliotou, C.~A.~Meegan, G.~J.~Fishman, N.~P.~Bhyat, M.~S.~Briggs, T.~M.~Koshut, W.~S.~Paciesas and G.~N.~Pendleton,
Astrophys. J. Lett. \textbf{413}, L101-104 (1993).
doi:10.1086/186969.

\bibitem{Fruchter:2006py}
A.~S.~Fruchter, A.~J.~Levan, L.~Strolger, P.~M.~Vreeswijk, S.~E.~Thorsett, D.~Bersier, I.~Burud, J.~M.~C.~Cero, n.~A.~Castro-Tirado and C.~Conselice, \textit{et al.}
Nature \textbf{441}, 463-468 (2006)
doi:10.1038/nature04787
[arXiv:astro-ph/0603537 [astro-ph]].

\bibitem{Janiuk2010}
A.~Janiuk, Y.~F.~Yuan, R.~Perna and T.~Di Matteo,
IAU Symp. \textbf{275}, 349 (2011).
doi:10.1017/S1743921310016388.
[arXiv:1010.0903 [astro-ph.HE]].

\bibitem{Liu2017}
T.~Liu, C.~Y.~Song, B.~Zhang, W.~M.~Gu and A.~Heger,
Astrophys. J. \textbf{852}, no.1, 20 (2018).
doi:10.3847/1538-4357/aa9e4f.
[arXiv:1710.00141 [astro-ph.HE]].

\bibitem{Evans2008}
P.~A.~Evans, A.~P.~Beardmore, K.~L.~Page, J.~P.~Osborne, P.~T.~O'Brien, R.~Willingale, R.~L.~C.~Starling, D.~N.~Burrows, O.~Godet and L.~Vetere, \textit{et al.}
Mon. Not. Roy. Astron. Soc. \textbf{397}, 1177 (2009).
doi:10.1111/j.1365-2966.2009.14913.x.
[arXiv:0812.3662 [astro-ph]].

\bibitem{Song:2018vup}
C.~Y.~Song and T.~Liu,
Astrophys. J. \textbf{871} (2019) no.1, 117
doi:10.3847/1538-4357/aaf6ae
[arXiv:1812.01708 [astro-ph.HE]].

\bibitem{Savchenko2017}
V.~Savchenko, C.~Ferrigno, E.~Kuulkers, A.~Bazzano, E.~Bozzo, S.~Brandt, J.~Chenevez, T.~J.~L.~Courvoisier, R.~Diehl and A.~Domingo, \textit{et al.}
Astrophys. J. Lett. \textbf{848}, no.2, L15 (2017)
doi:10.3847/2041-8213/aa8f94
[arXiv:1710.05449 [astro-ph.HE]].

\bibitem{Liu2012}
T.~Liu, E.~W.~Liang, W.~M.~Gu, S.~J.~Hou, W.~H.~Lei, L.~Lin, Z.~G.~Dai and S.~N.~Zhang,
Astrophys. J. \textbf{760}, 63 (2012)
doi:10.1088/0004-637X/760/1/63
[arXiv:1209.4522 [astro-ph.HE]].

\bibitem{Belczynski:2006br}
K.~Belczynski, R.~Perna, T.~Bulik, V.~Kalogera, N.~Ivanova and D.~Q.~Lamb,
Astrophys. J. \textbf{648}, 1110-1116 (2006)
doi:10.1086/505169
[arXiv:astro-ph/0601458 [astro-ph]].

\bibitem{Lee:2007js}
W.~H.~Lee and E.~Ramirez-Ruiz,
New J. Phys. \textbf{9}, 17 (2007)
doi:10.1088/1367-2630/9/1/017
[arXiv:astro-ph/0701874 [astro-ph]].

\bibitem{Liu:2015zta}
T.~Liu, Y.~Q.~Lin, S.~J.~Hou and W.~M.~Gu,
Astrophys. J. \textbf{806} (2015) no.1, 58
doi:10.1088/0004-637X/806/1/58
[arXiv:1504.02156 [astro-ph.HE]].

\bibitem{Fong:2013iia}
W.~F.~Fong and E.~Berger,
Astrophys. J. \textbf{776}, 18 (2013)
doi:10.1088/0004-637X/776/1/18
[arXiv:1307.0819 [astro-ph.HE]].

\bibitem{Giacomazzo:2012zt}
B.~Giacomazzo, R.~Perna, L.~Rezzolla, E.~Troja and D.~Lazzati,
Astrophys. J. Lett. \textbf{762}, L18 (2013)
doi:10.1088/2041-8205/762/2/L18
[arXiv:1210.8152 [astro-ph.HE]].

\bibitem{Ruiz:2016rai}
M.~Ruiz, R.~N.~Lang, V.~Paschalidis and S.~L.~Shapiro,
Astrophys. J. Lett. \textbf{824}, no.1, L6 (2016)
doi:10.3847/2041-8205/824/1/L6
[arXiv:1604.02455 [astro-ph.HE]].

\bibitem{Connaughton2016}
V.~Connaughton, E.~Burns, A.~Goldstein, M.~S.~Briggs, B.~B.~Zhang, C.~M.~Hui, P.~Jenke, J.~Racusin, C.~A.~Wilson-Hodge and P.~N.~Bhat, \textit{et al.}
Astrophys. J. Lett. \textbf{826}, no.1, L6 (2016)
doi:10.3847/2041-8205/826/1/L6
[arXiv:1602.03920 [astro-ph.HE]].

\bibitem{LIGOScientific:2016vlm}
B.~P.~Abbott \textit{et al.} [LIGO Scientific and Virgo],
Phys. Rev. Lett. \textbf{116} (2016) no.24, 241102
doi:10.1103/PhysRevLett.116.241102
[arXiv:1602.03840 [gr-qc]].

\bibitem{Janiuk2017}
A.~Janiuk, M.~Bejger, P.~Sukova and S.~Charzynski,
Galaxies \textbf{5}, no.1, 15 (2017)
doi:10.3390/galaxies5010015
[arXiv:1701.07753 [astro-ph.HE]].

\bibitem{Ali:2023zva}
S.~Ali,
JHEAp \textbf{38}, 58-65 (2023)
doi:10.1016/j.jheap.2023.05.001

\bibitem{Zhang:2016rli}
B.~Zhang,
Astrophys. J. Lett. \textbf{827}, no.2, L31 (2016)
doi:10.3847/2041-8205/827/2/L31
[arXiv:1602.04542 [astro-ph.HE]].

\bibitem{Perna:2016jqh}
R.~Perna, D.~Lazzati and B.~Giacomazzo,
Astrophys. J. Lett. \textbf{821} (2016) no.1, L18
doi:10.3847/2041-8205/821/1/L18
[arXiv:1602.05140 [astro-ph.HE]].

\bibitem{Loeb:2016fzn}
A.~Loeb,
Astrophys. J. Lett. \textbf{819} (2016) no.2, L21
doi:10.3847/2041-8205/819/2/L21
[arXiv:1602.04735 [astro-ph.HE]].

\bibitem{Liu2016}
T.~Liu, G.~E.~Romero, M.~L.~Liu and A.~Li,
Astrophys. J. \textbf{826}, no.1, 82 (2016).
doi:10.3847/0004-637X/826/1/82.
[arXiv:1602.06907 [astro-ph.HE]].

\bibitem{Woosley:2016nnw}
S.~E.~Woosley,
Astrophys. J. Lett. \textbf{824} (2016) no.1, L10
doi:10.3847/2041-8205/824/1/L10
[arXiv:1603.00511 [astro-ph.HE]].

\bibitem{Narayan2001}
R.~Narayan, T.~Piran and P.~Kumar,
Astrophys. J. \textbf{557} (2001), 949
doi:10.1086/322267
[arXiv:astro-ph/0103360 [astro-ph]].

\bibitem{Kohri:2005tq}
K.~Kohri, R.~Narayan and T.~Piran,
Astrophys. J. \textbf{629} (2005), 341-361
doi:10.1086/431354
[arXiv:astro-ph/0502470 [astro-ph]].

\bibitem{Chen:2006rra}
W.~X.~Chen and A.~M.~Beloborodov,
Astrophys. J. \textbf{657} (2007), 383-399
doi:10.1086/508923
[arXiv:astro-ph/0607145 [astro-ph]].

\bibitem{Kawanaka:2007sb}
N.~Kawanaka and S.~Mineshige,
Astrophys. J. \textbf{662} (2007), 1156-1166
doi:10.1086/517985
[arXiv:astro-ph/0702630 [astro-ph]].

\bibitem{Liu:2014xya}
T.~Liu, X.~F.~Yu, W.~M.~Gu and J.~F.~Lu,
Astrophys. J. \textbf{791} (2014) no.1, 69
doi:10.1088/0004-637X/791/1/69
[arXiv:1407.0790 [astro-ph.HE]].

\bibitem{Liu:2012qca}
T.~Liu, L.~Xue, W.~M.~Gu and J.~F.~Lu,
Astrophys. J. \textbf{762} (2013), 102
doi:10.1088/0004-637X/762/2/102
[arXiv:1211.2206 [astro-ph.HE]].

\bibitem{Blandford1977}
R.~D.~Blandford and R.~L.~Znajek,
Mon. Not. Roy. Astron. Soc. \textbf{179}, 433-456 (1977).
doi:10.1093/mnras/179.3.433.

\bibitem{Khan:2018e}
A.~Khan, V.~Paschalidis, M.~Ruiz and S.~L.~Shapiro,
Phys. Rev. D \textbf{97}, no.4, 044036 (2018)
doi:10.1103/PhysRevD.97.044036
[arXiv:1801.02624 [astro-ph.HE]].

\bibitem{Proga2006}
D.~Proga and B.~Zhang,
Mon. Not. Roy. Astron. Soc. \textbf{370}, L61-L65 (2006)
doi:10.1111/j.1745-3933.2006.00189.x
[arXiv:astro-ph/0601272 [astro-ph]].

\bibitem{deSouza:2009ne}
R.~d.~de Souza and R.~Opher,
JCAP \textbf{02}, 022 (2010)
doi:10.1088/1475-7516/2010/02/022
[arXiv:0910.5258 [astro-ph.HE]].

\bibitem{Zajacek:2019kla}
M.~Zaja\v{c}ek and A.~Tursunov,
[arXiv:1904.04654 [astro-ph.GA]].

\bibitem{Perna2019}
R.~Perna, D.~Lazzati and W.~Farr,
Astrophys. J. \textbf{875} (2019) no.1, 49
doi:10.3847/1538-4357/ab107b
[arXiv:1901.04522 [astro-ph.HE]].

\bibitem{Salafia2015}
O.~S.~Salafia, G.~Ghisellini, A.~Pescalli, G.~Ghirlanda and F.~Nappo,
Mon. Not. Roy. Astron. Soc. \textbf{450} (2015) no.4, 3549-3558
doi:10.1093/mnras/stv766
[arXiv:1502.06608 [astro-ph.HE]].

\bibitem{Hayes2020}
F.~Hayes, I.~S.~Heng, J.~Veitch and D.~Williams,
Astrophys. J. \textbf{891} (2020), 124
doi:10.3847/1538-4357/ab72fc
[arXiv:1911.04190 [astro-ph.HE]].

\bibitem{Misner1973}
C.~W.~Misner, K.~S.~Thorne and J.~A.~Wheeler,
W. H. Freeman, (1973).
ISBN 978-0-7167-0344-0, 978-0-691-17779-3.

\bibitem{Qiu:2021qrt}
J.~Qiu,
Eur. Phys. J. C \textbf{81}, no.12, 1094 (2021)
doi:10.1140/epjc/s10052-021-09890-3
[arXiv:2101.03034 [gr-qc]].

\bibitem{Heims:1962sjg}
S.~P.~Heims and E.~T.~Jaynes,
Rev. Mod. Phys. \textbf{34}, 143, no. 2(1962)
doi: 10.1103/RevModPhys.34.143

\bibitem{Frail:2001qp}
D.~A.~Frail, S.~R.~Kulkarni, R.~Sari, S.~G.~Djorgovski, J.~S.~Bloom, T.~J.~Galama, D.~E.~Reichart, E.~Berger, F.~A.~Harrison and P.~A.~Price, \textit{et al.}
Astrophys. J. Lett. \textbf{562} (2001), L55
doi:10.1086/338119
[arXiv:astro-ph/0102282 [astro-ph]].

\bibitem{Janiuk2017a}
A.~Janiuk, M.~Bejger, S.~Charzy\'nski and P.~Sukova,
New Astron. \textbf{51}, 7-14 (2017)
doi:10.1016/j.newast.2016.08.002
[arXiv:1604.07132 [astro-ph.HE]].

\bibitem{Morsony:2016upv}
B.~J.~Morsony, J.~C.~Workman and D.~M.~Ryan,
Astrophys. J. Lett. \textbf{825} (2016) no.2, L24
doi:10.3847/2041-8205/825/2/L24
[arXiv:1602.05529 [astro-ph.HE]].

\bibitem{Kim:2021hhl}
J.~Kim, M.~Im, G.~S.~H.~Paek, C.~U.~Lee, S.~L.~Kim, S.~W.~Chang, C.~Choi, S.~Hwang, W.~Kang and S.~Kim, \textit{et al.}
Astrophys. J. \textbf{916} (2021) no.1, 47
doi:10.3847/1538-4357/ac0446
[arXiv:2105.14902 [astro-ph.HE]].

\bibitem{Padovani:1999mw}
P.~Padovani,
Italian Phys. Soc. Proc. \textbf{65} (1999), 159-171
[arXiv:astro-ph/9901130 [astro-ph]].

\bibitem{Granot:2002za}
J.~Granot, A.~Panaitescu, P.~Kumar and S.~E.~Woosley,
Astrophys. J. Lett. \textbf{570} (2002), L61-L64
doi:10.1086/340991
[arXiv:astro-ph/0201322 [astro-ph]].

\bibitem{LIGOScientific:2017zic}
B.~P.~Abbott \textit{et al.} 
Astrophys. J. Lett. \textbf{848}, no.2, L13 (2017)
doi:10.3847/2041-8213/aa920c
[arXiv:1710.05834 [astro-ph.HE]].
 

\bibitem{ruffini1971}
R.~Ruffini and J.~A.~Wheeler,
Phys. Today \textbf{24} (1971) no.1, 30
doi:10.1063/1.3022513

\bibitem{Weber}   
   J.~Weber,
   ``General Relativity and gravitational waves,''
   Courier Corporation. (2004).
   
\bibitem{Einstein}
   A.~Einstein, N.~Rosen, 
   ``The particle problem in the general theory of relativity,''   
   Physical Review, 48(1), 73. (1935). 
   
\bibitem{John}
    John Lighton Synge
    ``Relativity: the special theory'' 
    Prabhat Prakash, (1964).
 
\bibitem{rindler1991}
  Rindler, Wolfgang
  ''Introduction to special relativity. 2,''
  ERIC, (1991).

\bibitem{Misner:1974qy} 
  C.~W.~Misner, K.~S.~Thorne and J.~A.~Wheeler,
  ``Gravitation,''
  San Francisco, 1279 p, (1973).

\bibitem{Stephani:2004ud} 
  H.~Stephani,
  ``Relativity: An introduction to special and general relativity,''
  Cambridge, UK: Univ. Pr. 396p, (2004)

\bibitem{Gary}
  G.~T.~Horowitz,
  ''Black Holes in Higher Dimensions''
  ISBN(1107013453,9781107013452)
  Cambridge University Press, (2012)
  
\bibitem{Heinicke:2015iva} 
  C.~Heinicke and F.~W.~Hehl,
  Int.\ J.\ Mod.\ Phys.\ D {\bf 24}, no. 02, 1530006 (2014).
  doi:10.1142/S0218271815300062
  [arXiv:1503.02172 [gr-qc]].

\bibitem{Hawking:1974rv} 
  S.~W.~Hawking,
  Nature {\bf 248}, 30 (1974).
  doi:10.1038/248030a0

\bibitem{Novikov:1989sz} 
  I.~D.~Novikov and V.~P.~Frolov,
  Fundam.\ Theor.\ Phys.\  {\bf 27} (1989).
  doi:10.1007/978-94-017-2651-1

\bibitem{Visser:2007fj} 
  M.~Visser,
  ``The Kerr spacetime: A Brief introduction,''
  arXiv:0706.0622 [gr-qc].

\bibitem{Hawking:1976de} 
  S.~W.~Hawking,
  Phys.\ Rev.\ D {\bf 13}, 191 (1976).
  doi:10.1103/PhysRevD.13.191

\bibitem{Davies:1978mf} 
  P.~C.~W.~Davies,
  Proc.\ Roy.\ Soc.\ Lond.\ A {\bf 353}, 499 (1977).
  doi:10.1098/rspa.1977.0047
  
\bibitem{Hawking:1982dh} 
  S.~W.~Hawking and D.~N.~Page,
  Commun.\ Math.\ Phys.\  {\bf 87}, 577 (1983).
  doi:10.1007/BF01208266

\bibitem{Pringle1981}
J.~E.~Pringle,
Ann. Rev. Astron. Astrophys. \textbf{19}, 137-160 (1981).
doi:10.1146/annurev.aa.19.090181.001033.

\bibitem{Meegan:2009qu}
C.~Meegan, G.~Lichti, P.~N.~Bhat, E.~Bissaldi, M.~S.~Briggs, V.~Connaughton, R.~Diehl, G.~Fishman, J.~Greiner and A.~S.~Hoover, \textit{et al.}
Astrophys. J. \textbf{702}, 791-804 (2009)
doi:10.1088/0004-637X/702/1/791
[arXiv:0908.0450 [astro-ph.IM]].

\bibitem{Li:2016iw}
X.~Li, F.~W.~Zhang, Q.~Yuan, Z.~P.~Jin, Y.~Z.~Fan, S.~M.~Liu and D.~M.~Wei,
Astrophys. J. Lett. \textbf{827} (2016) no.1, L16
doi:10.3847/2041-8205/827/1/L16
[arXiv:1602.04460 [astro-ph.HE]].

\bibitem{Parker:2009uva}
L.~E.~Parker and D.~Toms,
Cambridge University Press, 2009,
ISBN 978-0-521-87787-9, 978-0-521-87787-9, 978-0-511-60155-2
doi:10.1017/CBO9780511813924

\bibitem{Hawking:1974sw} 
  S.~W.~Hawking,
  Commun.\ Math.\ Phys.\  {\bf 43}, 199 (1975)
  Erratum: [Commun.\ Math.\ Phys.\  {\bf 46}, 206 (1976)].
  doi:10.1007/BF02345020, 10.1007/BF01608497
  
\bibitem{Hawking:1971tu} 
  S.~W.~Hawking,
  Phys.\ Rev.\ Lett.\  {\bf 26}, 1344 (1971).
  doi:10.1103/PhysRevLett.26.1344
  
\bibitem{Li:2018bny} 
  C.~Li, C.~Fang, M.~He, J.~Ding, P.~Li and J.~Deng,
  arXiv:1812.02567 [hep-th].

\bibitem{Wald:1999vt}
R.~M.~Wald,
Living Rev. Rel. \textbf{4}, 6 (2001)
doi:10.12942/lrr-2001-6
[arXiv:gr-qc/9912119 [gr-qc]].

\bibitem{jacobson1996}  
  Ted Jacobson,
  ''Introductory lectures on black hole thermodynamics''
  Given at Utrecht U. in, {\bf 26},, (1996).
 
\bibitem{Hartman:2008b} 
  T.~Hartman, K.~Murata, T.~Nishioka and A.~Strominger,
  JHEP {\bf 0904}, 019 (2009)
  doi:10.1088/1126-6708/2009/04/019
  [arXiv:0811.4393 [hep-th]].

\bibitem{Bekenstein:1973ur} 
  J.~D.~Bekenstein,
  Phys.\ Rev.\ D {\bf 7}, 2333 (1973).
  doi:10.1103/PhysRevD.7.2333

\bibitem{Natsuume:2014sfa} 
  M.~Natsuume,
  Lect.\ Notes Phys.\  {\bf 903}, pp.1 (2015)
  doi:10.1007/978-4-431-55441-7
  [arXiv:1409.3575 [hep-th]].

\bibitem{Hawking:1971vc}
S.~W.~Hawking,
Commun. Math. Phys. \textbf{25}, 152-166 (1972)
doi:10.1007/BF01877517

\bibitem{Mazur:2000pn}
P.~O.~Mazur,
`Black hole uniqueness theorems,''
[arXiv:hep-th/0101012 [hep-th]].

\bibitem{Saida:2011wj}
H.~Saida,
Entropy \textbf{13}, 1611-1647 (2011)
doi:10.3390/e13091611
[arXiv:1109.0842 [gr-qc]].

\bibitem{Hawking:1994ss}
S.~W.~Hawking,
[arXiv:hep-th/9409195 [hep-th]].

\bibitem{Penrose:1999vj}
R.~Penrose,
J. Astrophys. Astron. \textbf{20}, 233-248 (1999)
doi:10.1007/BF02702355

\bibitem{Penrose:1971uk}
R.~Penrose and R.~M.~Floyd,
Nature \textbf{229}, 177-179 (1971)
doi:10.1038/physci229177a0

\bibitem{Simpson:1973ua}
M.~Simpson and R.~Penrose,
Int. J. Theor. Phys. \textbf{7}, 183-197 (1973)
doi:10.1007/BF00792069

\bibitem{Christodoulou:2014yia} 
  M~Christodoulou and C~Rovelli,
  \textit{Phys.\ Rev.\ D }{\bf 91} no. 6, 064046 (2015).
  doi:10.1103/PhysRevD.91.064046
  [arciv:1411.2854 [gr-qc]].

\bibitem{Christodoulou:2016tuua} 
  M.~Christodoulou and T.~De Lorenzo,
  Phys.\ Rev.\ D {\bf 94}, no. 10, 104002 (2016)
  doi:10.1103/PhysRevD.94.104002
  [arXiv:1604.07222 [gr-qc]].

\bibitem{Hsu:2007dr}
S.~D.~H.~Hsu and D.~Reeb,
Phys. Lett. B \textbf{658}, 244-248 (2008)
doi:10.1016/j.physletb.2007.09.021
[arXiv:0706.3239 [hep-th]].

\bibitem{Dolan:2013ft}
B.~P.~Dolan, D.~Kastor, D.~Kubiznak, R.~B.~Mann and J.~Traschen,
Phys. Rev. D \textbf{87}, no.10, 104017 (2013)
doi:10.1103/PhysRevD.87.104017
[arXiv:1301.5926 [hep-th]].

\bibitem{Wang:2021llu}
P.~Wang and F.~Yao,
Nucl. Phys. B \textbf{976}, 115715 (2022)
doi:10.1016/j.nuclphysb.2022.115715
[arXiv:2107.14640 [gr-qc]].

\bibitem{Johnson:2019wcq}
C.~V.~Johnson, V.~L.~Martin and A.~Svesko,
Phys. Rev. D \textbf{101}, no.8, 086006 (2020)
doi:10.1103/PhysRevD.101.086006
[arXiv:1911.05286 [hep-th]].

\bibitem{Estabrook:1973ue}
F.~Estabrook, H.~Wahlquist, S.~Christensen, B.~DeWitt, L.~Smarr and E.~Tsiang,
Phys. Rev. D \textbf{7}, 2814-2817 (1973)
doi:10.1103/PhysRevD.7.2814

\bibitem{Cordero-Carrion:2001jpf}
I.~Cordero-Carrion, J.~M.~Ibanez and J.~A.~Morales-Lladosa,
J. Math. Phys. \textbf{52}, 112501 (2001)
doi:10.1063/1.3658864
[arXiv:1111.2717 [gr-qc]].

\bibitem{dInverno:1992gxs}
R.~d'Inverno,
``Introducing Einstein's relativity,''
ISBN: 9780198596868, https://inspirehep.net/literature/343530

\bibitem{Padmanabhan:2002sha}
T.~Padmanabhan,
Class. Quant. Grav. \textbf{19}, 5387-5408 (2002)
doi:10.1088/0264-9381/19/21/306
[arXiv:gr-qc/0204019 [gr-qc]].

\bibitem{Parikh:2005qs} 
  M~K~Parikh,
  \textit{Phys.\ Rev.\ D} {\bf 73} 124021 (2006).
 doi:10.1103/PhysRevD.73.124021
[hep-th/0508108].

\bibitem{Mann:2015luq}
R.~B.~Mann,
Springer, 2015,
ISBN 978-3-319-14495-5, 978-3-319-14496-2
doi:10.1007/978-3-319-14496-2

\bibitem{Hawking:1973uf}
S.~W.~Hawking and G.~F.~R.~Ellis,
Cambridge University Press, 2023,
ISBN 978-1-00-925316-1, 978-1-00-925315-4, 978-0-521-20016-5, 978-0-521-09906-6, 978-0-511-82630-6, 978-0-521-09906-6
doi:10.1017/9781009253161

\bibitem{Wiltshire:2009zza}
D.~L.~Wiltshire, M.~Visser and S.~M.~Scott,
Cambridge University Press, 2009,
ISBN 978-0-521-88512-6

\bibitem{Narlikar:1986kr}
J.~V.~Narlikar and T.~Padmanabhan,
Reidel, 1986,
doi:10.1007/978-94-009-4508-1

\bibitem{Banados:1992wn}
M.~Banados, C.~Teitelboim and J.~Zanelli,
Phys. Rev. Lett. \textbf{69}, 1849-1851 (1992)
doi:10.1103/PhysRevLett.69.1849
[arXiv:hep-th/9204099 [hep-th]].

\bibitem{Carlip:1995qv}
S.~Carlip,
Class. Quant. Grav. \textbf{12}, 2853-2880 (1995)
doi:10.1088/0264-9381/12/12/005
[arXiv:gr-qc/9506079 [gr-qc]].

\bibitem{Emparan:2020znc}
R.~Emparan, A.~M.~Frassino and B.~Way,
JHEP \textbf{11}, 137 (2020)
doi:10.1007/JHEP11(2020)137
[arXiv:2007.15999 [hep-th]].

\bibitem{Gukov:2003na}
S.~Gukov,
Commun. Math. Phys. \textbf{255}, 577-627 (2005)
doi:10.1007/s00220-005-1312-y
[arXiv:hep-th/0306165 [hep-th]].

\bibitem{Ong:2015tua}
Y.~C.~Ong,
JCAP \textbf{04}, 003 (2015)
doi:10.1088/1475-7516/2015/04/003
[arXiv:1503.01092 [gr-qc]].

\bibitem{Ali:2019icq}
S.~Ali, X.~Y.~Wang and W.~B.~Liu,
Commun. Theor. Phys. \textbf{71}, no.6, 718 (2019)
doi:10.1088/0253-6102/71/6/718

\bibitem{Ali:2018sqk}
S.~Ali, X.~Y.~Wang and W.~B.~Liu,
Int. J. Mod. Phys. A \textbf{33}, no.27, 1850159 (2018)
doi:10.1142/S0217751X18501592

\bibitem{Cvetic:2010jb}
M.~Cvetic, G.~W.~Gibbons, D.~Kubiznak and C.~N.~Pope,
Phys. Rev. D \textbf{84}, 024037 (2011)
doi:10.1103/PhysRevD.84.024037
[arXiv:1012.2888 [hep-th]].

\bibitem{Grumiller:2005zk}
D.~Grumiller,
J. Phys. Conf. Ser. \textbf{33}, 361-366 (2006)
doi:10.1088/1742-6596/33/1/044
[arXiv:gr-qc/0509077 [gr-qc]].

\bibitem{Ballik:2010rx}
W.~Ballik and K.~Lake,
[arXiv:1005.1116 [gr-qc]].

\bibitem{Gibbons:2012ac}
G.~W.~Gibbons,
AIP Conf. Proc. \textbf{1460}, no.1, 90-100 (2012)
doi:10.1063/1.4733363
[arXiv:1201.2340 [gr-qc]].

\bibitem{Ballik:2013uia}
W.~Ballik and K.~Lake,
Phys. Rev. D \textbf{88}, no.10, 104038 (2013)
doi:10.1103/PhysRevD.88.104038
[arXiv:1310.1935 [gr-qc]].

\bibitem{Finch:2012vli}
T.~K.~Finch,
Gen. Rel. Grav. \textbf{47}, no.5, 56 (2015)
doi:10.1007/s10714-015-1891-7
[arXiv:1211.4337 [gr-qc]].

\bibitem{Iliesiu:2021ari}
L.~V.~Iliesiu, M.~Mezei and G.~S\'arosi,
JHEP \textbf{07}, 073 (2022)
doi:10.1007/JHEP07(2022)073
[arXiv:2107.06286 [hep-th]].

\bibitem{Chew:2020twk}
X.~Y.~Chew and Y.~C.~Ong,
Phys. Rev. D \textbf{102}, no.6, 064055 (2020)
doi:10.1103/PhysRevD.102.064055
[arXiv:2005.01312 [gr-qc]].

\bibitem{Davidson:2010xe}
A.~Davidson and I.~Gurwich,
Int. J. Mod. Phys. D \textbf{19}, 2345-2351 (2010)
doi:10.1142/S0218271810018426
[arXiv:1007.1170 [gr-qc]].

\bibitem{DiNunno:2009cuq}
B.~S.~DiNunno and R.~A.~Matzner,
Gen. Rel. Grav. \textbf{42}, 63-76 (2010)
doi:10.1007/s10714-009-0814-x
[arXiv:0801.1734 [gr-qc]].

\bibitem{Bengtsson:2015zda}
I.~Bengtsson and E.~Jakobsson,
Mod. Phys. Lett. A \textbf{30}, no.21, 1550103 (2015)
doi:10.1142/S0217732315501035
[arXiv:1502.01907 [gr-qc]].

\bibitem{Haldar:2023pcv}
A.~Haldar,
New Astron. \textbf{104}, 102082 (2023)
doi:10.1016/j.newast.2023.102082

\bibitem{Wang:2019ear}
X.~Y.~Wang and W.~B.~Liu,
Nucl. Phys. B \textbf{943}, 114614 (2019)
doi:10.1016/j.nuclphysb.2019.114614

\bibitem{Biro:2019rms}
T.~S.~Bir\'o, V.~G.~Czinner, H.~Iguchi and P.~V\'an,
Phys. Lett. B \textbf{803}, 135344 (2020)
doi:10.1016/j.physletb.2020.135344
[arXiv:1912.04547 [gr-qc]].

\bibitem{Han:2018jnf}
S.~Z.~Han, J.~Z.~Yang, X.~Y.~Wang and W.~B.~Liu,
Int. J. Theor. Phys. \textbf{57}, no.11, 3429-3435 (2018)
doi:10.1007/s10773-018-3856-6

\bibitem{Wang:2018txl}
X.~Y.~Wang, S.~Z.~Han and W.~B.~Liu,
Phys. Lett. B \textbf{787}, 64-67 (2018)
doi:10.1016/j.physletb.2018.10.033

\bibitem{Haldar:2019buj}
A.~Haldar and R.~Biswas,
EPL \textbf{128}, no.3, 30007 (2019)
doi:10.1209/0295-5075/128/30007

\bibitem{Jiang:2020rxx}
J.~Jiang and S.~Z.~Han,
Phys. Lett. B \textbf{808}, 135684 (2020)
doi:10.1016/j.physletb.2020.135684

\bibitem{Ali:2020qkb}
S.~Ali, M.~A.~Kamran and M.~U.~Khan,
Phys. Scripta \textbf{97}, no.4, 045005 (2022)
doi:10.1088/1402-4896/ac5af3
[arXiv:2012.08136 [gr-qc]].

\bibitem{Zhang:2019pzd}
M.~Zhang,
Phys. Lett. B \textbf{790}, 205-210 (2019)
doi:10.1016/j.physletb.2019.01.032
[arXiv:1901.04128 [gr-qc]].

\bibitem{Maurya:2022vjd}
S.~Maurya, S.~Gutti and R.~Nigam,
Phys. Lett. B \textbf{833}, 137381 (2022)
doi:10.1016/j.physletb.2022.137381
[arXiv:2202.09543 [gr-qc]].

\bibitem{Zhang:2020gbv}
B.~Zhang and L.~You,
Commun. Theor. Phys. \textbf{72}, no.2, 025401 (2020)
doi:10.1088/1572-9494/ab6186
[arXiv:2002.09823 [gr-qc]].

\bibitem{Zhang:2016sjy}
B.~Zhang and L.~You,
Phys. Lett. B \textbf{765}, 226-230 (2017)
doi:10.1016/j.physletb.2016.12.027
[arXiv:1612.07865 [gr-qc]].

\bibitem{Bhaumik:2016sav}
N.~Bhaumik and B.~R.~Majhi,
Int. J. Mod. Phys. A \textbf{33}, no.02, 1850011 (2018)
doi:10.1142/S0217751X18500112
[arXiv:1607.03704 [gr-qc]].

\bibitem{Egan:2009yy}
C.~A.~Egan and C.~H.~Lineweaver,
Astrophys. J. \textbf{710}, 1825-1834 (2010)
doi:10.1088/0004-637X/710/2/1825
[arXiv:0909.3983 [astro-ph.CO]].

\bibitem{Jacobson:2005kr}
T.~Jacobson, D.~Marolf and C.~Rovelli,
Int. J. Theor. Phys. \textbf{44}, 1807-1837 (2005)
doi:10.1007/s10773-005-8896-z
[arXiv:hep-th/0501103 [hep-th]].

\bibitem{Frolov:2018awz}
V.~P.~Frolov,
doi:10.1142/9789811203961\_0004
[arXiv:1805.04562 [gr-qc]].

\bibitem{Wald:2002mon}
R.~M.~Wald,
NATO Sci. Ser. II \textbf{60}, 477-522 (2002)
doi:10.1007/978-94-010-0347-6\_20

\bibitem{Srednicki:1993im}
M.~Srednicki,
Phys. Rev. Lett. \textbf{71}, 666-669 (1993)
doi:10.1103/PhysRevLett.71.666
[arXiv:hep-th/9303048 [hep-th]].

\bibitem{Frampton:2008mw}
P.~Frampton, S.~D.~H.~Hsu, D.~Reeb and T.~W.~Kephart,
Class. Quant. Grav. \textbf{26}, 145005 (2009)
doi:10.1088/0264-9381/26/14/145005
[arXiv:0801.1847 [hep-th]].

\bibitem{Bekenstein:1972tm}
J.~D.~Bekenstein,
  Lett.\ Nuovo Cim.\  {\bf 4}, 737 (1972).
  doi:10.1007/BF02757029

\bibitem{tHooft:1984kcu}
G.~'t Hooft,
Nucl. Phys. B \textbf{256}, 727-745 (1985)
doi:10.1016/0550-3213(85)90418-3

\bibitem{Hawking:1975vcx}
S.~W.~Hawking,
Commun. Math. Phys. \textbf{43}, 199-220 (1975)
[erratum: Commun. Math. Phys. \textbf{46}, 206 (1976)]
doi:10.1007/BF02345020

\bibitem{Almheiri:2020cfm}
A.~Almheiri, T.~Hartman, J.~Maldacena, E.~Shaghoulian and A.~Tajdini,
Rev. Mod. Phys. \textbf{93}, no.3, 035002 (2021)
doi:10.1103/RevModPhys.93.035002
[arXiv:2006.06872 [hep-th]].

\bibitem{Strominger:1996sh}
A.~Strominger and C.~Vafa,
Phys. Lett. B \textbf{379}, 99-104 (1996)
doi:10.1016/0370-2693(96)00345-0
[arXiv:hep-th/9601029 [hep-th]].

\bibitem{Marolf:2017jkr}
D.~Marolf,
Rept. Prog. Phys. \textbf{80}, no.9, 092001 (2017)
doi:10.1088/1361-6633/aa77cc
[arXiv:1703.02143 [gr-qc]].

\bibitem{Zhang:2015gda}
B.~Zhang,
Phys. Rev. D \textbf{92}, no.8, 081501 (2015)
doi:10.1103/PhysRevD.92.081501
[arXiv:1510.02182 [gr-qc]].

\bibitem{Hawking:1996ny}
S.~W.~Hawking and R.~Penrose,
Sci. Am. \textbf{275}, 44-49 (1996)
doi:10.1038/scientificamerican0796-60

\bibitem{Penrose:1969pc}
R.~Penrose,
Riv. Nuovo Cim. \textbf{1}, 252-276 (1969)
doi:10.1023/A:1016578408204

\bibitem{Rindler:1956yx}
W.~Rindler,
Mon. Not. Roy. Astron. Soc. \textbf{116}, 662-677 (1956)
doi:10.1023/A:1015347106729

\bibitem{Ng:1993jb}
Y.~J.~Ng and H.~Van Dam,
Mod. Phys. Lett. A \textbf{9}, 335-340 (1994)
doi:10.1142/S0217732394000356

\bibitem{Hawking:1976ra}
S.~W.~Hawking,
Phys. Rev. D \textbf{14}, 2460-2473 (1976)
doi:10.1103/PhysRevD.14.2460

\bibitem{Caticha:2005qd}
A.~Caticha,
AIP Conf. Proc. \textbf{803}, no.1, 355-365 (2005)
doi:10.1063/1.2149814
[arXiv:gr-qc/0508108 [gr-qc]].

\bibitem{Ashtekar:1984zz}
A.~Ashtekar and A.~Magnon,
Class. Quant. Grav. \textbf{1}, L39-L44 (1984)
doi:10.1088/0264-9381/1/4/002

\bibitem{Gibbons:2004uw}
G.~W.~Gibbons, H.~Lu, D.~N.~Page and C.~N.~Pope,
J. Geom. Phys. \textbf{53}, 49-73 (2005)
doi:10.1016/j.geomphys.2004.05.001
[arXiv:hep-th/0404008 [hep-th]].

\bibitem{Myers:1986un}
R.~C.~Myers and M.~J.~Perry,
Annals Phys. \textbf{172}, 304 (1986)
doi:10.1016/0003-4916(86)90186-7

\bibitem{Tangherlini:1963bw}
F.~R.~Tangherlini,
Nuovo Cim. \textbf{27}, 636-651 (1963)
doi:10.1007/BF02784569

\bibitem{Kawai:2015uya}
H.~Kawai and Y.~Yokokura,
Phys. Rev. D \textbf{93}, no.4, 044011 (2016)
doi:10.1103/PhysRevD.93.044011
[arXiv:1509.08472 [hep-th]].

\bibitem{Yang:2018arj}
J.~Z.~Yang and W.~B.~Liu,
Phys. Lett. B \textbf{782}, 372-374 (2018)
doi:10.1016/j.physletb.2018.05.050

\bibitem{Zhang:2019abv}
M.~Zhang,
Phys. Lett. B \textbf{799}, 135063 (2019)
doi:10.1016/j.physletb.2019.135063
[arXiv:1909.13629 [gr-qc]].

\bibitem{Wang:2020fgz}
X.~Y.~Wang, Y.~R.~Wang and W.~B.~Liu,
Int. J. Mod. Phys. D \textbf{29}, no.07, 2050048 (2020)
doi:10.1142/S0218271820500480

\bibitem{Wang:2018dvo}
X.~Y.~Wang, J.~Jiang and W.~B.~Liu,
Class. Quant. Grav. \textbf{35}, no.21, 215002 (2018)
doi:10.1088/1361-6382/aae276
[arXiv:1803.09649 [gr-qc]].

\bibitem{Wang:2019dpk}
X.~Y.~Wang and W.~B.~Liu,
Eur. Phys. J. C \textbf{79}, no.5, 416 (2019)
doi:10.1140/epjc/s10052-019-6936-8

\bibitem{Wang:2019ake}
X.~Y.~Wang and W.~B.~Liu,
Phys. Lett. B \textbf{788}, 464-467 (2019)
doi:10.1016/j.physletb.2018.11.056

\bibitem{Ali:2021kdu}
S.~Ali, M.~Ullah and J.~Zeb,
The International Conference on BSM: From Theory To Experiment, pp. 1-6, (2021),
Andromeda Publishers.
doi:10.31526/ACP.BSM-2021.36.

\bibitem{Ali:2020olc}
S.~Ali, P.~Wen and W.~B.~Liu,
Int. J. Theor. Phys. \textbf{59}, no.4, 1206-1213 (2020)
doi:10.1007/s10773-020-04400-9

\bibitem{Wen:2020thi}
P.~Wen, X.~Y.~Wang and W.~B.~Liu,
Int. J. Mod. Phys. A \textbf{35}, no.30, 2050194 (2020)
doi:10.1142/S0217751X20501948

\bibitem{landsberg:1989} 
  T~P~Landsberg and A~De Vos,
 \textit{Journal of Physics A: Mathematical and General}, \textbf{22} 8, pp.1073, (1989).
 doi:10.1088/0305-4470/22/8/021.  

\bibitem{Zhang:2015}
  B.~Zhang,
  Phys.\ Rev.\ D {\bf 92}, no. 8, 081501 (2015)
  doi:10.1103/PhysRevD.92.081501
  [arXiv:1510.02182 [gr-qc]].

\bibitem{Kerr:2007dk}
R.~P.~Kerr,
``Discovering the Kerr and Kerr-Schild metrics,''
[arXiv:0706.1109 [gr-qc]].

\bibitem{Banados:1992gq}
M.~Banados, M.~Henneaux, C.~Teitelboim and J.~Zanelli,
Phys. Rev. D \textbf{48}, 1506-1525 (1993)
[erratum: Phys. Rev. D \textbf{88}, 069902 (2013)]
doi:10.1103/PhysRevD.48.1506
[arXiv:gr-qc/9302012 [gr-qc]].

\bibitem{Birmingham:2001dt}
D.~Birmingham, I.~Sachs and S.~Sen,
Int. J. Mod. Phys. D \textbf{10}, 833-858 (2001)
doi:10.1142/S0218271801001207
[arXiv:hep-th/0102155 [hep-th]].

\bibitem{Hossenfelder:2012jw}
S.~Hossenfelder,
Living Rev. Rel. \textbf{16}, 2 (2013)
doi:10.12942/lrr-2013-2
[arXiv:1203.6191 [gr-qc]].

\bibitem{Li:2002xb}
X.~Li,
Phys. Lett. B \textbf{540}, 9-13 (2002)
doi:10.1016/S0370-2693(02)02123-8
[arXiv:gr-qc/0204029 [gr-qc]].

\bibitem{Hawking:2005kf}
S.~W.~Hawking,
Phys. Rev. D \textbf{72}, 084013 (2005)
doi:10.1103/PhysRevD.72.084013
[arXiv:hep-th/0507171 [hep-th]].

\bibitem{Michel:2014zsa}
F.~Michel and R.~Parentani,
Phys. Rev. D \textbf{90}, no.4, 044033 (2014)
doi:10.1103/PhysRevD.90.044033
[arXiv:1404.7482 [gr-qc]].

\bibitem{Shi:2021nkx}
Y.~H.~Shi, R.~Q.~Yang, Z.~Xiang, Z.~Y.~Ge, H.~Li, Y.~Y.~Wang, K.~Huang, Y.~Tian, X.~Song and D.~Zheng, \textit{et al.}
Nature Commun. \textbf{14}, no.1, 3263 (2023)
doi:10.1038/s41467-023-39064-6
[arXiv:2111.11092 [quant-ph]].

\bibitem{Robertson:2012ku}
S.~J.~Robertson,
J. Phys. B \textbf{45}, 163001 (2012)
doi:10.1088/0953-4075/45/16/163001
[arXiv:1508.02569 [gr-qc]].

\bibitem{Unruh:1976db}
W.~G.~Unruh,
Phys. Rev. D \textbf{14}, 870 (1976)
doi:10.1103/PhysRevD.14.870

\bibitem{Wald:1975kc}
R.~M.~Wald,
Commun. Math. Phys. \textbf{45}, 9-34 (1975)
doi:10.1007/BF01609863

\bibitem{MunozdeNova:2018fxv}
J.~R.~Mu\~noz de Nova, K.~Golubkov, V.~I.~Kolobov and J.~Steinhauer,
Nature \textbf{569}, no.7758, 688-691 (2019)
doi:10.1038/s41586-019-1241-0
[arXiv:1809.00913 [gr-qc]].

\bibitem{Brout:1995wp}
R.~Brout, S.~Massar, R.~Parentani and P.~Spindel,
Phys. Rev. D \textbf{52}, 4559-4568 (1995)
doi:10.1103/PhysRevD.52.4559
[arXiv:hep-th/9506121 [hep-th]].

\bibitem{Vanzo:2011wq}
L.~Vanzo, G.~Acquaviva and R.~Di Criscienzo,
Class. Quant. Grav. \textbf{28}, 183001 (2011)
doi:10.1088/0264-9381/28/18/183001
[arXiv:1106.4153 [gr-qc]].

\bibitem{Deng:2016qua}
Y.~Deng and G.~Cleaver,
Int. J. Theor. Phys. \textbf{56}, no.3, 741-750 (2017)
doi:10.1007/s10773-016-3215-4
[arXiv:1602.06035 [gr-qc]].

\bibitem{Kraus:1994fh}
P.~Kraus and F.~Wilczek,
Mod. Phys. Lett. A \textbf{9}, 3713-3719 (1994)
doi:10.1142/S0217732394003567
[arXiv:gr-qc/9406042 [gr-qc]].

\bibitem{Iso:2006ut}
S.~Iso, H.~Umetsu and F.~Wilczek,
Phys. Rev. D \textbf{74}, 044017 (2006)
doi:10.1103/PhysRevD.74.044017
[arXiv:hep-th/0606018 [hep-th]].

\bibitem{Sakalli:2014sea}
I.~Sakalli and A.~Ovgun,
EPL \textbf{110}, no.1, 10008 (2015)
doi:10.1209/0295-5075/110/10008
[arXiv:1409.5539 [gr-qc]].

\bibitem{Parikh:1999mf}
M.~K.~Parikh and F.~Wilczek,
Phys. Rev. Lett. \textbf{85}, 5042-5045 (2000)
doi:10.1103/PhysRevLett.85.5042
[arXiv:hep-th/9907001 [hep-th]].

\bibitem{Li:2020qqa}
G.~Q.~Li and Y.~W.~Zhuang,
Turk. J. Phys. \textbf{44}, no.5, 458-464 (2020)
doi:10.3906/fiz-2005-1

\bibitem{Sakalli:2015raa}
I.~Sakalli and H.~Gursel,
Eur. Phys. J. C \textbf{76}, no.6, 318 (2016)
doi:10.1140/epjc/s10052-016-4158-x
[arXiv:1505.01990 [gr-qc]].

\bibitem{Chen:2014jwq}
P.~Chen, Y.~C.~Ong and D.~h.~Yeom,
Phys. Rept. \textbf{603}, 1-45 (2015)
doi:10.1016/j.physrep.2015.10.007
[arXiv:1412.8366 [gr-qc]].

\bibitem{Witten:1998qj}
E.~Witten,
Adv. Theor. Math. Phys. \textbf{2}, 253-291 (1998)
doi:10.4310/ATMP.1998.v2.n2.a2
[arXiv:hep-th/9802150 [hep-th]].

\bibitem{Sotiriou:2008rp}
T.~P.~Sotiriou and V.~Faraoni,
Rev. Mod. Phys. \textbf{82}, 451-497 (2010)
doi:10.1103/RevModPhys.82.451
[arXiv:0805.1726 [gr-qc]].

\bibitem{Witten:1998zw}
E.~Witten,
Adv. Theor. Math. Phys. \textbf{2}, 505-532 (1998)
doi:10.4310/ATMP.1998.v2.n3.a3
[arXiv:hep-th/9803131 [hep-th]].
 
\bibitem{Capozziello:2011et}
S.~Capozziello and M.~De Laurentis,
Phys. Rept. \textbf{509}, 167-321 (2011)
doi:10.1016/j.physrep.2011.09.003
[arXiv:1108.6266 [gr-qc]].
 
\bibitem{Shaikh:2023wsc}
A.~Y.~Shaikh,
Chin. J. Phys. \textbf{86}, 628-646 (2023)
doi:10.1016/j.cjph.2023.09.020
 
\bibitem{Nojiri:2008nt}
S.~Nojiri and S.~D.~Odintsov,
TSPU Bulletin \textbf{N8(110)}, 7-19 (2011)
[arXiv:0807.0685 [hep-th]].
 
\bibitem{Stelle:1976gc}
K.~S.~Stelle,
Phys. Rev. D \textbf{16}, 953-969 (1977)
doi:10.1103/PhysRevD.16.953
 
\bibitem{Utiyama:1962sn}
R.~Utiyama and B.~S.~DeWitt,
J. Math. Phys. \textbf{3}, 608-618 (1962)
doi:10.1063/1.1724264
 
\bibitem{CANTATA:2021ktz}
E.~N.~Saridakis \textit{et al.} [CANTATA],
Springer, 2021,
ISBN 978-3-030-83714-3, 978-3-030-83717-4, 978-3-030-83715-0,
doi:10.1007/978-3-030-83715-0,
[arXiv:2105.12582 [gr-qc]].
 
\bibitem{Faraoni:2008mf}
V.~Faraoni,
[arXiv:0810.2602 [gr-qc]].
 
\bibitem{delaCruz-Dombriz:2006kob}
A.~de la Cruz-Dombriz and A.~Dobado,
Phys. Rev. D \textbf{74}, 087501 (2006)
doi:10.1103/PhysRevD.74.087501,
[arXiv:gr-qc/0607118 [gr-qc]].
 
\bibitem{Dunsby:2010wg}
P.~K.~S.~Dunsby, E.~Elizalde, R.~Goswami, S.~Odintsov and D.~S.~Gomez,
Phys. Rev. D \textbf{82}, 023519 (2010)
doi:10.1103/PhysRevD.82.023519
[arXiv:1005.2205 [gr-qc]].
 
\bibitem{Nojiri:2019dio}
S.~Nojiri, S.~D.~Odintsov and V.~K.~Oikonomou,
Phys. Dark Univ. \textbf{28}, 100541 (2020)
doi:10.1016/j.dark.2020.100541
[arXiv:1911.07329 [gr-qc]].
 
\bibitem{Nojiri:2021mxf}
S.~Nojiri, S.~D.~Odintsov, V.~K.~Oikonomou and A.~A.~Popov,
Nucl. Phys. B \textbf{973}, 115617 (2021)
doi:10.1016/j.nuclphysb.2021.115617
[arXiv:2111.09457 [gr-qc]].
 
\bibitem{Khoury:2003aq}
J.~Khoury and A.~Weltman,
Phys. Rev. Lett. \textbf{93}, 171104 (2004)
doi:10.1103/PhysRevLett.93.171104
[arXiv:astro-ph/0309300 [astro-ph]].
 
\bibitem{Sokolowski:2008kf}
L.~M.~Sokolowski,
Acta Phys. Polon. B \textbf{39}, 2879-2901 (2008)
[arXiv:0810.2554 [gr-qc]].
 
\bibitem{Appleby:2009uf}
S.~A.~Appleby, R.~A.~Battye and A.~A.~Starobinsky,
JCAP \textbf{06}, 005 (2010)
doi:10.1088/1475-7516/2010/06/005
[arXiv:0909.1737 [astro-ph.CO]].
 
\bibitem{Song:2006ej}
Y.~S.~Song, W.~Hu and I.~Sawicki,
Phys. Rev. D \textbf{75}, 044004 (2007)
doi:10.1103/PhysRevD.75.044004
[arXiv:astro-ph/0610532 [astro-ph]].
 
\bibitem{Martinelli:2021hir}
M.~Martinelli and S.~Casas,
Universe \textbf{7}, no.12, 506 (2021)
doi:10.3390/universe7120506
[arXiv:2112.10675 [astro-ph.CO]].
 
\bibitem{deMartino:2015zsa}
I.~de Martino, M.~De Laurentis and S.~Capozziello,
Universe \textbf{1}, no.2, 123-157 (2015)
doi:10.3390/universe1020123
[arXiv:1507.06123 [gr-qc]].
 
\bibitem{Capozziello:2011gw}
S.~Capozziello and S.~Vignolo,
Int. J. Geom. Meth. Mod. Phys. \textbf{9}, 1250006 (2012)
doi:10.1142/S0219887812500065
[arXiv:1103.2302 [gr-qc]].
 
\bibitem{Lanahan-Tremblay:2007sxd}
N.~Lanahan-Tremblay and V.~Faraoni,
Class. Quant. Grav. \textbf{24}, 5667-5680 (2007)
doi:10.1088/0264-9381/24/22/024
[arXiv:0709.4414 [gr-qc]].
 
\bibitem{Schmidt:2006jt}
H.~J.~Schmidt,
eConf \textbf{C0602061}, 12 (2006)
doi:10.1142/S0219887807001977
[arXiv:gr-qc/0602017 [gr-qc]].

\bibitem{Motohashi:2018wdq}
H.~Motohashi and M.~Minamitsuji,
Phys. Lett. B \textbf{781}, 728-734 (2018)
doi:10.1016/j.physletb.2018.04.041
[arXiv:1804.01731 [gr-qc]].
 
\bibitem{Corda:2009re}
C.~Corda,
Int. J. Mod. Phys. D \textbf{18}, 2275-2282 (2009)
doi:10.1142/S0218271809015904
[arXiv:0905.2502 [gr-qc]].
 
\bibitem{LIGOScientific:2017ync}
B.~P.~Abbott \textit{et al.}, 
Astrophys. J. Lett. \textbf{848}, no.2, L12 (2017)
doi:10.3847/2041-8213/aa91c9
[arXiv:1710.05833 [astro-ph.HE]].

\bibitem{DeLaurentis:2016jfs}
M.~De Laurentis, O.~Porth, L.~Bovard, B.~Ahmedov, and A.~Abdujabbarov,
Phys. Rev. D \textbf{94}, no.12, 124038 (2016)
doi:10.1103/PhysRevD.94.124038
[arXiv:1611.05766 [gr-qc]].

\bibitem{Chamblin:1999tk} 
  A.~Chamblin, R.~Emparan, C.~V.~Johnson and R.~C.~Myers,
  Phys.\ Rev.\ D {\bf 60}, 064018 (1999)
  doi:10.1103/PhysRevD.60.064018
  [hep-th/9902170].
 
\bibitem{Chamblin:1999hg} 
  A.~Chamblin, R.~Emparan, C.~V.~Johnson and R.~C.~Myers,
  Phys.\ Rev.\ D {\bf 60}, 104026 (1999)
  doi:10.1103/PhysRevD.60.104026
  [hep-th/9904197].

\bibitem{Kubiznak:2012wp} 
  D.~Kubiznak and R.~B.~Mann,
  JHEP {\bf 1207}, 033 (2012)
  doi:10.1007/JHEP07(2012)033
  [arXiv:1205.0559 [hep-th]].
  
\bibitem{Gunasekaran:2012dq} 
  S.~Gunasekaran, R.~B.~Mann and D.~Kubiznak,
  JHEP {\bf 1211}, 110 (2012)
  doi:10.1007/JHEP11(2012)110
  [arXiv:1208.6251 [hep-th]].
  
\bibitem{Altamirano:2013ane} 
  N.~Altamirano, D.~Kubiznak and R.~B.~Mann,
  Phys.\ Rev.\ D {\bf 88}, no. 10, 101502 (2013)
  doi:10.1103/PhysRevD.88.101502
  [arXiv:1306.5756 [hep-th]].
  
\bibitem{Wei:2014hba} 
  S.~W.~Wei and Y.~X.~Liu,
  Phys.\ Rev.\ D {\bf 90}, no. 4, 044057 (2014)
  doi:10.1103/PhysRevD.90.044057
  [arXiv:1402.2837 [hep-th]].
  
\bibitem{Frassino:2014pha} 
  A.~M.~Frassino, D.~Kubiznak, R.~B.~Mann and F.~Simovic,
  JHEP {\bf 1409}, 080 (2014)
  doi:10.1007/JHEP09(2014)080
  [arXiv:1406.7015 [hep-th]].
  
\bibitem{Altamirano:2013uqa} 
  N.~Altamirano, D.~Kubizňák, R.~B.~Mann and Z.~Sherkatghanad,
  Class.\ Quant.\ Grav.  {\bf 31}, 042001 (2014)
  doi:10.1088/0264-9381/31/4/042001
  [arXiv:1308.2672 [hep-th]].
  
\bibitem{Belhaj:2012bg} 
  A.~Belhaj, M.~Chabab, H.~El Moumni and M.~B.~Sedra,
  Chin.\ Phys.\ Lett.  {\bf 29}, 100401 (2012)
  doi:10.1088/0256-307X/29/10/100401
  [arXiv:1210.4617 [hep-th]].
  
\bibitem{Hendi:2012um} 
  S.~H.~Hendi and M.~H.~Vahidinia,
  Phys.\ Rev.\ D {\bf 88}, no. 8, 084045 (2013)
  doi:10.1103/PhysRevD.88.084045
  [arXiv:1212.6128 [hep-th]].
  
\bibitem{Altamirano:2014tva} 
  N.~Altamirano, D.~Kubiznak, R.~B.~Mann and Z.~Sherkatghanad,
  Galaxies {\bf 2}, 89 (2014)
  doi:10.3390/galaxies2010089
  [arXiv:1401.2586 [hep-th]].
  
\bibitem{Dolan:2014jva} 
  B.~P.~Dolan,
  Mod.\ Phys.\ Lett.\ A {\bf 30}, no. 03n04, 1540002 (2015)
  doi:10.1142/S0217732315400027
  [arXiv:1408.4023 [gr-qc]].
  
\bibitem{Bhattacharya:2017hfj}
K.~Bhattacharya and B.~R.~Majhi,
Phys. Rev. D \textbf{95}, no.10, 104024 (2017)
doi:10.1103/PhysRevD.95.104024
[arXiv:1702.07174 [gr-qc]].

\bibitem{Majhi:2016txt}
B.~R.~Majhi and S.~Samanta,
Phys. Lett. B \textbf{773}, 203-207 (2017)
doi:10.1016/j.physletb.2017.08.038
[arXiv:1609.06224 [gr-qc]].

\bibitem{Ghodrati:2020mtx}
M.~Ghodrati and D.~Gregoris,
Int. J. Mod. Phys. A \textbf{37}, no.34, 2250202 (2022)
doi:10.1142/S0217751X22502025
[arXiv:2003.04412 [hep-th]].

\bibitem{Chougule:2018cny}
S.~Chougule, S.~Dey, B.~Pourhassan and M.~Faizal,
Eur. Phys. J. C \textbf{78}, no.8, 685 (2018)
doi:10.1140/epjc/s10052-018-6172-7
[arXiv:1809.00868 [gr-qc]].

\bibitem{Upadhyay:2019hyw}
S.~Upadhyay, Nadeem-ul-islam and P.~A.~Ganai,
JHAP \textbf{2}, no.1, 25-48 (2022)
doi:10.22128/jhap.2021.454.1004
[arXiv:1912.00767 [gr-qc]].

\bibitem{Bousso:2002ju}
R.~Bousso,
Rev. Mod. Phys. \textbf{74}, 825-874 (2002)
doi:10.1103/RevModPhys.74.825
[arXiv:hep-th/0203101 [hep-th]].

\bibitem{Abdalla:2001as}
E.~Abdalla and L.~A.~Correa-Borbonet,
Phys. Rev. D \textbf{65}, 124011 (2002)
doi:10.1103/PhysRevD.65.124011
[arXiv:hep-th/0109129 [hep-th]].

\bibitem{Ramallo:2013bua}
A.~V.~Ramallo,
Springer Proc. Phys. \textbf{161}, 411-474 (2015)
doi:10.1007/978-3-319-12238-0\_10
[arXiv:1310.4319 [hep-th]].

\bibitem{Hubeny:2014bla}
V.~E.~Hubeny,
Class. Quant. Grav. \textbf{32}, no.12, 124010 (2015)
doi:10.1088/0264-9381/32/12/124010
[arXiv:1501.00007 [gr-qc]].

\bibitem{Gleiser:2013mga}
M.~Gleiser and D.~Sowinski,
Phys. Lett. B \textbf{727}, 272-275 (2013)
doi:10.1016/j.physletb.2013.10.005
[arXiv:1307.0530 [hep-th]].

\bibitem{Gleiser:2015rwa}
M.~Gleiser and N.~Jiang,
Phys. Rev. D \textbf{92}, no.4, 044046 (2015)
doi:10.1103/PhysRevD.92.044046
[arXiv:1506.05722 [gr-qc]].

\bibitem{Casadio:2016aum}
R.~Casadio and R.~da Rocha,
Phys. Lett. B \textbf{763}, 434-438 (2016)
doi:10.1016/j.physletb.2016.10.072
[arXiv:1610.01572 [hep-th]].

\bibitem{Lee:2017ero} 
  C.~O.~Lee,
  Phys.\ Lett.\ B {\bf 772}, 471 (2017)
  doi:10.1016/j.physletb.2017.07.013
  [arXiv:1705.09047 [gr-qc]].

\bibitem{Sadeghi:2016dvc}
J.~Sadeghi, B.~Pourhassan and M.~Rostami,
Phys. Rev. D \textbf{94}, no.6, 064006 (2016)
doi:10.1103/PhysRevD.94.064006
[arXiv:1605.03458 [gr-qc]].

\bibitem{shannon:1948}
  C.~E.~Shannon,
  Bell System Technical Journal 27.3 (1948): 379-423. 
  Wiley Online Lib.
  doi:10.1145/584091.584093

\bibitem{Wei:2015iwa} 
  S.~W.~Wei and Y.~X.~Liu,
  Phys.\ Rev.\ Lett. \textbf{115}, no. 11, 111302 (2015)
  doi:10.1103/PhysRevLett.116.169903, 10.1103/PhysRevLett.115.111302
  [arXiv:1502.00386 [gr-qc]].
  Erratum: [Phys.\ Rev.\ Lett.\  {\bf 116}, no. 16, 169903 (2016)]
  
\bibitem{Shah:2023vxf}
N.~A.~Shah, A.~A.~Naqash, A.~K.~Khan, R.~F.~Shah and S.~A.~Lone,
JHAP \textbf{3}, no.2, 17-30 (2023)
doi:10.22128/jhap.2023.565.1028

\bibitem{Pessoa:2022fwo}
P.~Pessoa, B.~Arderucio Costa and S.~Press\'e,
[arXiv:2210.00324 [gr-qc]].

\bibitem{Casadio:2022pla}
R.~Casadio, R.~da Rocha, P.~Meert, L.~Tabarroni and W.~Barreto,
Class. Quant. Grav. \textbf{40}, no.7, 075014 (2023)
doi:10.1088/1361-6382/acbe89
[arXiv:2206.10398 [gr-qc]].

\bibitem{Correa:2015vka}
R.~A.~C.~Correa and R.~da Rocha,
Eur. Phys. J. C \textbf{75}, no.11, 522 (2015)
doi:10.1140/epjc/s10052-015-3735-8
[arXiv:1502.02283 [hep-th]].
  
\bibitem{Barreto:2022ohl}
W.~Barreto and R.~da Rocha,
Phys. Rev. D \textbf{105}, no.6, 064049 (2022)
doi:10.1103/PhysRevD.105.064049
[arXiv:2201.08324 [hep-th]].

\bibitem{daRocha:2021jzn}
R.~da Rocha,
Phys. Lett. B \textbf{823}, 136729 (2021)
doi:10.1016/j.physletb.2021.136729
[arXiv:2108.13484 [gr-qc]].

\bibitem{Braga:2016wzx} 
N. ~R. ~F. ~Braga and R. ~da Rocha, 
Phys. Lett. B \textbf{767}, 386-391 (2017), 
doi:10.1016/j.physletb.2017.02.031, 
[arXiv:1612.03289 [hep-th]]. 

\bibitem{Niu:2011tb}
C.~Niu, Y.~Tian and X.~N.~Wu,
Phys. Rev. D \textbf{85}, 024017 (2012)
doi:10.1103/PhysRevD.85.024017
[arXiv:1104.3066 [hep-th]].

\bibitem{Cai:2007wz}
R.~G.~Cai, S.~P.~Kim and B.~Wang,
Phys. Rev. D \textbf{76}, 024011 (2007)
doi:10.1103/PhysRevD.76.024011
[arXiv:0705.2469 [hep-th]].

\bibitem{Cai:2007vv}
R.~G.~Cai, L.~M.~Cao and Y.~W.~Sun,
JHEP \textbf{11}, 039 (2007)
doi:10.1088/1126-6708/2007/11/039
[arXiv:0709.3568 [hep-th]].

\bibitem{Eune:2013qs}
M.~Eune, W.~Kim and S.~H.~Yi,
JHEP \textbf{03}, 020 (2013)
doi:10.1007/JHEP03(2013)020
[arXiv:1301.0395 [gr-qc]].

\bibitem{Chen:2013ce} 
  S.~Chen, X.~Liu, C.~Liu and J.~Jing,
  Chin.\ Phys.\ Lett.\  {\bf 30}, 060401 (2013)
  doi:10.1088/0256-307X/30/6/060401
  [arXiv:1301.3234 [gr-qc]].

\bibitem{Mo:2016sel} 
  J.~X.~Mo and G.~Q.~Li,
  Phys.\ Rev.\ D {\bf 92}, no. 2, 024055 (2015)
  doi:10.1103/PhysRevD.92.024055
  [arXiv:1604.07931 [gr-qc]].

\bibitem{Wei:2019gur}
S.~W.~Wei and Y.~X.~Liu,
Phys. Rev. D \textbf{100}, no.6, 064004 (2019)
doi:10.1103/PhysRevD.100.064004
[arXiv:1905.12187 [gr-qc]].

\bibitem{Wei:2014qwa}
S.~W.~Wei and Y.~X.~Liu,
Phys. Rev. D \textbf{91}, no.4, 044018 (2015)
doi:10.1103/PhysRevD.91.044018
[arXiv:1411.5749 [hep-th]].

\bibitem{Nojiri:2003ft}
S.~Nojiri and S.~D.~Odintsov,
Phys. Rev. D \textbf{68}, 123512 (2003)
doi:10.1103/PhysRevD.68.123512
[arXiv:hep-th/0307288 [hep-th]].

\bibitem{Atazadeh:2008zj}
K.~Atazadeh, M.~Farhoudi and H.~R.~Sepangi,
Phys. Lett. B \textbf{660}, 275-281 (2008)
doi:10.1016/j.physletb.2007.12.057
[arXiv:0801.1398 [gr-qc]].

\bibitem{Balasubramanian:2001nb}
V.~Balasubramanian, J.~de Boer and D.~Minic,
Phys. Rev. D \textbf{65}, 123508 (2002)
doi:10.1103/PhysRevD.65.123508
[arXiv:hep-th/0110108 [hep-th]].

\bibitem{Moon:2011hq} 
  T.~Moon, Y.~S.~Myung and E.~J.~Son,
  Gen.\ Rel.\ Grav.  {\bf 43}, 3079 (2011)
  doi:10.1007/s10714-011-1225-3
  [arXiv:1101.1153 [gr-qc]].

\bibitem{Ovgun:2017bgx}
A.~\"Ovg\"un,
Adv. High Energy Phys. \textbf{2018}, 8153721 (2018)
doi:10.1155/2018/8153721
[arXiv:1710.06795 [gr-qc]].

\bibitem{Khan:2020zpz}
Y.~H.~Khan, P.~A.~Ganai and S.~Upadhyay,
PTEP \textbf{2020}, no.10, 103 (2020)
doi:10.1093/ptep/ptaa135
[arXiv:2008.11012 [gr-qc]].

\bibitem{Upadhyay:2018bqy} 
S. ~Upadhyay, S. ~Soroushfar and R. ~Saffari, 
Mod. Phys. Lett. A \textbf{36}, no.29, 2150212 (2021) 
doi:10.1142/S0217732321502126 
[arXiv:1801.09574 [gr-qc]]. 

\bibitem{Upadhyay:2018gee}
S.~Upadhyay,
Gen. Rel. Grav. \textbf{50}, no.10, 128 (2018)
doi:10.1007/s10714-018-2459-0
[arXiv:1810.01283 [gr-qc]].


\end{thebibliography}
\end{document}